\documentclass[aps,prx,twocolumn,superscriptaddress,longbibliography]{revtex4-2}

\usepackage{amsmath}
\usepackage{graphicx}
\usepackage{amsfonts}
\usepackage{amssymb}
\usepackage[usenames, dvipsnames, x11names]{xcolor}
\usepackage{dsfont}
\usepackage{amsmath} % or simply amstext
\usepackage{xcolor}
\usepackage{hyperref}
\usepackage{soul}

\newcommand{\aop}{\hat{a}}
\newcommand{\aopd}{\hat{a}^\dagger}

\newcommand{\sigmamu}{\hat{\sigma}^\mu}
\newcommand{\sigmanu}{\hat{\sigma}^\nu}

\newcommand{\sigmaz}{\sigma^z}

\newcommand{\Hint}{\hat{H}_{\mathrm{int}}}
\newcommand{\Hdr}{\hat{H}_{\mathrm{mod}}}
\newcommand{\hata}{\hat{a}}
\newcommand{\omegakl}{\omega_{k,l}}
\newcommand{\omegalk}{\omega_{l,k}}
\newcommand{\varepsilonk}{\varepsilon_{k}}
\newcommand{\Usq}{\hat{U}_{1,\mathrm{q}}}

\newcommand{\Deltakljpm}{\Delta_{kl,\pm}^{(j)}}

\newcommand{\tramp}{t_{\mathrm{ramp}}}

\newcommand{\Ugq}{\hat{U}_{\mathrm{G,q}}}

\newcommand{\rmd}{\mathrm{d}}
\newcommand{\Herr}{\hat{H}_{\mathrm{err}}}

\newcommand{\Hstirap}{\hat{H}_{\mathrm{0}}}

\newcommand{\Deltakl}{\Delta_{kl,\sigma}^{(j)}}

\newcommand{\gdc}{g^{\mathrm{dc}}}
\newcommand{\gac}{g^{\mathrm{ac}}}

\newcommand{\omegamodj}{\omega_{\mathrm{mod},j}}

\newcommand{\omegatmodj}{\tilde{\omega}_{\mathrm{mod},j}}
\hypersetup{colorlinks = true, urlcolor = blue, linkcolor = blue, citecolor = blue}

\begin{document}
%%%%%%%%%%%%%%%%%%%%%%%%%%%%
\title{Fast and Robust Geometric Two-Qubit Gates for Superconducting Qubits and beyond}

\author{F. Setiawan}\email{setiawan@uchicago.edu}\email{Current Affiliation: Riverlane Research Inc., Cambridge, MA 02142, USA}
\affiliation{Pritzker School of Molecular Engineering, University of Chicago, 5640 South Ellis Avenue, Chicago, Illinois 60637, USA}
\author{Peter Groszkowski}\email{Current Affiliation: National Center for Computational Sciences, Oak Ridge National Laboratory, TN 37831, USA}
\affiliation{Pritzker School of Molecular Engineering, University of Chicago, 5640 South Ellis Avenue, Chicago, Illinois 60637, USA}
\author{Aashish A. Clerk}
\affiliation{Pritzker School of Molecular Engineering, University of Chicago, 5640 South Ellis Avenue, Chicago, Illinois 60637, USA}
\date{\today}

\begin{abstract}
    Quantum protocols based on adiabatic evolution are remarkably robust against imperfections of control pulses and system uncertainties. While adiabatic protocols have been successfully implemented for quantum operations such as quantum state transfer and single-qubit gates, their use for geometric two-qubit gates remains a challenge. In this paper, we propose a general scheme to realize robust geometric two-qubit gates in multilevel qubit systems where the interaction between the qubits is mediated by an auxiliary system (such as a bus or coupler). While our scheme utilizes Stimulated Raman Adiabatic Passage (STIRAP), it is substantially simpler than STIRAP-based gates that have been proposed for atomic platforms, requiring fewer control tones and ancillary states, as well as utilizing only a generic dispersive interaction. We also show how our gate can be accelerated using a shortcuts-to-adiabaticity approach, 
    allowing one to achieve a gate that is both fast and relatively robust.  We present a comprehensive theoretical analysis of the performance of our two-qubit gate in a parametrically-modulated superconducting circuits comprising two fluxonium qubits coupled to an auxiliary system.
\end{abstract}

\maketitle
\section{Introduction}

Geometric quantum gates~\cite{Zanardi1999,Pachos1999,Unayan1999Laser,duan2001geometric, Moller2008Quantum,Kis2002Qubit,Faoro2003Non,Solinas2003Holonomic,frees2019adiabatic,Zeng2019Geometric,Dridi2020Optimal,Laforgue2022Optimal,Laforgue2022Optimalb} are robust against a range of parameter uncertainties and pulse imperfections. One powerful technique that allows one to construct such geometric gates is adiabatic evolution.
While single-qubit gates based on adiabatic evolution have been implemented in a number of platforms, including quantum dots~\cite{Wu2013Geometric}, trapped ions~\cite{Toyoda2013Realization} and nitrogen-vacancy centers in diamond~\cite{Huang2019Experimental}, the implementation of two-qubit gates based on geometric phases and adiabatic evolution remains a challenge.  A powerful adiabatic protocol well suited for operations on isolated qubit levels is stimulated Raman adiabatic passage (STIRAP)~\cite{Vitanov2017Stimulated}.  While typically used for state transfer, STIRAP can also be exploited for single-qubit gates \cite{duan2001geometric,Kis2002Qubit,Ribeiro2019Accelerated,Setiawan2021Analytic}.  
STIRAP has also been proposed as a way of realizing geometric two-qubit gates in atomic systems, both for trapped ions~\cite{duan2001geometric} and Rydberg atoms~\cite{Moller2008Quantum}.  These protocols however use extremely specific kinds of qubit-qubit interactions (e.g., phonon-sideband processes in Ref.~\cite{duan2001geometric} and the Rydberg blockade in Ref.~\cite{Moller2008Quantum}) as well as large numbers of control fields and ancillary states, making them ill suited for other kinds of qubit platforms such as superconducting circuits.

\begin{figure}[t!]
\centering
\includegraphics[width=\linewidth]{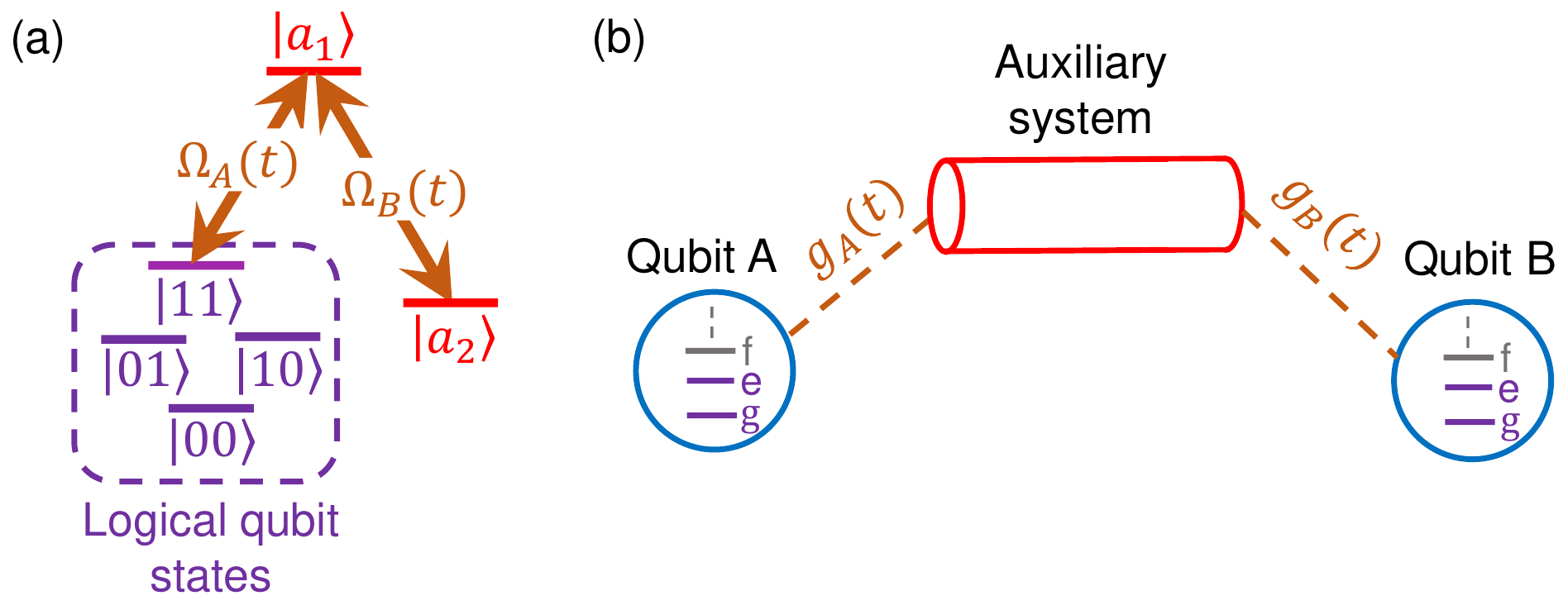}
\caption{
(a) $\Lambda$-system formed by the logical qubit state $|11\rangle$ and ancillary states $|a_1\rangle$ and $|a_2\rangle$.  Depicted transitions are realized via parametrically modulated couplings with envelopes $\Omega_A(t)$ and $\Omega_B(t)$.  A double-STIRAP protocol in this system generates a geometric phase, forming the basis of a two-qubit gate. (b) Schematic setup for implementing our STIRAP-based gate: qubits $A$ and $B$ are coupled via effectively tunable couplings  $g_A(t)$ and $g_B(t)$ to a single mode of an auxiliary system. The tunable couplings could be realized directly, or indirectly by modulating the auxiliary system (or qubit) frequencies.  }\label{fig:qubitbus} 
\end{figure}

In this paper, we present an alternate, platform-agnostic approach
for designing geometric two-qubit gates using STIRAP that is considerably simpler than the proposals of Refs.~\cite{duan2001geometric,Moller2008Quantum}, and that does not require a special form of qubit-qubit interaction. 
Furthermore, our approach is directly compatible with shortcuts to adiabaticity methods~\cite{demirplak2003adiabatic,demirplak2005assisted,berry2009transitionless,Ibanez2012Multiple,Guery2019Shortcuts,Ribeiro2019Accelerated}, and hence can be much faster than a naive adiabatic gate.   
We explicitly show that this acceleration can be performed while still retaining some of the robustness that makes adiabatic protocols so attractive.  ur approach is general and hence realizable in a variety of systems.  It is especially well suited to setups comprising two remote qubits that interact in a tunable manner with a common auxiliary system (see Fig.~\ref{fig:qubitbus}), something that has been realized in numerous experiments (see, e.g., Refs.~\cite{Chen2014Qubit,zhong2019violating,Magnard2020Microwave,zhong2021deterministic,Yan2022Entanglement,McKay2016Universal,reagor2018demonstration,Mundada2019Suppression,Ganzhorn2019Gate,Yuan2020High,Ganzhorn2020Benchmarking,abrams2020implementation,Stehlik2021Tunable,Sung2021Realization,leung2019deterministic,Hong2020Demonstration}). 
 
To demonstrate the effectiveness of our approach, we explore a superconducting circuit implementation consisting of two fluxonium qubits~\cite{manucharyan2009fluxonium,Earnest2018Realization,Nguyen2019High,Helin2021Universal} tunably coupled to a common auxiliary system (representing a bus or coupler). We consider cases where the required tuning is realized either (1) directly by using tunable couplings~\cite{Chen2014Qubit,zhong2019violating,Magnard2020Microwave,zhong2021deterministic,Yan2022Entanglement} between the auxiliary system and the qubits,  or by (2) using static couplings to the qubits, but frequency modulating the auxiliary mode~\cite{McKay2016Universal,reagor2018demonstration,Mundada2019Suppression,Ganzhorn2019Gate,Yuan2020High,Ganzhorn2020Benchmarking}.
Using experimentally realistic parameters, we demonstrate, via full master-equation simulations that take into account both non-rotating-wave approximation (non-RWA) effects and dissipation, that our  
accelerated two-qubit gates yield a competitive gate fidelity. 
For a direct scheme of realizing tunable interactions using time-dependent couplings, we obtain a gate fidelity of 
approximately 0.9995 for gate times in the range of $t_g = 45$--$60$ ns; this is achieved without any fine-tuning of $t_g$, though we did not model the internal structure of the couplers. 
If we instead use static couplings but parametrically modulate the auxiliary system frequency, we obtain a gate fidelity of 
approximately 0.999 at a gate time $t_g = 130$ ns (where here we model all key elements of the system). Our proposed accelerated adiabatic controlled-$Z$ (CZ) gates together with its corresponding arbitrary one-qubit gate proposed in Ref.~\cite{Setiawan2021Analytic} pave the way towards universal quantum computation that are fast and robust against imperfections in the control fields. 

The paper is organized as follows. We begin in Sec.~\ref{sec:quantumgate} by briefly reviewing the established idea of the STIRAP geometric gate \cite{duan2001geometric,Kis2002Qubit} and then outlining our general approach to implementing two-qubit geometric gates.  In Sec.~\ref{sec:coherror}, we review the acceleration protocols as well as methods to minimize the effects of coherent non-RWA errors.  We show the robustness and fundamental performance of our geometric gates in Sec.~\ref{sec:gateperformance}.  Readers who are already familiar with the STIRAP protocol as well as its accelerated version and who are interested in the details of their application to implementing two-qubit gates can skip directly to Sec.~\ref{sec:fluxonium}. In that section, we explore the implementation and performance of our STIRAP gates in realistic superconducting circuits comprising two fluxonium qubits connected via an auxiliary system.  We conclude in Sec.~\ref{sec:conclusions}.

\section{Two-qubit STIRAP quantum gates}~\label{sec:quantumgate}

In this section we outline  the general concept behind our geometric two-qubit gate that complements the geometric one-qubit gate proposed in Refs.~\cite{duan2001geometric,Kis2002Qubit} and its accelerated version~\cite{Ribeiro2019Accelerated,Setiawan2021Analytic}. To set the stage, in what follows we first discuss the basic physics behind our adiabatic two-qubit gate by focusing on a simplified RWA Hamiltonian. Following Ref.~\cite{Setiawan2021Analytic}, we show in the next section how our adiabatic geometric gate can be accelerated and enhanced for its implementation in realistic multilevel qubit settings where non-RWA effects cannot be neglected. 

\subsection{Double-STIRAP two-qubit gates in an ideal $\Lambda$ system}
The basic building block of our two qubit gate is the canonical state transfer protocol STIRAP~\cite{Vitanov2017Stimulated} that utilizes a $\Lambda$-level structure [see Fig.~\ref{fig:qubitbus}(a)]. An ideal $\Lambda$ system comprises three energy levels, two of which ($|q\rangle$ and $|a_2\rangle$) are  resonantly coupled to a common (typically) excited state $|a_1\rangle$ via tunable couplings. Denoting these (in general complex) tunable couplings as $\Omega_{j}(t)$ ($j=A,B$), the Hamiltonian of the $\Lambda$ system within the RWA is ($\hbar = 1$)
\begin{align}\label{eq:Hstirap}
    \Hstirap(t) &= \frac{1}{2} \left[ \Omega_A(t)|a_1\rangle \langle q| + \Omega_B(t) |a_1\rangle \langle a_2| + \mathrm{H.c.} \right].
\end{align}
In this paper, we always take state $|q\rangle$ to be a target logical qubit state that we would like to print a geometric phase on, while $|a_1\rangle$ and $|a_2\rangle$ are the ancillary states that are utilized for the gate operations. 

%%%%%%%%%%%%%%%%%%%%%%%%%%%%%%%%%%%%%%%

An adiabatic geometric gate is realized by using a cyclic evolution of the zero-energy adiabatic eigenstate of $\Hstirap(t)$ to generate a nontrivial geometric phase on the qubit state $|q\rangle$. This can be understood more clearly by first writing the control pulses as~\cite{Ribeiro2019Accelerated}
\begin{subequations}\label{eq:omegadrive}
\begin{align}
\Omega_A(t) &= \Omega_0 \sin [\theta(t)],\\
\Omega_B(t) &= \Omega_0 \cos[\theta(t)]e^{i\gamma(t)}\label{eq:omegaae}.
\end{align} 
\end{subequations}
The relative magnitudes and phases between the two pulses are controlled by the time-dependent angles $\theta(t)$ and $\gamma(t)$, respectively. The overall scale of the pulse amplitude $\Omega_0$ (which we set to be a constant for the duration of the protocol and zero otherwise) determines the instantaneous adiabatic gap of $\Hstirap(t)$:
\begin{equation}
    \Omega_{\mathrm{ad}}(t) \equiv \frac{1}{2}\sqrt{|\Omega_A(t)|^2 + |\Omega_B(t)|^2 } = \frac{\Omega_0}{2}.
    \label{eq:AdiabaticGap}
\end{equation}
This gap separates the instantaneous zero-energy dark state (which is orthogonal to the upper level $|a_1\rangle$) of Hamiltonian $\Hstirap(t)$ from its instantaneous bright states with energy $\pm \Omega_0/2$. 

By performing a ``double-STIRAP protocol" on the $\Lambda$ system, one evolves the dark state 
\begin{equation}\label{eq:d2}
|\rmd(t)\rangle = \cos[\theta(t)]|q\rangle - e^{i\gamma(t)}\sin[\theta(t)]|a_2\rangle
\end{equation}
cyclically as $|q\rangle \rightarrow |a_2\rangle \rightarrow |q\rangle$.  This corresponds to a cyclic variation of the pulse  parameter $\theta(t)$~(see Ref.~\cite{Ribeiro2019Accelerated} or Appendix~\ref{sec:tripodgate} for details) that results in a geometric phase being written on the qubit state $|q\rangle$.  If we pick the phase of our pulse $\gamma(t)$ to be
\begin{equation}\label{eq:gamma}
\gamma(t) = \gamma_0 \Theta \left(t - \frac{t_g}{2} \right),
\end{equation}
with $\Theta(t)$ being the Heaviside step function, the geometric phase is simply $\gamma_0$~\cite{Ribeiro2019Accelerated}. In this paper, we denote the initial and final gate times as $t=0$ and $t = t_g$, respectively. Since the pulse $\Omega_B(t) =0 $ at $t_g/2$, the discontinuity of $\gamma(t)$ at $t_g/2$ [Eq.~\eqref{eq:gamma}] does not affect the adiabaticity condition.

\begin{figure}[t!]
\centering
\includegraphics[width=0.9\linewidth]{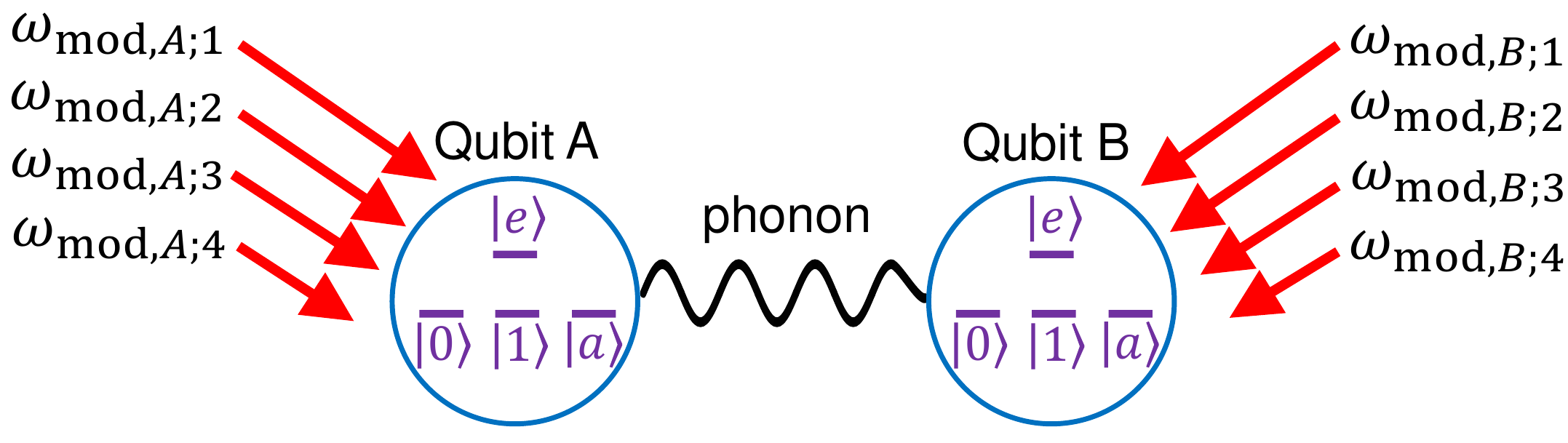}
\caption{A STIRAP-based two-qubit gate proposed in Ref.~\cite{duan2001geometric} for trapped-ion settings. This scheme is more complicated than our scheme (Fig.~\ref{fig:qubitbus}). It requires a $\Lambda$-level structure for each qubit, which consists of the logical qubit state $|1\rangle$ and an ancillary state $|a\rangle$ controllably coupled to another ancillary state $|e\rangle$. It also requires a total of eight control tones (shown by red arrows); four for each qubit. This scheme also utilizes a phonon-mediated interaction, a resource that is not available in most quantum computing platforms including superconducting qubit setups. }\label{fig:Duan}
\end{figure}

If the computational state $|q\rangle$ is a single-qubit state then the double-STIRAP protocol will implement a single-qubit phase gate~\cite{duan2001geometric,Kis2002Qubit,Ribeiro2019Accelerated,Setiawan2021Analytic}. As pointed out in Ref.~\cite{duan2001geometric}, if, instead, the computational state $|q\rangle$ is chosen to be a logical two-qubit state $|11\rangle$, the double STIRAP protocol then implements a two-qubit gate (assuming that no other qubit states are affected by the dynamics). For the adiabatic limit [$\dot{\theta}(t)/\Omega_0 \rightarrow 0$], the resulting gate unitary in the logical qubit subspace is~\cite{Ribeiro2019Accelerated}
\begin{align}\label{eq:UQubitAdiabatic}
\Ugq= |00\rangle \langle 00| + |01\rangle \langle 01| + |10\rangle \langle 10| + e^{i\gamma_0} |11\rangle\langle 11|.
\end{align}
As an example, the conditional-$Z$ (CZ) gate is obtained by setting the angle parameter $\gamma_0 = \pi$. 

As shown by Eq.~\eqref{eq:UQubitAdiabatic}, for the double-STIRAP protocol to realize a two-qubit gate, it is crucial that the geometric phase is written on only one of the two-qubit states (e.g., the logical state $|11\rangle$) and not on other logical qubit states ($|00\rangle$, $|01\rangle$, and $|10\rangle$).
For most gate schemes, this is accomplished using a static qubit-qubit interaction; this is the approach taken in previous proposals for STIRAP gates in atomic systems \cite{duan2001geometric,Moller2008Quantum}.
However, such static interactions are often problematic as the qubits continue to interact even after all control pulses are turned off. Surprisingly, as we show below, our two-qubit gate setup does not suffer from similar drawbacks as it does not require a static qubit-qubit interaction. 

We stress that there are several crucial differences (both conceptual and practical) between the two-qubit gate scheme we present, and the STIRAP gates proposed in Ref.~\cite{duan2001geometric} for trapped ions and Ref.~\cite{Moller2008Quantum} for Rydberg atoms.  As we will see, our scheme requires only two ancillary states, and two independent control tones.  In contrast, the previous atomic proposals require four ancillary states and either  four \cite{Moller2008Quantum} or eight \cite{duan2001geometric} control tones (cf.~Fig.~\ref{fig:Duan}).  Our approach also ultimately requires only a generic dispersive interaction, as opposed to the more specific interactions required in the atomic proposals (phonon-mediated interactions~\cite{duan2001geometric} or the Rydberg blockade~\cite{Moller2008Quantum}).  

\subsection{Generic realization of STIRAP two-qubit gates using time-dependent couplings}\label{sec:badlambda}
\begin{figure}[t!]
\centering
\includegraphics[width=0.9\linewidth]{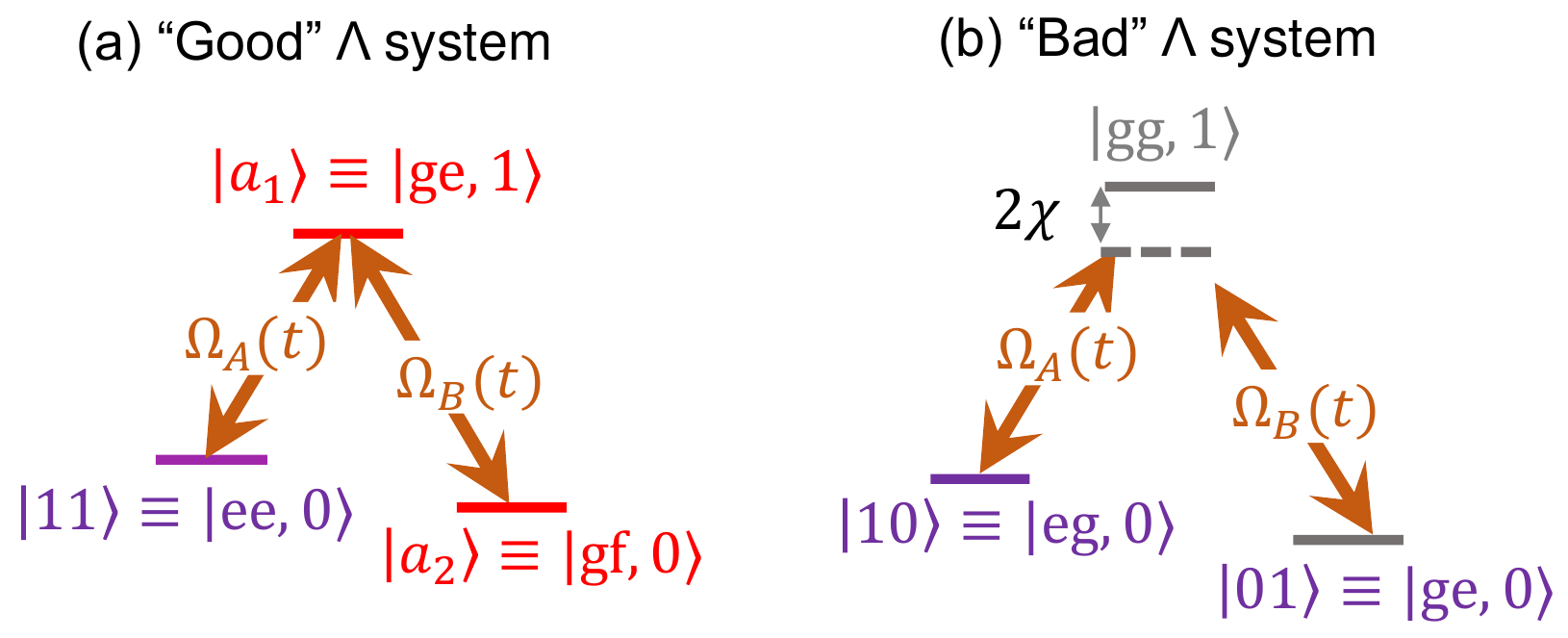}
\caption{ (a) The ``good"' $\Lambda$ system as in Fig.~\ref{fig:qubitbus}(a) containing the target computational state $|11\rangle \equiv |ee,0\rangle$, which ideally acquires a geometric phase.  (b) The ``bad" $\Lambda$ system that has the computational state $|10\rangle \equiv |eg,0\rangle$; we do not want any phase to be written on this state.  To realize this selectivity and a two-qubit gate, the transition $|eg,0\rangle \leftrightarrow |gg,1\rangle$ in the left-arm of the $\Lambda_{\rm bad}$ system needs to be detuned from the left modulation tone $\Omega_A(t)$. This can be realized by introducing a dispersive interaction $\chi$ between qubit $B$ and the auxiliary mode.}
\label{fig:badlambda} 
\end{figure}

In this section, we discuss in detail our geometric two-qubit gate design based on the STIRAP protocol. Our scheme is conceptually different from the  scheme in Refs.~\cite{duan2001geometric,Moller2008Quantum} and is compatible with remotely connected qubit setups~\cite{leung2019deterministic,zhong2019violating,Magnard2020Microwave,zhong2021deterministic,Yan2022Entanglement}. The basic working principle of our two-qubit gate relies on time-dependent couplings~\cite{Chen2014Qubit,Yan2018Tunable,zhong2019violating,Magnard2020Microwave,Foxen2020Demonstrating,Li2020Tunable,zhong2021deterministic,Yan2022Entanglement} that mediate the excitation transfer between the qubits and a common auxiliary system that serves as a bus or coupler~[see Fig.~\ref{fig:qubitbus}(b)]. The $\Lambda$-level structure required for our double-STIRAP protocol is formed by utilizing two  levels in qubit $A$ (levels $g$ and $e$),  qubit $B$ (levels $e$ and $f$), and the auxiliary system (levels 0 and 1). In the order of increasing energy, we label the energy levels in each qubit by $g, e, f, h,\cdots$ and those in the auxiliary system by $0,1,2,\cdots$.  In particular,  we use the state of composite qubit-auxiliary system $ |ge,1\rangle \equiv |a_1\rangle$ for the upper level and $|ee,0\rangle \equiv |11\rangle$ and $|gf,0\rangle \equiv |a_2\rangle$ for the two lower levels of the $\Lambda$ system~[see Fig.~\ref{fig:badlambda}(a)]. (Note that here the words ``upper" and ``lower" states are used to refer to the schematic diagram of the $\Lambda$ system in Fig.~\ref{fig:badlambda} and do not necessarily reflect the energies of the levels.) We use the notation $|a b,c\rangle$ to denote the composite state of two qubits coupled to a common auxiliary system where $a, b$, and $c$ are the states of qubit $A$, qubit $B$, and the auxiliary system, respectively. 

Our protocol assumes that the $g \rightarrow e$ transition in qubit $A$, the $0 \rightarrow 1$ transition in the auxiliary system, and the $e\rightarrow f$ transition in qubit $B$ have nondegenerate transition frequencies. As a result, static couplings between the qubits and auxiliary system cause at most weak hybridization. The gate is activated by modulating the couplings such that the  modulation frequencies $\omega_{\mathrm{mod},A} \equiv \varepsilon_{ge1} - \varepsilon_{ee0} $ and  $\omega_{\mathrm{mod},B} \equiv \varepsilon_{ge1} - \varepsilon_{gf0}$ resonantly drive  the left and right transitions of the $\Lambda$ system, respectively. Here, $\varepsilon_{k}$ denotes the energy of state $|k\rangle$. This modulation can be described by the Hamiltonian of the $\Lambda$ system,
\begin{align}
\hat{H}_{\Lambda}(t) = \hat{H}_{\mathrm{static},\Lambda} + \hat{H}_{\mathrm{mod},\Lambda}(t),
\end{align}
where $\hat{H}_{\mathrm{static},\Lambda}$ represents the static (time-independent) contributions and  
\begin{align}
\hat{H}_{\mathrm{mod},\Lambda}(t) = &\frac{1}{2} \biggl[\gac_A(t)e^{-i\omega_{\mathrm{mod},A}t}n_{A,eg}\hat{\Pi}_{A,eg}\hat{a} + \nonumber\\
&\qquad\gac_B(t)e^{-i\omega_{\mathrm{mod},B}t}n_{B,fe} \hat{\Pi}_{B,fe}\hat{a}\biggr] + \mathrm{H.c.}
\end{align}
captures the modulation of the couplings. Here, $\gac_{j}(t)$ is the strength of the tunable coupling between qubit $j$ and the auxiliary system, $\omega_{\mathrm{mod},j}$ is the frequency of the modulation tone $j$, $n_{j,kl}$ is the matrix element for the $|l\rangle_j \rightarrow |k\rangle_j$ transition of qubit $j$, $\hat{\Pi}_{j,kl}$ is the outer product $|k\rangle_j {}_j\langle l|$ for states in qubit $j$, and $\hat{a}$ is the annihilation operator of the auxiliary mode. Note that many choices are possible for the auxiliary system; for notational simplicity, we take it to be a (possibly nonlinear) bosonic mode with annihilation operator $\hat{a}$. Moreover, while we assume modulated couplings here to illustrate the basic physics, we show below that one can also realize our gate using static couplings and an auxiliary system whose transition frequency is modulated.

For a double-STIRAP protocol to implement a two-qubit gate, one needs to ensure that the protocol imprints a geometric phase onto only one of the qubit states ($|ee,0\rangle \equiv |11\rangle$) and not onto other qubit states. To this end, we use the left-resonant modulation tone to move the excitation in qubit $A$ to the auxiliary system only if qubit $B$ is in state $|e\rangle$, and not in state $|g\rangle$.  Formally, this means that the transition $|eg,0\rangle \leftrightarrow |gg,1\rangle$ has to be energetically detuned from  the $|ee,0\rangle \leftrightarrow |ge,1\rangle$ transition. Note that the transitions $|eg,0\rangle \leftrightarrow |gg,1\rangle$ and $|ge,0\rangle \leftrightarrow |gg,1\rangle$ form another $\Lambda$ configuration [Fig.~\ref{fig:badlambda}(b)], where the modulation introduces population leakage and unwanted phase to be imprinted on the computational states $|eg,0\rangle$ and $|ge,0\rangle$.   Since this $\Lambda$ system is not part of our ideal gate protocol, we call it a ``bad" $\Lambda$ system (labeled $\Lambda_{\rm bad}$ throughout), to differentiate it from the ``good" $\Lambda$ system that is used to implement the gate.

The degeneracy between the $|ee,0\rangle \leftrightarrow |ge,1\rangle$ and $|eg,0\rangle \leftrightarrow |gg,1\rangle$ transitions can be lifted by introducing a dispersive shift $\chi$ between states in the auxiliary system and qubit $B$, where the required dispersive interaction has the form
\begin{equation}
\hat{H}_{\mathrm{disp}} = \chi (\hat{\Pi}_{B,ee}- \hat{\Pi}_{B,gg}) \hat{a}^\dagger\hat{a}.
\end{equation}
This dispersive shift can be realized by introducing a nonresonant static coupling $\gdc_{BC}$ between qubit $B$ and the auxiliary system. Ultimately, our gate is protected from the coherent errors arising from the $\Lambda_{\mathrm{bad}}$ system by the energetic detuning ($2\chi$) between the left-arm transitions of the $\Lambda$ and $\Lambda_{\rm bad}$ systems.  As such, the dispersive interaction strength $\chi$ sets a limit to the gate speed.  We stress that $\chi$ is not a qubit-qubit interaction, but rather an effective interaction just between the auxiliary system and qubit $B$.  Thus, this speed limit on our gate is very different from more conventional approaches, where the gate speed is typically limited by the size of a direct qubit-qubit interaction.  As mentioned, the lack of any static qubit-qubit interaction provides us with another advantage: when control pulses are off, there is in principle no residual qubit-qubit interaction.   

\subsection{Schemes to implement effective tunable interactions}\label{sec:ZZerror}

\begin{figure}[t!]
\centering
\includegraphics[width=0.8\linewidth]{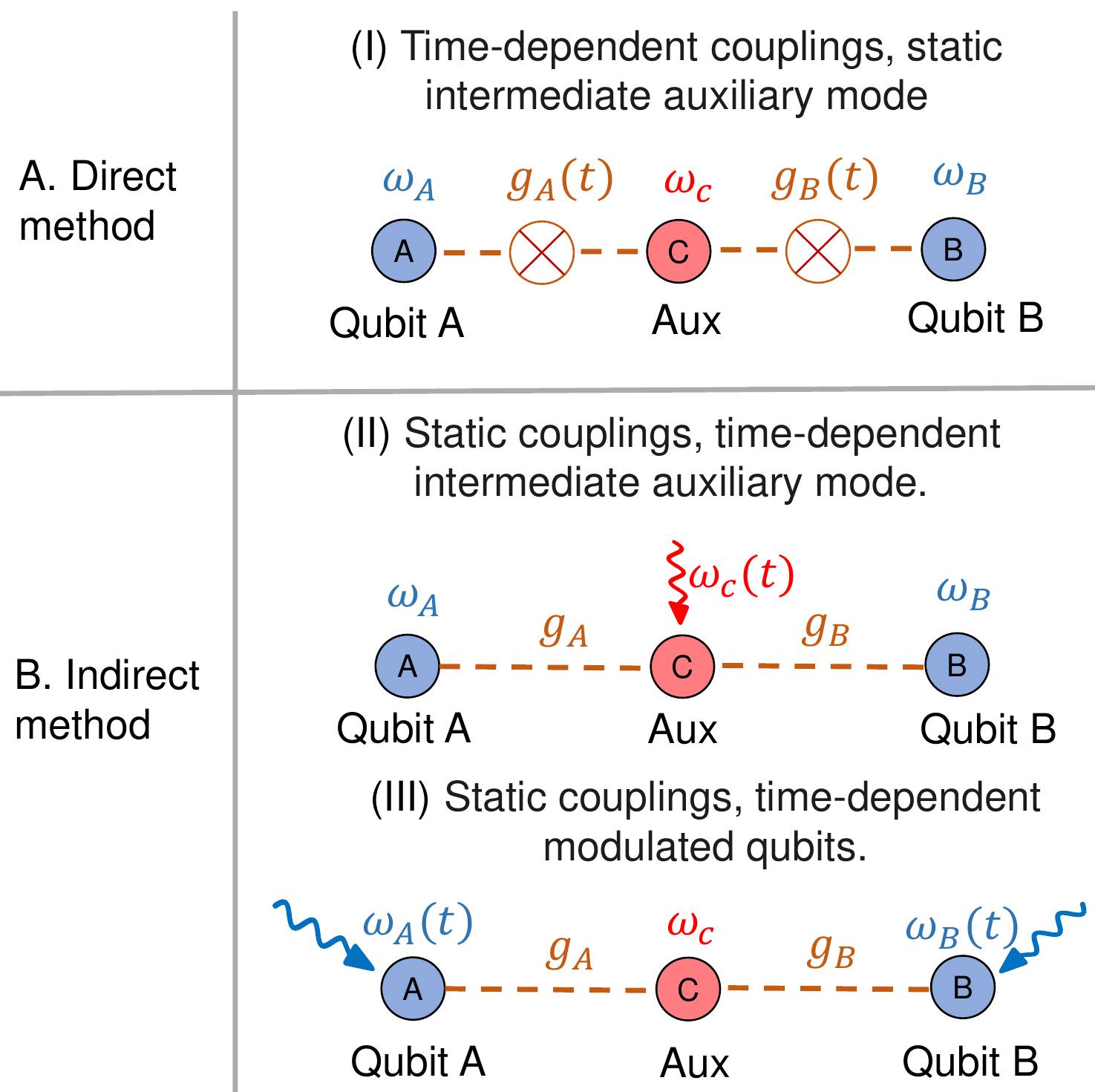}
\caption{
    Different schemes for realizing tunable interactions for our two-qubit gate: A. Direct method: (I) time-dependent couplings, static intermediate auxiliary mode. B. Indirect method: (II) static couplings, time-dependent intermediate auxiliary mode, and (III) static couplings, time-dependent modulated qubits.
}
\label{fig:scheme}
\end{figure}

As discussed above, our protocol uses appropriately engineered tunable interactions
between the auxiliary system and the qubits; these can be implemented in various ways. The most direct method was already introduced in Sec.~\ref{sec:badlambda}:  use explicitly time-modulated coupling elements (tunable couplings) between each qubit and the auxiliary system [bus or coupling mode; see method (I) of Fig.~\ref{fig:scheme}]. Such tunable couplings have been realized in superconducting-circuit experiments involving transmon qubits~\cite{Chen2014Qubit,zhong2019violating,Magnard2020Microwave,zhong2021deterministic,Yan2022Entanglement} and fluxonium qubits~\cite{moskalenko2022high}. This direct method is favorable, since it requires only one static qubit-bus coupling, and hence there is in principle no residual qubit-qubit coupling when the gate is off.    

In addition to the direct method [method (I) of Fig.~\ref{fig:scheme}], one can also \textit{indirectly} realize effective tunable interactions 
by combining static qubit-auxiliary system interactions with either
parametric modulation of the auxiliary system [method (II) of Fig.~\ref{fig:scheme}],  and/or the qubits themselves [method (III) of Fig.~\ref{fig:scheme}]. The auxiliary coupler here can be a tunable transmon, fluxonium,  etc. An advantage of this scheme is that the modulation of frequency-tunable couplers have been demonstrated in a number of experiments~\cite{McKay2016Universal,reagor2018demonstration,Mundada2019Suppression,Ganzhorn2019Gate,Yuan2020High,abrams2020implementation,Ganzhorn2020Benchmarking}. Note that in contrast to the  direct method, these indirect schemes require two static qubit-auxiliary system couplings. As a result, there is a mediated static qubit-qubit interaction that remains on even when the gate is over.  This is common to most existing protocols using couplers.  As we show in our detailed simulations, through judicious parameter choice and circuit design, this spurious coupling can be minimized and even in some cases eliminated. 

\section{Mitigation of Coherent Errors}\label{sec:coherror}
In this section, we discuss accelerating our gate using the \textit{shortcuts-to-adiabaticity} (STA) protocol based on the superadiabatic transitionless driving (SATD) method~\cite{Baksic2016Speeding,Ribeiro2019Accelerated}. Moreover, we go beyond the RWA settings and use a two-pronged analytical approach proposed in Ref.~\cite{Setiawan2021Analytic} to enhance the performance of our protocol in the presence of the following types of errors.
\begin{enumerate}
\item Nonadiabatic errors that arise when the gates are accelerated.
\item Non-RWA errors due to modulation crosstalk as well as spurious couplings between computational and noncomputational states.
\end{enumerate}
\subsection{Nonadiabatic errors}\label{sec:accelerated}
In the adiabatic limit where the protocol time becomes much longer than the inverse adiabatic gap $1/\Omega_0$ [cf.~Eq.~\eqref{eq:AdiabaticGap}], the geometric gate approaches a perfect gate. However, any dissipation of the lower $\Lambda$-system levels makes long protocol times incompatible with high fidelity.  If one naively accelerates the adiabatic protocol without additional pulse shaping, the resulting nonadiabatic transitions will result in substantial errors. To alleviate this problem, one can turn to STA approaches (see, e.g., Refs.~\cite{demirplak2003adiabatic,demirplak2005assisted,berry2009transitionless,Baksic2016Speeding,Ribeiro2019Accelerated,demirplak2008consistency}). 

Following Ref.~\cite{Ribeiro2019Accelerated}, we use the SATD shortcuts method \cite{Baksic2016Speeding,Ribeiro2019Accelerated} that allows fast operation without nonadiabatic errors, by having the system evolution follow a dressed adiabatic eigenstate (see Appendix~\ref{sec:SATD}). The accelerated protocol is implemented through pulse shaping of the original control fields~\cite{Baksic2016Speeding,Zhou2017Accelerated,Ribeiro2017Systematic,Ribeiro2019Accelerated,roque2020engineering}. One appealing feature of the SATD method is that this pulse modification can be described \textit{analytically}, with a modification of the form~\cite{Ribeiro2019Accelerated}
\begin{subequations}\label{eq:omegaSATD}
\begin{align}
\Omega_A(t) &\rightarrow \tilde{\Omega}_A(t) \equiv \Omega_0 \left[\sin[\theta(t)] + 4 \frac{\cos[\theta(t)]\ddot{\theta}(t)}{\Omega_0^2 + 4 \dot{\theta}^2(t)}\right],\\
\Omega_B(t) &\rightarrow \tilde{\Omega}_B(t) \equiv \Omega_0 e^{i\gamma(t)}\left[\cos[\theta(t)] - 4 \frac{\sin[\theta(t)]\ddot{\theta}(t)}{\Omega_0^2 + 4 \dot{\theta}^2(t)}\right],
\end{align}
\end{subequations}
where the angle $\gamma(t)$ [Eq.~\eqref{eq:gamma}] remains the same as the adiabatic version. One can show~\cite{Ribeiro2019Accelerated} that the accelerated
protocol derived using Eqs.~\eqref{eq:omegaSATD} gives rise to the same unitary $\hat{U}_{\mathrm{G},\mathrm{q}}$ in the qubit subspace as in the adiabatic limit [cf.~Eq.~\eqref{eq:UQubitAdiabatic}]. 
For an ideal (RWA) Hamiltonian of the $\Lambda$ system [Eq.~\eqref{eq:Hstirap}] and a fixed gate time, our SATD method gives an infinite number of perfect gate protocols (each with a different pulse shape, characterized by a different value of $\Omega_0$). We can use this degeneracy as a resource to mitigate non-RWA errors in realistic systems as discussed in the following subsection and Ref.~\cite{Setiawan2021Analytic}.

\subsection{Non-RWA errors}
\label{sec:chirp}
Coherent errors also arise from nonresonant couplings that would be neglected within the RWA.  We can partially mitigate their effects by using the strategy introduced in Ref.~\cite{Setiawan2021Analytic}. This strategy consists of two steps~(see Appendix~\ref{sec:enhancedSATD} for details): (1) using the power optimal $\Omega_0 = \Omega_{\mathrm{opt}}$ where $\Omega_{\mathrm{opt}}/2\pi \simeq 1.135/t_g$, and (2) frequency chirping the control fields.

%%%%%%%%%%%%%%%%%%%%%%%%%%%%%%%%%%%%%%%%%%%%%%%%%%%%%%%%
%%%%%%%%%%%%%%%%%%%%%%%%%%%%%%%%%%%%%%%%%%%%%%%%%%%%%%%%
\section{Gate performance}\label{sec:gateperformance}
%%%%%%%%%%%%%%%%%%%%%%%%%%%%%%%%%%%%%%%%%%%%%%%%%%%%%%%%
%%%%%%%%%%%%%%%%%%%%%%%%%%%%%%%%%%%%%%%%%%%%%%%%%%%%%%%%
We start by considering the fundamental advantages and limitations of our two-qubit gate.
The system evolution generated by our control pulses, including coherent and possibly dissipative errors, corresponds to a quantum map $\mathcal{M}$.  We wish to quantify how close this is to the ideal unitary two-qubit gate we are interested in. 
We do this via the state-averaged gate fidelity given by~\cite{nielsen2002simple,cabrera2007average}
\begin{widetext}
\begin{align}\label{eq:fidelity}
\bar{F} &= \frac{1}{4}+\frac{1}{80} \sum_{\mu,\nu = \{0,x,y,z\} |\{ \sigmamu_A \otimes \sigmanu_B\neq \mathds{1} \otimes\mathds{1}\}}   \mathrm{tr}\bigg\{\left[\hat{U}_{\mathrm{g}}((\sigmamu_A \otimes \sigmanu_B)\oplus \mathbf{0}_{\mathrm{q}^c})\hat{U}^{\dagger}_{\mathrm{g}} \right] \mathcal{M}\left[ (\sigmamu_A \otimes \sigmanu_B)\oplus \mathbf{0}_{\mathrm{q}^c} \right] \bigg\}, 
\end{align}
\end{widetext}
where $\sigmamu_A \otimes \sigmanu_B $ (with $\mu,\nu = \{0,x,y,z\}$) are the Pauli matrices acting on qubits $A$ and $B$ within the qubit subspace (namely, the $4\times 4$  block) and $\mathbf{0}_{\mathrm{q}^c}$ is a zero matrix in the nonqubit subspace.
Here, 
\begin{equation}
\hat{U}_{\mathrm{g}} =  (\Usq  \Ugq) \oplus \mathds{1}_{\mathrm{q}^c}
\end{equation}
is the ideal target unitary two-qubit gate operation that performs a unitary $\Ugq$ [cf.~Eq.~\eqref{eq:UQubitAdiabatic}] in the qubit subspace up to a trivial single-qubit phase gate $\Usq = \mathrm{exp}\left( i\phi_A\right) \left(\sigmaz_A \otimes \mathds{1}_{B}\right) + \mathrm{exp}\left( i\phi_B\right) \left( \mathds{1}_A\otimes \sigmaz_B \right)$. Throughout this paper, the gate fidelity is calculated by optimizing over the trivial single-qubit phases $\phi_A$ and $\phi_B$  which is achieved by choosing the single-qubit $Z$ gates on qubits $A$ and $B$ that maximize the gate fidelity. From the fidelity, we can calculate the state-averaged gate error using $\bar{\varepsilon} = 1-\bar{F}$.

\subsection{Robustness advantages of accelerated geometric gates}\label{sec:robustness}
To understand the advantages of our accelerated STIRAP two-qubit gate against parameter and pulse uncertainties, we compare it against the simplest corresponding gate based on dynamical phases.  This latter approach would simply employ a static  $ZZ$ interaction to imprint a dynamical phase on state $|ee,0\rangle \equiv |11\rangle$. Up to innocuous global and single-qubit phases, the dynamical $ZZ$ gate can be described using a RWA Hamiltonian
\begin{equation}\label{eq:ZZHamilton}
\hat{H}_{\mathrm{dyn}}^{ZZ} = \frac{\Omega_0}{4} \hat{\sigma}_A^z \otimes \hat{\sigma}_B^z,
\end{equation}
where $\Omega_0/4$ is the $ZZ$ interaction strength. 

%%%%%%%%%%%%%%%%%
\begin{figure}[t!]
\centering
\includegraphics[width=\linewidth]{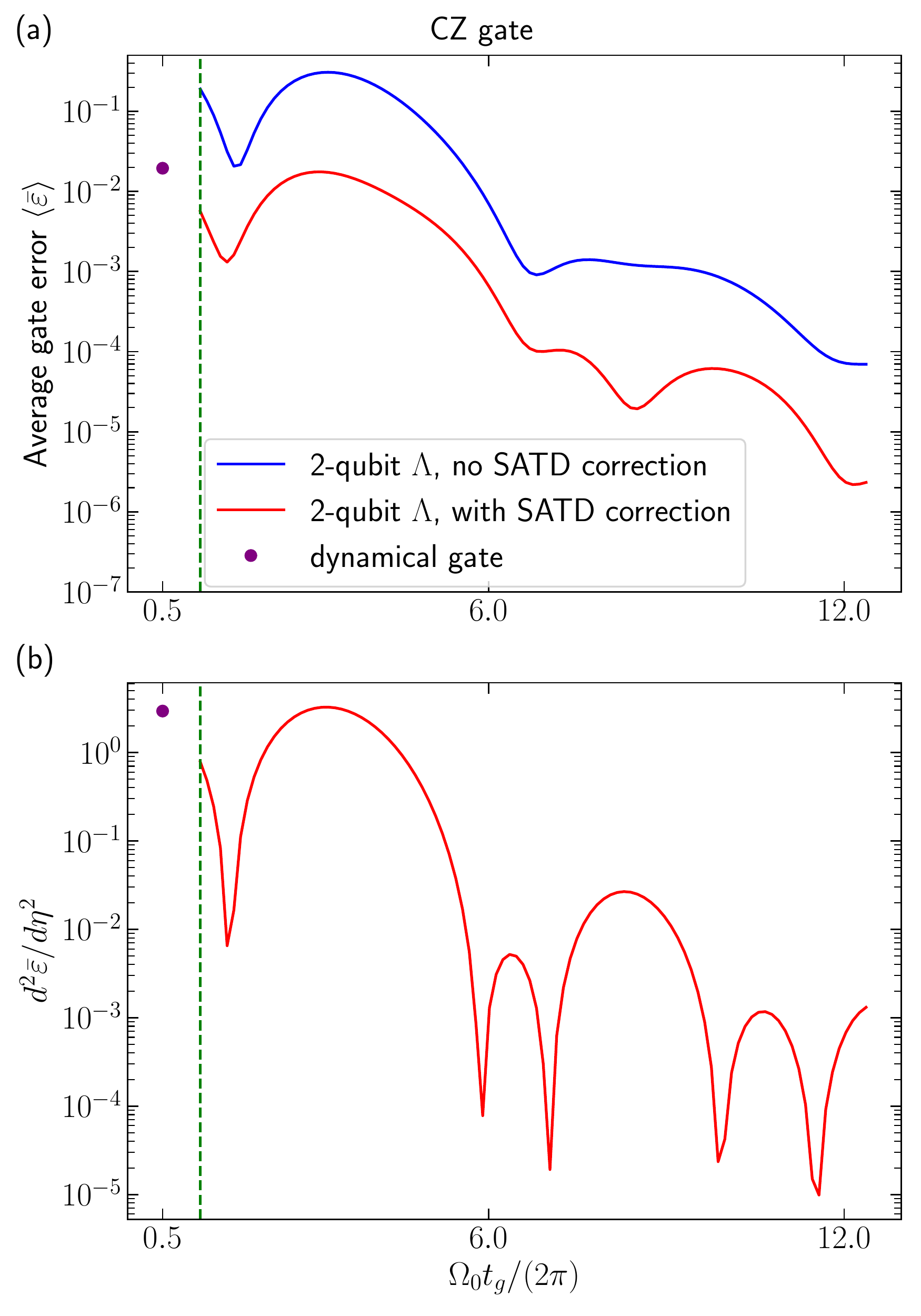}
\caption{Comparison of the gate robustness between different CZ gates. (a) The state-averaged gate errors $\langle\bar{\varepsilon} \rangle$ are averaged over a 20\% uncertainty ($\eta \in [-0.2,0.2]$) in the pulse amplitude. The plots are for the uncorrected adiabatic protocol (blue curve), SATD-accelerated adiabatic protocol (red curve), and dynamical gate (purple dot) as a function of $\Omega_0 t_g/2\pi$. (b) Differential sensitivity of the state-averaged gate error with respect to small pulse magnitude uncertainty ($\eta$) as a function of $\Omega_0 t_g/2\pi$. The plots are for the SATD-accelerated adiabatic protocol (red curve), and dynamical gate (purple dot). For both panels, $\Omega_0 t_g = \pi$ for a dynamical CZ gate and the green dashed line denotes the SATD power-optimal protocol ($\Omega_0 = \Omega_{\mathrm{opt}}$).}
\label{fig:robustness}
\end{figure} 
%%%%%%%%%%%%%%%%%

An appealing feature of our geometric gate compared to dynamical gates is its robustness against imperfections in control pulses; this is inherited from the underlying purely adiabatic gate.  For example, imagine a situation where there is an overall uncertainty in the scale of applied pulse amplitudes (due, e.g., to some unknown attenuation).  This would result in pulse envelopes $\Omega_{A,B}(t) \rightarrow \Omega_{A,B}(t)(1+\eta)$, where $\eta$ parameterizes overall uncertainty in the scale of the amplitudes.  We could introduce a corresponding uncertainty in the dynamical gate, by letting  $\Omega_{0} \rightarrow \Omega_{0} (1+\eta)$.  
Given this uncertainty, we consider two metrics for characterizing the robustness of a given gate.
\begin{enumerate}
\item In the limit of small uncertainties $\eta$, the gate error will have a quadratic dependence on $\eta$.  The coefficient of this quadratic term
(i.e.,~$d^2\bar{\varepsilon}/d\eta^2 \equiv \xi$)
is thus a measure of robustness.  
\item  For larger uncertainties (with $\eta$ being described by some probability distribution), we could characterize robustness by averaging the gate error over $\eta$.
For concreteness, we take $\eta$ to be described by a uniform probability distribution over the interval $[-\eta_0,\eta_0]$; the average gate error is
\begin{equation}
\langle \bar{\varepsilon}(\Omega_0 t_g)\rangle\equiv\frac{1}{2\eta_0} \int_{-\eta_0}^{\eta_0} d\eta \,\bar{\varepsilon}(\Omega_0t_g,\eta).
\end{equation}
\end{enumerate}

To calculate the gate performance, we evolve the initial states using the Schr\"{o}dinger equation with the RWA Hamiltonian given by Eq.~\eqref{eq:Hstirap} for our STIRAP gate and by Eq.~\eqref{eq:ZZHamilton} for the dynamical gate. In this paper, we choose the lowest-order polynomial [Eq.~\eqref{eq:Px}] that gives a smooth variation of $\theta(t)$ [Eq.~\eqref{eq:theta}] for our STIRAP gate. The simulations here and throughout this paper are done numerically using the Python package QuTiP~\cite{johansson2012qutip,johansson2012qutip2}.

Figure~\ref{fig:robustness} shows the robustness of CZ gates
in the presence of pulse-amplitude uncertainties.  Results are shown for the uncorrected adiabatic (blue curve), accelerated adiabatic (red curve), and dynamical-gate protocols (purple dot). For panel (a), we consider large uncertainties, and consider the $\eta$-averaged gate error with $\eta_0 = 0.2$. We see that the accelerated STIRAP gate provides, as expected, a robustness advantage over the dynamical gate.  While this improvement is modest if one uses power-optimized STIRAP (green dashed line), it can be improved significantly by using slightly larger values of $\Omega_0$.  

Figure~\ref{fig:robustness}(b) shows the differential sensitivity of the gate to small values of $\eta$.  
We again see advantages compared to the dynamical gate.  The sensitivity to $\eta$ can be reduced by orders of magnitude compared to the dynamical gate, by using SATD with power levels slightly higher than the power-optimal case.  
For the power-optimal case, the differential sensitivity of the SATD gate does not change significantly (i.e., $\xi\simeq 0.802$ for $t_g = 45$ ns) even if we consider the realistic system as in Sec.~\ref{sec:methodone} where we include the $38$ lowest-energy levels with all the non-RWA dynamics.

\subsection{Comparison against Raman-type gates}

The accelerated STIRAP gate also had advantages over Raman-style two-qubit gates that use an auxiliary system (coupler) to facilitate virtual transitions.  For the Raman gate, the qubit-auxiliary system couplings are modulated with frequencies that are detuned from the qubit-auxiliary mode transition frequencies and the gate operates by using an effective qubit-qubit coupling obtained via higher-order, coupler-mediated processes.  As a result of being perturbative, the effective interaction here is small, giving rise to a slower gate compared to our SATD protocol (for a fixed modulation amplitude) that uses resonant modulations. This ultimately results in the SATD gate being more tolerant of dissipation than the Raman approach (see, e.g., Appendix H in Ref.~\cite{Setiawan2021Analytic} for details on the comparison with the Raman protocol).

\subsection{Fundamental limitation of the STIRAP two-qubit gate performance:  $\Lambda_{\rm bad}$ system}\label{sec:gatelimit}
The performance of our gate is fundamentally limited by the coherent error resulting from the undesired driving of the $\Lambda_{\rm bad}$ system. This error is ultimately determined by the size of the dispersive coupling $\chi$ between qubit $B$ and the auxiliary system (see Fig.~\ref{fig:badlambda}).
As described in Sec~\ref{sec:badlambda}, this nonresonant driving of the $\Lambda_{\rm bad}$ system can cause both leakage and phase errors. To understand these errors better, we simulate our gate protocol using the idealized RWA Hamiltonian [Eq.~\eqref{eq:Hstirap}] with only one additional spurious term: the nonresonant drive of the left arm of the $\Lambda_{\rm bad}$ system, namely the $|ge,0\rangle \leftrightarrow |gg,1\rangle$ transition driven by $\Omega_A(t)$ (see Fig.~\ref{fig:badlambda}).

In our simulation, we use the power-optimal SATD protocol ($\Omega_0 = \Omega_{\mathrm{opt}}$) that minimizes the coherent error due to the driving of the  $\Lambda_{\rm bad}$ system.  We do not chirp the modulation frequency, as that is not effective against bad-$\Lambda$ errors.  
One simple error arising from the nonresonant driving of the $\Lambda_{\rm bad}$ transition  is a modification of the effective two-qubit phase by $\phi_{ZZ} \equiv \phi_{00} + \phi_{11} -\phi_{01} - \phi_{10}$ where $\phi_{k}$ is the phase on the logical qubit state $|k\rangle$. We can correct this error by simply shifting the geometric phase $\gamma_0$ that our gate implements by $\phi_{ZZ}$, i.e.,
\begin{equation}\label{eq:gammazero}
\gamma_0 \rightarrow \gamma_0 - \phi_{ZZ}.
\end{equation} 
Once this phase error is corrected, the dominant remaining error is leakage from state $|ge,0\rangle$ to $|gg,1\rangle$. In the large dispersive regime ($2\chi t_g/2\pi \gtrsim 10$), the gate error as a function of gate time can be numerically fitted to (not shown)
\begin{align}\label{eq:badscaling}
\bar{\varepsilon} = \begin{cases}
c_{\mathrm{phase}}(\chi t_g)^{-2} & \text{without $ZZ$ phase correction}, \\[1em]
c_{\mathrm{leak}}(\chi t_g)^{-4} & \text{with $ZZ$ phase correction},
\end{cases}
\end{align}
where $c_{\mathrm{phase}} \simeq 0.05$ and $c_{\mathrm{leak}} \simeq 0.26$. We note that the above quadratic and quartic scaling of the phase and leakage errors in the small parameter $\chi t_g$ are consistent with the analytic arguments based on the Magnus expansion of the coherent errors (see Refs.~\cite{Setiawan2021Analytic,roque2020engineering}). For the case where the $ZZ$ correction is applied, the numerical fit to obtain the scaling of the gate error [Eq.~\eqref{eq:badscaling}] is done by neglecting the fast oscillations of the gate error with respect to the gate time $t_g$ that arise due to the shuttling of the population between states $|gg,1\rangle$ and $|ge,0\rangle$.

\section{Fluxonium qubits coupled to a common auxiliary system}\label{sec:fluxonium}
\subsection{Basic setup}

We now analyze realizations of our gate in a concrete physical setting: two fluxonium qubits~\cite{manucharyan2009fluxonium,Earnest2018Realization,Nguyen2019High,Helin2021Universal} coupled to a common auxiliary system (such as a cavity, transmon, fluxonium, or a particular mode of a waveguide).  Our analysis and modeling will consider gate performance in the presence of both coherent and dissipative error mechanisms. 
We note that the topic of fluxonium two-qubit gates has been the subject of considerable recent activity, both  theoretical~\cite{chen2021fast,Nesterov2021Proposal,moskalenko2021tunable,nesterov2022controlled,Cai2021All,Nesterov2018Microwave,nguyen2022scalable,weiss2022fast} and experimental~\cite{dogan2022demonstration,Ficheux2021Fast,Xiong2022Arbitrary,Bao2022Fluxonium,moskalenko2022high}.  As we discuss below, the combination of parametric modulation and accelerated adiabatic evolution gives our approach a number of unique potential advantages over other approaches.

\begin{figure}[t!]
\centering
\includegraphics[width=1\linewidth]{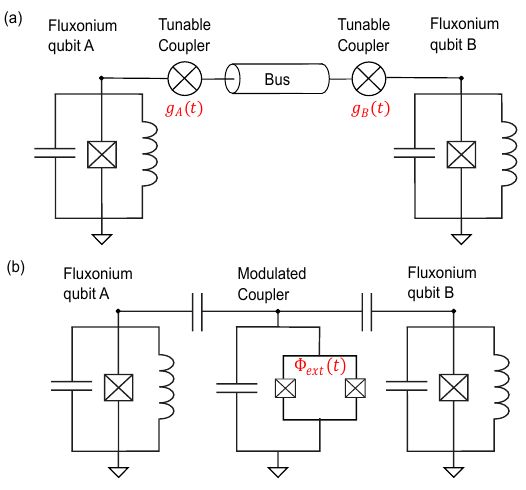}
\caption{Circuit schematics for systems consisting of two fluxonium qubits $A$ and $B$, (a) coupled via tunable couplings to an auxiliary system (bus) or (b) statically coupled to a frequency-modulated auxiliary system (which we here consider to be a frequency-tunable transmon).}
\label{fig:physicalsetup}
\end{figure}

Fluxonium circuits are attractive qubit platforms as they possess large nonlinearities and long relaxation times (which can reach milliseconds~\cite{Earnest2018Realization,Lin2018Demonstration,Nguyen2019High,Helin2021Universal,somoroff2021millisecond}).  They can also exhibit first-order insensitivity to $1/f$ flux noise dephasing. 
A fluxonium qubit consists of a Josephson junction (typically shunted by a capacitance) that forms a loop with a superinductance [see Fig.~\ref{fig:physicalsetup}].  The Hamiltonian is 
\begin{equation}\label{eq:Hf}
\hat{H}_{j} = 4 E_{C,j} \hat{n}_j^2 - E_{J,j} \cos \hat{\varphi}_j + \frac{1}{2} E_{L,j}(\hat{\varphi}_j- 2\pi \Phi_{\mathrm{ext},j}/\Phi_0)^2,
\end{equation}
where $E_{C,j}$, $E_{J,j}$, and $E_{L,j}$ are the capacitive, Josephson, and inductive energies of qubit $j=A,B$; $\hat{n}_j$ and $\hat{\varphi}_j$ are the qubit-$j$ charge and phase operators, obeying $[\hat{\varphi}_j, \hat{n}_j] = i$. 
The loop formed by the Josephson junction and the shunting inductance is threaded by an external magnetic flux $\Phi_{\mathrm{ext},j}$ and $\Phi_0 = h/2e$ is the flux quantum. 

In our analysis below, we consider two scenarios where either (1) the fluxonia are coupled to a fixed-frequency auxiliary system (e.g., a bus) via directly tunable couplings, [Fig.~\ref{fig:physicalsetup}(a)] 
or (2) the qubits are statically coupled to a frequency-modulated auxiliary system (coupler) [Fig.~\ref{fig:physicalsetup}(b)]. 
We model the bus in the first scheme [Fig.~\ref{fig:physicalsetup}(a)] as a harmonic oscillator and the coupler in the second scheme [Fig.~\ref{fig:physicalsetup}(b)] as a weakly anharmonic oscillator
(e.g.,~a transmon)
with Hamiltonian
\begin{equation}\label{eq:HC}
\hat{H}_C = \omega_C \hat{a}^\dagger\hat{a} - U \hat{a}^\dagger\hat{a}^\dagger\hat{a}\hat{a},
\end{equation}
where $\hat{a}$ is the coupler lowering operator, $\omega_C$ is the harmonic frequency, and $U$ is the nonlinearity.

The total Hamiltonian of the composite system (fluxonium qubits $A$ and $B$, auxiliary system $C$) is 
\begin{equation}\label{eq:Htotal}
\hat{H}(t) = \hat{H}_A + \hat{H}_B + \hat{H}_C + \Hint + \Hdr(t).
\end{equation}
The static interaction term $\Hint$ is given by
\begin{equation}
\Hint = \sum_{j=A,B}\gdc_{jC} \hat{n}_j (\hat{a}^{\dagger}+\hat{a}) + \gdc_{AB} \hat{n}_A \hat{n}_B,
\end{equation}
where $\gdc_{jC}$ is the static coupling strength between qubit $j$ and the auxiliary system, and $\gdc_{AB}$ is the static qubit-qubit coupling strength.  The term $\Hdr(t)$ describes the temporal modulation that is applied to perform the gate (see Fig.~\ref{fig:scheme}), and is described further below.  While we consider capacitive couplings here, our basic gate physics is also applicable to systems with inductive couplings. 

Diagonalizing the 
time-independent Hamiltonian $\hat{H}_{\mathrm{static}} = \hat{H}(t) - \Hdr(t)$
results in dressed eigenstates of the form $|ab,c\rangle$, where the labels $a$, $b$ and $c$ refer to the states of qubit $A$, qubit $B$, and the auxiliary system, respectively. We always operate in a strongly detuned regime, where the hybridization between the auxiliary system and the two qubits is weak (hence, the dressed eigenstates each have a strong overlap with a single uncoupled eigenstate). In this paper, we take the logical qubit states to be  $|gg,0\rangle \equiv |00\rangle$, $|ge,0\rangle \equiv |01\rangle$, $|eg,0\rangle \equiv |10\rangle$, $|ee,0\rangle \equiv |11\rangle$. These four qubit states together with the two ancillary states ($|ge,1\rangle\equiv |a_1\rangle$ and $|gf,0\rangle\equiv |a_2\rangle$) of the $\Lambda$ system form the computational subspace of our STIRAP two-qubit gate [see Figs.~\ref{fig:qubitbus}(a) and~\ref{fig:badlambda}(a)].  

Each of the three modulation schemes depicted in Fig.~\ref{fig:scheme} corresponds to a different $\Hdr(t)$ in Eq.~\eqref{eq:Htotal}.  Here, we only analyze the first two:  the direct tunable coupling scheme [method (I)] and  the indirect frequency-modulated auxiliary mode scheme [method (II)].  Method (III) is also a possible implementation route, i.e.,~use static qubit-auxiliary system couplings and modulate the frequencies of both qubits $A$ and $B$.  

\subsection{Method (I): time-dependent couplings, static intermediate auxiliary mode}
\label{sec:methodone}

We begin by first considering method (I) of Fig.~\ref{fig:scheme}, where the needed resonant interactions between the qubits and the auxiliary mode are induced by modulating tunable couplings. Such tunable couplings have been implemented in several transmon-based setups~\cite{Chen2014Qubit,zhong2019violating,Magnard2020Microwave,zhong2021deterministic,Yan2022Entanglement}. There are also recent theoretical studies~\cite{Huang2018Universal,moskalenko2021tunable,weiss2022fast} and an experiment ~\cite{moskalenko2022high} studying tunable couplings for fluxonium qubits.
The focus here is the operation of our gate, and not the internal workings of a specific tunable coupling device.  As such, we simply model these elements via time-dependent matrix elements, leading to
\begin{equation}
\Hdr(t) = 
\frac{1}{2}\sum_{j=A,B} ( \gac_j(t) e^{-i\int_0^t\omegatmodj(t') dt'}+\mathrm{c.c.}) \hat{n}_j (\hat{a}^{\dagger}+\hat{a}),
\end{equation}
where we have written the time-dependent coupling between qubit $j$ and the auxiliary system in terms of a complex envelope function $\gac_j(t)$, and a chirped modulation frequency $\omegatmodj(t)$ [Eq.~\eqref{eq:modomegaapp}].
The amplitude $\gac_j(t)$ is directly related to our effective pulse amplitudes $\tilde{\Omega}_{j}(t)$ [Eqs.~\eqref{eq:omegaSATD}] by 
\begin{subequations}\label{eq:gtac}
\begin{align}
\gac_A(t) &= \tilde{\Omega}_{A}(t)/\langle ge,1|\hat{n}_A (\hat{a}^{\dagger}+\hat{a}) |ee,0 \rangle, \\
\gac_B(t) &= \tilde{\Omega}_{B}(t)/\langle ge,1|\hat{n}_B (\hat{a}^{\dagger}+\hat{a}) |gf,0\rangle, 
\end{align}
\end{subequations}
where the transitions $ |ee,0 \rangle\leftrightarrow |ge,1 \rangle$ and $ |gf,0 \rangle\leftrightarrow |ge,1 \rangle$ correspond to the left and right arms of the $\Lambda$ system, respectively [see Fig.~\ref{fig:badlambda}(a)]. 

\begin{table}[t!]
\begin{tabular}{c @{\hskip 0.5in} c } 
\hline
\hline
Qubit $A$  & Qubit $B$ \\
\hline
$E_{J,A}/h$ = 5.5 GHz & $E_{J,B}/h$ = 5.7 GHz\\
$E_{C,A}/h$ = 1.5 GHz &  $E_{C,B}/h$ = 1.2 GHz\\
$E_{L,A}/h$ = 1.0 GHz & $E_{L,B}/h$ = 1.0 GHz\\
$\Phi_{\mathrm{ext},A} = 0.5\Phi_0$ & $\Phi_{\mathrm{ext},B} = 0.5\Phi_0$\\
$\omega_A/2\pi$ = 0.606 GHz& $\omega_B/2\pi$ = 0.354 GHz\\
\hline
\hline
\\
\hline
\hline
Auxiliary system & Static couplings\\
\hline
 $\omega_C/2\pi$ = 7.5 GHz & $\gdc_{AC}/h$ = 0\\
 $U$ = 0 & $\gdc_{BC}/h$ = 0.8 GHz\\
  & $\gdc_{AB}/h$ = 0\\
\hline
\hline
\end{tabular}
\caption{Circuit parameters used for simulating the method (I) implementation of our gate 
(time-dependent couplings, static
intermediate auxiliary mode). Here $\omega_A$, $\omega_B$, and $\omega_C$ are the  frequency differences between the lowest and second lowest bare  energy levels of the uncoupled qubit $A$, qubit $B$ and auxiliary system, respectively. Parameters for the two qubits match those used in Ref.~\cite{Nesterov2018Microwave}, while the parameters for the auxiliary system and static couplings are chosen to mitigate spurious non-RWA processes.  }
\label{table:paramsone}
\end{table}

For our simulations, we choose realistic system parameters that yield reasonable coupling strengths and avoid spurious resonances; see Table~\ref{table:paramsone}.  The resulting transition frequencies and matrix elements for the $\Lambda$ and $\Lambda_{\rm bad}$ transitions (calculated using the scqubits  package \cite{groszkowski2021scqubits}) are shown in Table~\ref{table:frequencyone}. 
While our modulation tones are relatively high frequency (i.e., $\omega_{\mathrm{mod},A}/2\pi = 6.94$ GHz and $\omega_{\mathrm{mod},B}/2\pi = 2.86$ GHz), similar high-frequency modulations have been employed in recent experiments with flux-tunable couplings (see e.g.,~Ref.~\cite{Yao2017Universal}).

 For the method (I) implementation scheme, we require only a single static coupling,  between qubit $B$ and the auxiliary system (i.e., $\gdc_{BC}\ne 0$).  This has several desirable features.    The tunable coupling operator $ \hat{n}_j(\hata^\dagger + \hata)$ connects only transitions involving qubit $j$.  This implies that there is zero modulation crosstalk for this scheme (i.e., modulation tones designed to drive qubit $A$ transitions do not accidentally drive qubit $B$ transitions and vice versa).  Furthermore, the use of only a single static coupling also eliminates any qubit-qubit $ZZ$ interactions when the gate is inactive. 

\begin{table}[t!]
\begin{tabular}{c c c  c  c } 
\multicolumn{5}{c}{(a) $\Lambda$ system} \\
\hline
\hline
$|k\rangle$ & $|l\rangle$ & $\omegakl/2\pi$ & $|\langle k|\hat{n}_A(\hata^\dagger+\hata)|l\rangle|$ & $|\langle k|\hat{n}_B(\hata^\dagger+\hata)|l\rangle|$ \\
\hline
$|ee,0\rangle$& $|ge,1\rangle$ &  6.94 GHz & 0.12& 0.0\\
$|gf,0\rangle$ & $|ge,1\rangle$ & 2.86 GHz& 0.0 & 0.55\\
\hline
\hline
\\
\multicolumn{5}{c}{(b) $\Lambda_{\rm bad}$ system} \\
\hline
\hline
$|k\rangle$ & $|l\rangle$ & $\omegakl/2\pi$ & $|\langle k|\hat{n}_A(\hata^\dagger+\hata)|l\rangle|$ & $|\langle k|\hat{n}_B(\hata^\dagger+\hata)|l\rangle|$ \\
\hline
$|eg,0\rangle$ & $|gg,1\rangle$ & 6.54 GHz& 0.09& 0.0\\
$|ge,0\rangle$ & $|gg,1\rangle$&  6.80 GHz& 0.0&0.04\\
\hline
\hline
\end{tabular}
\caption{The frequencies ($\omegakl \equiv \varepsilon_{l} - \varepsilon_{k}$) and matrix elements [$|\langle k|\hat{n}_{A,B}(\hata^\dagger+\hata)|l\rangle|$] of $|k\rangle \rightarrow |l\rangle$ transitions for (a) $\Lambda$ system and (b) $\Lambda_{\rm bad}$ system, for the method (I) simulations. The modulation frequencies of the tunable couplings are $\omega_{\mathrm{mod},A}/2\pi = 6.94$ GHz and $\omega_{\mathrm{mod},B}/2\pi = 2.86$ GHz. The detuning of the $\Lambda_{\rm bad}$ system from the $\Lambda$ system are $2\chi/2\pi \equiv |\omega_{ee0,ge1}-\omega_{eg0,gg1}|/2\pi = 0.4$ GHz (for the left arm) and $|\omega_{gf0,ge1}-\omega_{ge0,gg1}|/2\pi = 3.94$ GHz (for the right arm), respectively. }\label{table:frequencyone}
\end{table}

In the following, we discuss the performance of a CZ gate implemented using this direct scheme. To this end, we simulated the evolution of the system's density matrix, keeping the 38 lowest-energy levels (a number sufficient for reaching convergence). 

\subsubsection{Gate performance with coherent errors only}\label{sec:coherent}

\begin{figure}[t!]
\centering
\includegraphics[width=\linewidth]{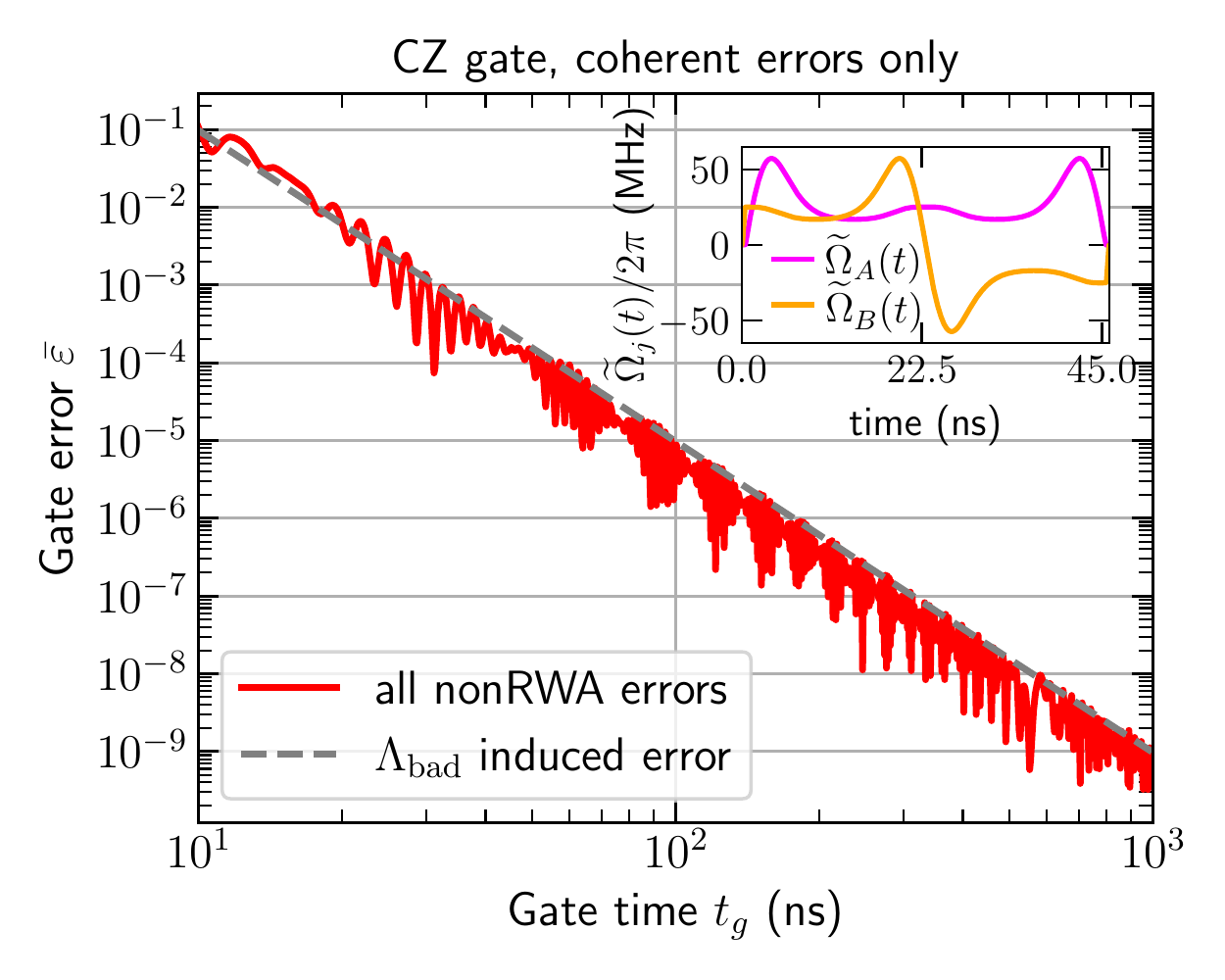}
\caption{State-averaged gate error $\bar{\varepsilon} = 1-\bar{F}$ of a CZ gate as a function of gate time $t_g$, in the absence of dissipation, for the direct tunable coupling scheme [method (I) of Fig.~\ref{fig:scheme}]. Results are calculated using 38 lowest-energy levels, including all non-RWA processes.  We use power-optimized SATD pulses to mitigate both non-RWA and adiabatic errors, as discussed in the main text (see the inset for an example of the SATD modulation tone envelope $\Omega_j(t)$ at $t_g$ = 45 ns).
The gate error of our full multilevel simulations (red curve) follows closely the scaling 
$c_{\mathrm{leak}}(\chi t_g)^{-4}$ [gray dashed line; cf.~Eq.~\eqref{eq:badscaling}] of a reduced model that considers only \textit{one} non-RWA process, i.e., the nonresonant drive of the left arm of the $\Lambda_{\rm{bad}}$ system.  This indicates that this is the dominant coherent error channel.   The circuit parameters used are given in Table~\ref{table:paramsone}.
}
\label{fig:Err_direct_coh}
\end{figure}

We first discuss gate performance by taking into account all coherent errors, but neglecting dissipation (this is treated in the next section).  
Coherent errors are mitigated as discussed above:  we use the SATD protocol with power optimal $\Omega_0$, i.e., $\Omega_0 = \Omega_{\mathrm{opt}}$, (see Appendix~\ref{subsec:RMSvoltage} for details), and chirped tone frequencies as per Eq.~\eqref{eq:modomegaapp} (see Appendix~\ref{sec:Magnuscorrection} for details). 
We also cancel unwanted $ZZ$ phases by optimizing the choice of the adiabatic phase $\gamma_0$ that determines our pulse envelopes; cf.~Eq.~\eqref{eq:gammazero}.
Using the parameters in Table~\ref{table:paramsone}, we can derive the CZ gate pulse shapes (i.e., modulation envelopes and chirped frequency profiles) for our SATD gate (see the inset of Fig.~\ref{fig:Err_direct_coh} and Appendix~\ref{sec:pulse} for an example power-optimal SATD pulse shape). 

Fig.~\ref{fig:Err_direct_coh} shows the
state-averaged gate error $\bar{\varepsilon}$ [Eq.~\eqref{eq:fidelity}] for a CZ gate as a function of gate time $t_g$. As expected, errors decrease with increasing $t_g$, due to the corresponding decrease in instantaneous power (which reduces the amplitude of non-RWA processes).  
We find that the errors here are dominated by leakage in the $\Lambda_{\rm bad}$ system, i.e.,~leakage from $|eg,0\rangle$ to $|gg,1\rangle$ (see Fig.~\ref{fig:badlambda}).   The oscillation of the coherent error with respect to the gate time is due to the detuned Rabi oscillations that shuttle the population back and forth between states $|eg,0\rangle$ and $|gg,1\rangle$. Ignoring the fast oscillation, we show in Fig.~\ref{fig:Err_direct_coh} that the scaling of the total gate error of the full system matches closely the scaling of leakage error  $\bar{\varepsilon}= c_{\mathrm{leak}}(\chi t_g)^{-4}$ [Eq.~\eqref{eq:badscaling}; dashed gray line] of the reduced model in Sec.~\ref{sec:gatelimit}, where $c_{\mathrm{leak}}\simeq 0.26$ and $\chi$ is the dispersive shift of the left-arm of the $\Lambda_{\mathrm{bad}}$ system.

\subsubsection{Gate performance with $T_1$ dissipation}\label{sec:fnoise}

We now add dissipation to our simulations.  As our qubits are operated at their sweet spot, the system is first-order insensitive to dephasing due to $1/f$ flux noise. The most dominant dissipations are thus $T_1$ decay processes. We consider $T_1$ relaxation that arises from dielectric loss in the circuit capacitors, as it is usually the dominant contribution~\cite{Nguyen2019High,Helin2021Universal}. The $T_1$ relaxation rate for the transition $|k\rangle\rightarrow|l\rangle$ due to the capacitor $C_j$ ($j = A,B,C$)  is~\cite{pop2014coherent,smith2020superconducting}
\begin{equation}\label{eq:T1diel}
1/(T_1)_{j;kl} = \frac{ 8 \omegalk E_{C_j}}{ |\omegalk|Q_{\mathrm{diel}}} \left[\coth\left(\frac{\omegalk}{2k_{\mathrm{B}} T} \right)+1\right] |\langle l|\hat{n}_j|k\rangle|^2.
\end{equation}
Here, $Q_{\mathrm{diel}}$ is the dielectric quality factor, $\omegalk \equiv \varepsilon_{k} - \varepsilon_{l}$ is the transition energy between the composite-system eigenstates $|l\rangle$ and $|k\rangle$, and $T$ is the temperature.
The total relaxation rate for a given transition is 
\begin{equation}
1/(T_1)_{kl} = \sum_{j=A,B,C} 1/(T_1)_{j;kl}. 
\end{equation}
Table~\ref{table:relaxation} shows the most dominant relaxation times involving any one of the six computational states $|k\rangle$ (four logical qubit states together with two ancillary states in the $\Lambda$ system). They are calculated for zero temperature $(T=0)$ and with  $Q_{\mathrm{diel}} = 10^{6}$.

\begin{table}[t]
\begin{tabular}{c c c } 
\hline
\hline
$|k\rangle$ & $|l\rangle$ & $(T_1)_{kl}$ ($\mu$s)\\
\hline
$|ge,1\rangle$ & $|ge,0\rangle$& 21.51 \\
$|gf,0\rangle$ & $|ge,0\rangle$ & 24.73 \\
\hline
\hline
\end{tabular}
\caption{ The $T_1$ relaxation times for the most dominant relaxation processes involving computational states, for the method (I) simulations. The relaxation times are calculated at zero temperature ($T = 0$)  and for a dielectric quality factor $Q_{\mathrm{diel}} = 10^6$. }\label{table:relaxation}
\end{table}

To simulate our gate including the effects of $T_1$ relaxation, we use the Lindblad master equation
\begin{equation}\label{eq:Lindblad}
    \partial_t\hat{\rho}(t) = -i [\hat{H}(t), \hat{\rho}(t)] + \sum_{l< k}\left(
        \hat{Z}_{kl}\hat{\rho}(t)\hat{Z}_{kl} - \frac{1}{2}\{ \hat{Z}^2_{kl},\hat{\rho}(t)\} \right).
\end{equation}
Here,  $\hat{\rho}(t)$ is the density matrix of our composite qubit-auxiliary system,  $\hat{H}(t)$ is the Hamiltonian used to realize the two-qubit gate [Eq.~\eqref{eq:Htotal}],
and the $\hat{Z}_{kl}$ are jump operators associated with each relaxation transition: 
\begin{equation}
\hat{Z}_{kl} = \sqrt{1/(T_1)_{kl}} |l\rangle\langle k|.   
\end{equation}

\begin{figure}[t!]
\centering
\includegraphics[width=\linewidth]{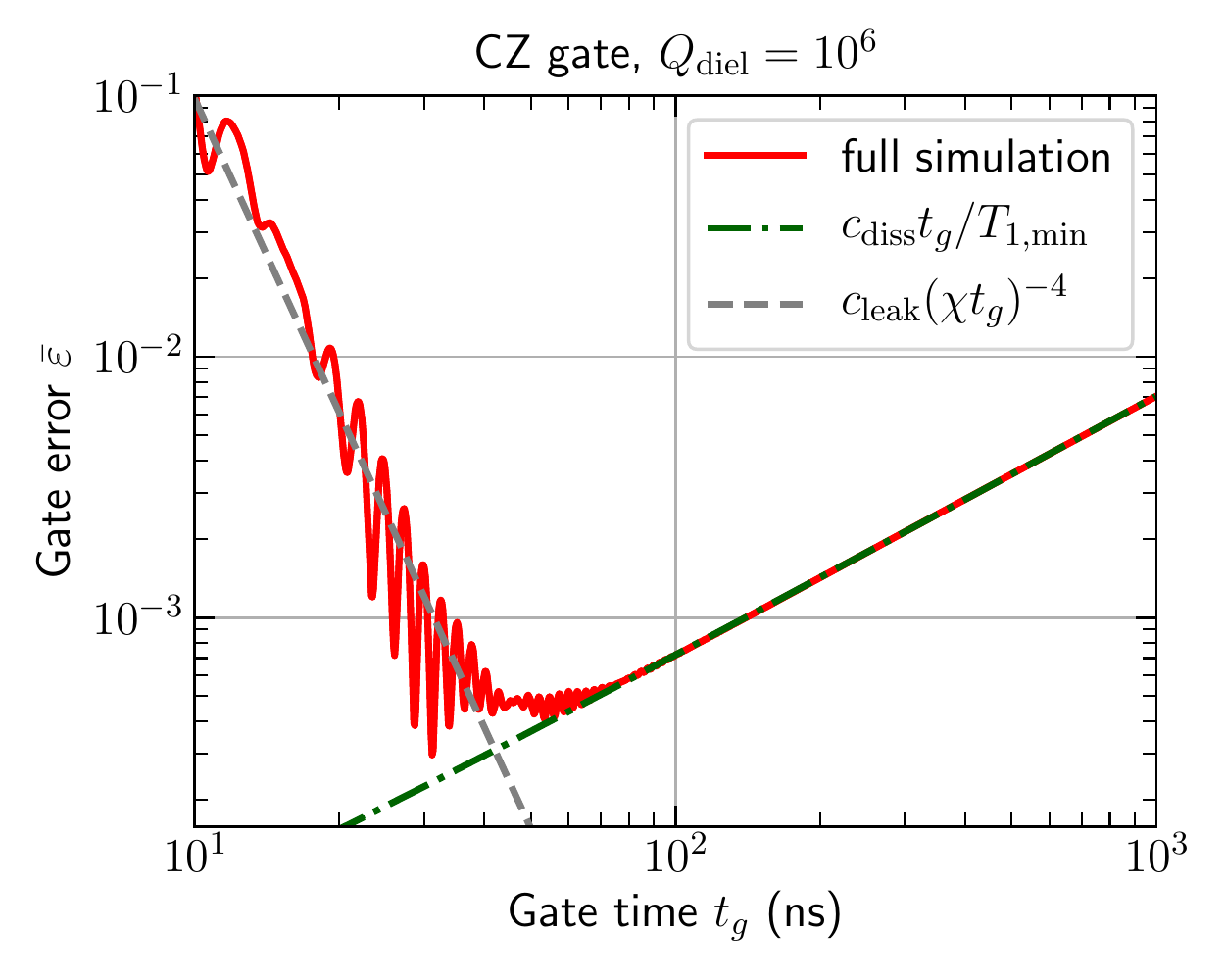}
\caption{State-averaged gate error $\bar{\varepsilon} = 1-\bar{F}$ of a CZ gate as a function of gate time $t_g$ for the direct tunable coupling scheme [method (I) of Fig.~\ref{fig:scheme}],
including both coherent and dissipative error channels.  Simulations treat 38 lowest-energy levels, all non-RWA processes, and $T_1$ relaxation due to dielectric capacitor loss with $Q_{\mathrm{diel}}=10^{6}$ (cf. Refs.~\cite{Wang2019Cavity,smith2020superconducting}). As before, power-optimized, chirped SATD pulses are used to realize the gate.  
In the short-gate-time regime, the gate error of the full model (red curve) follows the scaling of leakage error $\bar{\varepsilon}_{\mathrm{leak}} = c_{\mathrm{leak}}(\chi t_g)^{-4}$ [gray dashed line; Eq.~\eqref{eq:badscaling}] of the reduced model in Sec.~\ref{sec:badlambda}, where $c_{\mathrm{leak}} \simeq 0.26$. In the long-gate-time limit, dissipation dominates the gate error where the error scales with the numerically fitted results of $\bar{\varepsilon}_{\mathrm{diss}}  = c_{\mathrm{diss}} t_g/T_{1,\rm{min}}$ (green dash-dot line), where $c_{\mathrm{diss}} \simeq 0.152$.  We find that an error $\bar{\varepsilon} < 10^{-3}$ can be achieved over a range of gate times $45  \lesssim t_g \lesssim 60 {\,\rm ns}$.
Circuit parameters are listed in Table~\ref{table:paramsone}.}\label{fig:Err_direct_diss}
\end{figure}

We numerically simulate the above master equation to assess the impact of both non-RWA errors and dissipation on our gate performance; results are shown in Fig.~\ref{fig:Err_direct_diss}. The gate error is a nonmonotonic function of the gate time $t_g$. For short $t_g$, non-RWA errors limit performance, with the errors decreasing generally as $t_g$ increases.   As discussed in Sec.~\ref{sec:coherent}, the coherent errors for the full multilevel model follow closely the gate-time scaling of leakage errors due to the $\Lambda_{\mathrm{bad}}$ system alone, i.e., they match the error scaling of the reduced model in Sec.~\ref{sec:badlambda}.   This scaling is indicated by a gray dashed line in Fig.~\ref{fig:Err_direct_diss}.   
For longer $t_g$, the errors are dominated by dissipation and increase linearly with increasing $t_g$. Letting $T_{1,\rm{min}}$ be the shortest relaxation time in the system, we fit the dissipation-induced error to $\bar{\varepsilon}_{\mathrm{diss}}= c_{\mathrm{diss}} t_g/T_{1,\rm{min}}$, as shown by the green dash-dot line, where $c_{\mathrm{diss}} \simeq 0.152$.

As we understand the error scaling with $t_g$ in both the coherent and dissipation limited regimes, we can make estimates for how the minimum error and optimal gate time scale with system parameters.  Note that the coherent error is controlled by the dispersive coupling $\chi$ (which controls the detuning of the $\Lambda_{\mathrm{bad}}$ system), whereas the dissipation is controlled by $T_{1,\rm{min}}$.  By adding the two error scalings $\bar{\varepsilon}_{\mathrm{leak}}$ and $\bar{\varepsilon}_{\mathrm{diss}}$ in quadrature and then minimizing with respect to $t_g$, we find a minimum error that scales like
\begin{equation}
    \bar{\varepsilon}_{\rm{min}} = c_{\mathrm{min}}(\chi T_{1,\rm{min}})^{-4/5},
\end{equation}
at an optimal gate time $t_g \simeq 7.96  (T_{1,\rm{min}}/\chi^4)^{1/5}$, where $c_{\mathrm{min}} \simeq 1.35$.
Our numerics reveal that an extremely high gate fidelity $> 0.9995$ can be achieved for gate times in the range $t_g = 45$--$60$ ns; this is in good agreement with the above estimate.  
For the optimal gate time of $t_g = 45$ ns, the modulation amplitudes [cf. Eq.~\eqref{eq:gtac}]  have maximum values of $g^{\mathrm{ac}}_{A,\mathrm{max}}/2\pi \simeq 478.6$ MHz and $g^{\mathrm{ac}}_{B,\mathrm{max}}/2\pi \simeq 104.4$ MHz.
While the gate performance is extremely promising, we note that we have not modeled the internal workings of the modulated tunable couplings.  Nonetheless, these results show that method (I) is an extremely promising implementation strategy.

\subsection{Method (II): static couplings, time-dependent intermediate auxiliary mode}

We next consider an alternate implementation strategy for our gate, where one uses static interactions between the auxiliary system and both qubits, and time modulates the frequency of the auxiliary system [method (II) of Fig.~\ref{fig:scheme}]~\cite{McKay2016Universal,reagor2018demonstration,Mundada2019Suppression,Ganzhorn2019Gate,Yuan2020High,abrams2020implementation,Ganzhorn2020Benchmarking,Stehlik2021Tunable,Sung2021Realization}. This corresponds to choosing the modulation term in Eq.~\eqref{eq:Htotal} to be
\begin{equation}
\Hdr(t) =  \delta \omega_C(t) \aopd\aop,
\end{equation}
where $\delta \omega_C(t)=  \frac{1}{2}\sum_{j=A,B} \left( \gac_{j}(t) e^{-i\int_0^t\omegatmodj(t') dt'} + \textrm{c.c.} \right)$. For our SATD gate, the modulation envelopes $\gac_j(t)$ are related to the SATD pulse amplitudes $\tilde{\Omega}_{j}(t)$ [Eqs.~\eqref{eq:omegaSATD}] by 
\begin{subequations}\label{eq:gac}
\begin{align}
\gac_A(t) &= \tilde{\Omega}_{A}(t)/\langle ge,1|\hat{a}^{\dagger}\hat{a} |ee,0 \rangle, \\
\gac_B(t) &= \tilde{\Omega}_{B}(t)/\langle ge,1|\hat{a}^{\dagger}\hat{a}|gf,0\rangle, 
\end{align}
\end{subequations}
respectively corresponding to the left and right arms of the $\Lambda$ system [see Fig.~\ref{fig:badlambda}(a)]. 

\begin{table}[t!]
\begin{tabular}{c @{\hskip 0.5in}c } 
\hline
\hline
Qubit $A$  & Qubit $B$ \\
\hline
$E_{J,A}/h$ = 4.5 GHz & $E_{J,B}/h$ = 3.5 GHz\\
$E_{C,A}/h$ = 1.8 GHz & $E_{C,B}/h$ = 1.1 GHz\\
$E_{L,A}/h$ = 1.5 GHz & $E_{L,B}/h$ = 1.0 GHz\\
$\Phi_{\mathrm{ext},A} = 0.5\Phi_0$&  $\Phi_{\mathrm{ext},B} = 0.5\Phi_0$\\
$\omega_A/2\pi$ = 1.79 GHz&  $\omega_B/2\pi$ = 0.86 GHz\\
\hline
\hline
\\
\hline
\hline
Auxiliary system & Static couplings\\
\hline
 $\omega_C/2\pi$ = 1.11 GHz& $\gdc_{AC}/h$ = 0.63 GHz\\
 $U$ = 5 MHz& $\gdc_{BC}/h$ = 0.6 GHz\\
  & $\gdc_{AB}/h$ = 0.04 GHz\\
\hline
\hline
\end{tabular}
\caption{Circuit parameters for the method (II) implementation of our gate (static couplings, time-dependent auxiliary mode frequency). Here $\omega_A$, $\omega_B$, $\omega_C$ are the frequency differences between the lowest and second lowest bare energy levels of the uncoupled qubit $A$, qubit $B$ and auxiliary system (coupler), respectively. }
\label{table:paramstwo}
\end{table}

\begin{table}[h]
\begin{tabular}{c c c  c } 
\multicolumn{4}{c}{(a) $\Lambda$ system} \\
\hline
\hline
$|k\rangle$ & $|l\rangle$ & $\omegakl/2\pi$ & $|\langle k|\hata^\dagger\hata|l\rangle|$  \\
\hline
$|ee,0\rangle$& $|ge,1\rangle$ &  -0.84 GHz & 0.194\\
$|gf,0\rangle$ & $|ge,1\rangle$ & -2.43 GHz & 0.111\\
\hline
\hline
\\
\multicolumn{4}{c}{(b) $\Lambda_{\rm bad}$ system} \\
\hline
\hline
$|k\rangle$ & $|l\rangle$ & $\omegakl/2\pi$& $|\langle k|\hata^\dagger\hata|l\rangle|$  \\
\hline
$|eg,0\rangle$ & $|gg,1\rangle$ & -0.71 GHz & 0.181\\
$|ge,0\rangle$ & $|gg,1\rangle$&  0.31 GHz & 0.366\\
\hline
\hline
\end{tabular}
\caption{The frequencies 
($\omegakl \equiv \varepsilon_{l} - \varepsilon_{k}$) and matrix elements ($\langle k|\hata^\dagger \hata |l\rangle$) of $|k\rangle \rightarrow |l\rangle$ transitions for (a) $\Lambda$ system and (b) $\Lambda_{\rm bad}$ system, for the method (II) simulations. The modulation frequencies of the auxiliary system are $\omega_{\mathrm{mod},A}/2\pi = 0.84$ GHz and $\omega_{\mathrm{mod},B}/2\pi = 2.43$ GHz. The detuning of the left arm of the $\Lambda_{\rm bad}$ system from the left arm of the $\Lambda$ system is $2\chi/2\pi \equiv||\omega_{ee0,ge1}|-|\omega_{eg0,gg1}||/2\pi = 0.13$ GHz. The detuning of the right arm of the $\Lambda_{\rm bad}$ system from the left arm of the $\Lambda$ system is $||\omega_{gf0,ge1}|-|\omega_{eg0,gg1}||/2\pi = 0.53$ GHz.}\label{table:frequencytwo}
\end{table}

For this scheme, we need nonzero static couplings between the auxiliary system and each qubit, i.e., $\gdc_{AC},\gdc_{BC} \neq 0$.  This will generically lead to a static $ZZ$ interaction between the qubits. While the resulting phase error $\phi_{ZZ}$ that arises during the gate implementation can be corrected by simply shifting the phase $\gamma_0$ of our control tones, the bigger issue is that this $ZZ$ interaction remains on even when the gate is turned off.  By judiciously choosing the parameters of our circuits and introducing a direct static interqubit coupling ($\gdc_{AB}$) with a suitably chosen strength, one can in principle completely eliminate the static $ZZ$ interaction; analogous strategies have been used in other systems; see, e.g.,~Ref.~\cite{Sung2021Realization}.  For our setup (two fluxonium qubits connected via a frequency-tunable transmon coupler), the $ZZ$ interactions can be eliminated if we pick parameters so that the transmon-coupler frequency is in between the two qubit frequencies. This motivates our choice of having asymmetric parameters for qubits $A$ and $B$ (see Table~\ref{table:paramstwo}).

Optimal gate performance also requires parameter choices that minimize non-RWA errors.  After an approximate numerical optimization, we pick parameters that mitigate both the static $ZZ$ interaction and non-RWA error channels, while still remaining compatible with experiments. The resulting parameters are given in Table~\ref{table:paramstwo}. For these values,  the $ZZ$ interaction strength is negligible, i.e., $(\varepsilon_{11}+\varepsilon_{00}-\varepsilon_{01}-\varepsilon_{10})/h \simeq 6.57$ kHz; it could in principle be minimized even further with more parameter fine tuning.  The frequencies and matrix elements of the $\Lambda$ and $\Lambda_{\rm bad}$ system transitions for our system parameters are shown in Table~\ref{table:frequencytwo}.

\begin{figure}[t!]
\centering
\includegraphics[width=\linewidth]{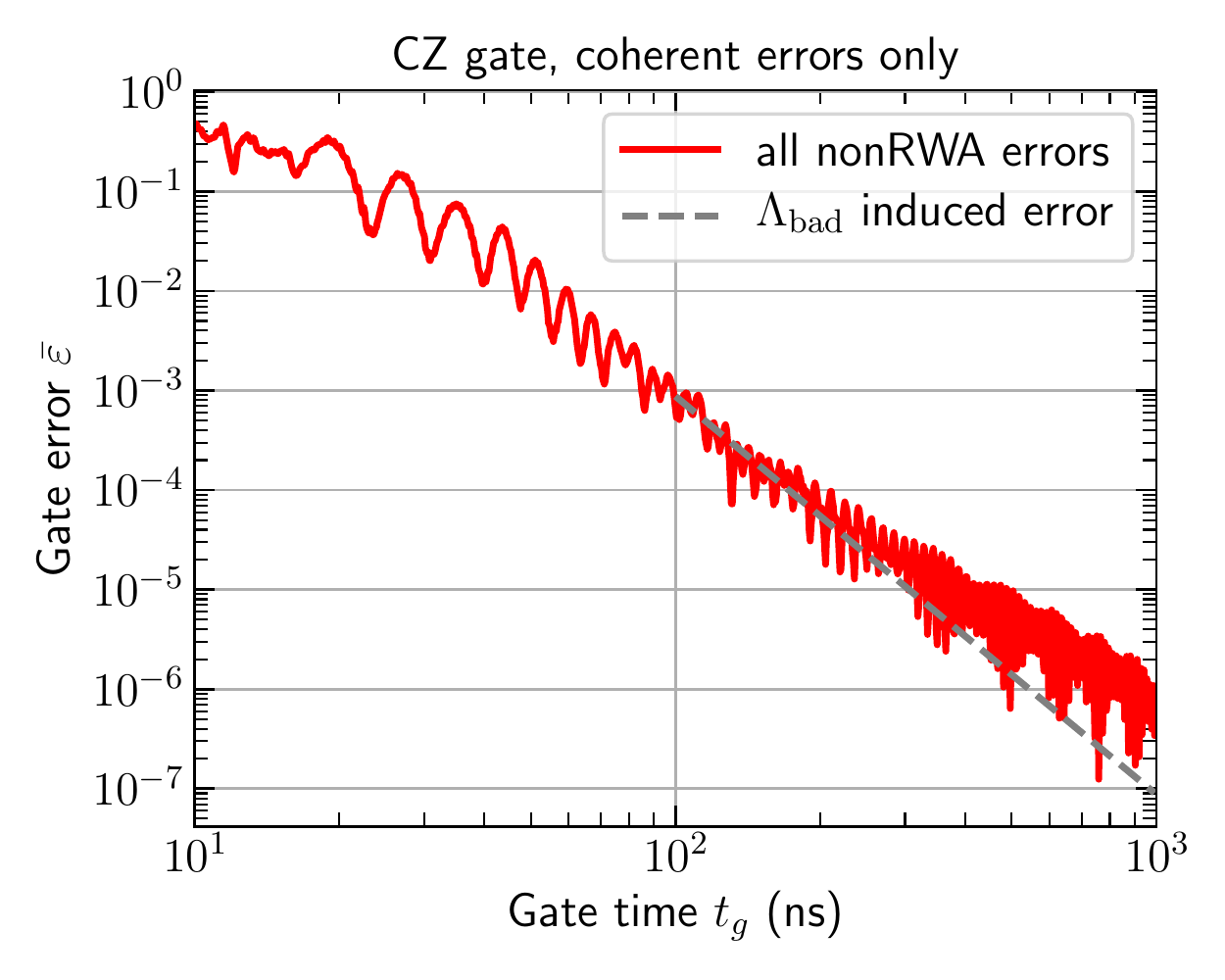}
\caption{
    State-averaged gate error $\bar{\varepsilon} = 1-\bar{F}$ of a CZ gate as a function of gate time $t_g$, for the method (II) implementation of our gate (cf.~Fig.~\ref{fig:scheme}), in the absence of dissipation.  Results are calculated using 20 lowest-energy levels and all non-RWA processes.  
    We use power-optimized SATD pulses with frequency chirping (as discussed in the main text), and optimize the adiabatic phase $\gamma_0$ to cancel spurious $ZZ$ interaction terms.  
    The coherent error has contributions from several processes, and hence is slightly above the scaling that would be expected if the only error was $\Lambda_{\mathrm{bad}}$ system leakage, 
    i.e.,~the scaling $c_{\mathrm{leak}}(\chi t_g)^{-4}$ [gray dashed line; see Eq.~\eqref{eq:badscaling}]
    Circuit parameters are given in Table~\ref{table:paramstwo}. }\label{fig:Err_indirect_coh}
\end{figure} 

\begin{figure}[t!]
\centering
\includegraphics[width=\linewidth]{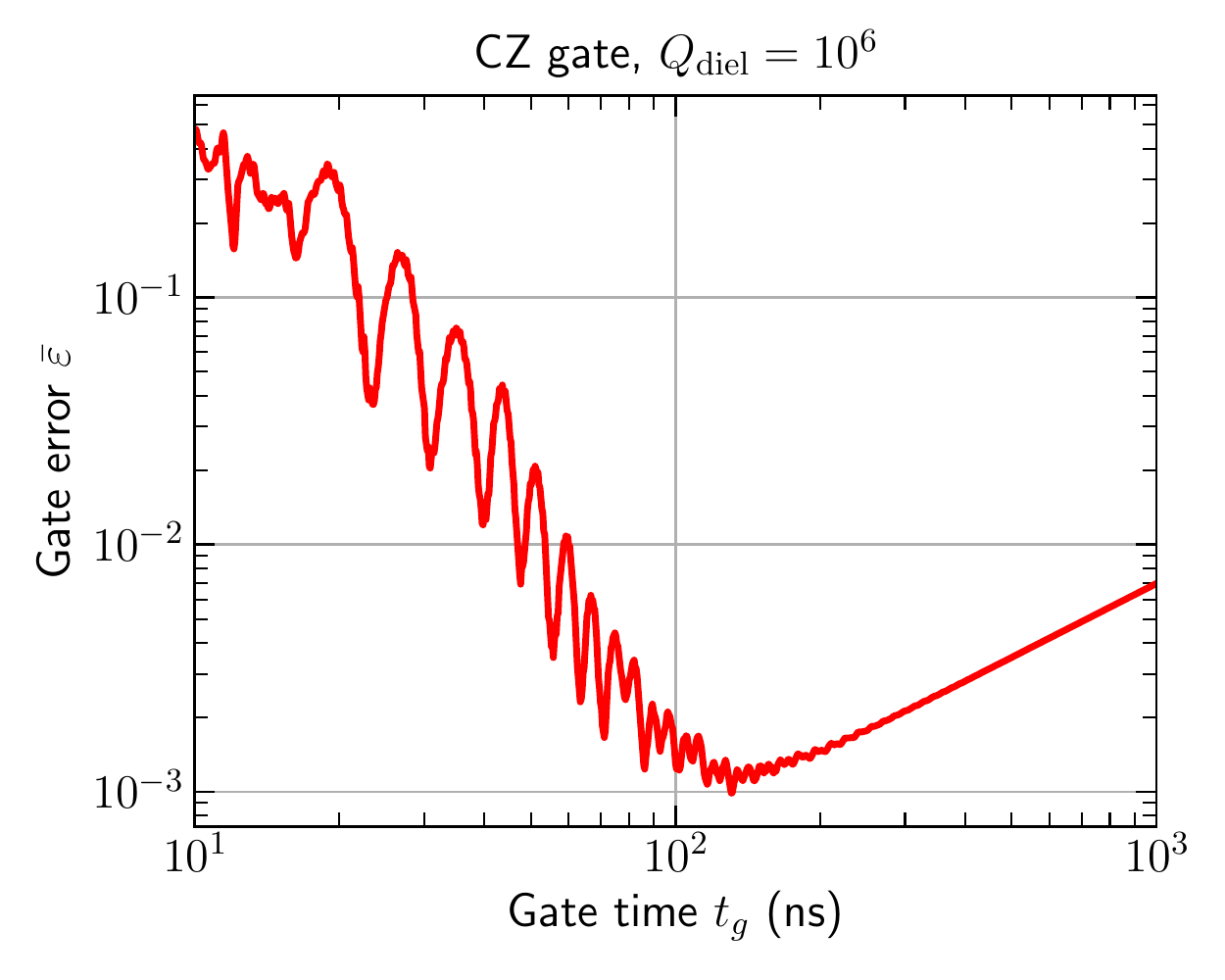}
\caption{State-averaged gate error $\bar{\varepsilon} = 1-\bar{F}$ for the CZ gate as a function of gate time $t_g$ for the method (II) implementation of our scheme (cf.~Fig.~\ref{fig:scheme}), including both coherent and dissipative errors.  Results are calculated using 20 lowest-energy levels, all non-RWA processes, and $T_1$ relaxation due to dielectric capacitor loss (with  $Q_{\mathrm{diel}}=10^{6}$). We use power-optimized SATD pulse shapes with frequency chirping (see the main text), and correct for unwanted $ZZ$ interaction phase shifts by optimizing the adiabatic phase shift $\gamma_0$ 
[cf.~Eq.~\eqref{eq:gammazero}]. 
Circuit parameters are given in Table~\ref{table:paramstwo}.}\label{fig:Err_indirect_diss}
\end{figure} 
Using the optimal parameter set as shown in Table~\ref{table:paramstwo}, we simulated our accelerated CZ gate in the Hilbert space spanned by the 20 lowest-energy levels of the system. Figures~\ref{fig:Err_indirect_coh} and~\ref{fig:Err_indirect_diss} show the gate errors as functions of gate time, calculated with all non-RWA error channels for the case without and with $T_1$ dissipation, respectively. 

We find that there are several equally dominant non-RWA error channels.  
In addition to leakage from $|eg,0\rangle$ to  $|gg,1\rangle$ (arising from nonresonant driving of the left arm of the $\Lambda_{\rm bad}$ system), there is also, e.g., the nonresonant driving of the right arm of the $\Lambda_{\rm bad}$ system by the modulation tones $\Omega_A(t)$.  This causes leakage from  $|ge,0\rangle$ to $|gg,1\rangle$.  While these coherent errors generally decrease with increasing gate time, the total gate error no longer follows the gate-time scaling $c_{\mathrm{leak}}(\chi t_g)^{-4}$ [gray dashed line; cf.~Eq.~\eqref{eq:badscaling}] of the gate error obtained from the reduced model  that considers only the nonresonant drive of the left arm of the $\Lambda_{\rm bad}$ system.  Again, this signifies the presence of several non-RWA error channels.  

If dissipation is taken into account as in  Fig.~\ref{fig:Err_indirect_diss}, it will dominate the gate errors in the long-gate-time regime (see $t_g \gtrsim 200$ ns in Fig.~\ref{fig:Err_indirect_diss}) and wash away the oscillations in the gate error. In this dissipation-dominated regime, the gate error increases linearly with increasing gate time. As shown in Fig.~\ref{fig:Err_indirect_diss}, at an intermediate gate time regime  ($t_g \approx 130$ ns), where neither the coherent errors nor dissipation dominates, we can obtain a gate fidelity of approximately 0.999 (even with the inclusion of all non-RWA error channels and $T_1$ dissipation  with dielectric quality factor $Q_{\mathrm{diel}} = 10^6$, where the dominant $T_1$ relaxation times are shown in Table~\ref{table:relaxationindirect}). For the optimal gate time of $t_g = 130$ ns, the modulation amplitudes [cf. Eq.~\eqref{eq:gac}]  have maximum values  of $g^{\mathrm{ac}}_{A,\mathrm{max}}/2\pi \simeq 102.5$ MHz and $g^{\mathrm{ac}}_{B,\mathrm{max}}/2\pi \simeq 179.1$ MHz.

\begin{table}[h]
\begin{tabular}{c c c } 
\hline
\hline
$|k\rangle$ & $|l\rangle$ & $(T_1)_{kl}$ ($\mu$s)\\
\hline
$|gf,0\rangle$ & $|ge,0\rangle$ & 39.55 \\
$|ge,1\rangle$ & $|ge,0\rangle$& 92.67 \\
$|gf,0\rangle$ & $|gg,1\rangle$& 99.38 \\
$|ee,0\rangle$ & $|ge,0\rangle$ & 103.61 \\
$|eg,0\rangle$ & $|gg,0\rangle$ & 103.98 \\
\hline
\hline
\end{tabular}
\caption{The $T_1$ relaxation times for the most dominant processes involving computational states
for the method (II) simulations. These are calculated for dielectric capacitor loss with a dielectric quality factor $Q_{\mathrm{diel}} = 10^6$ and for zero temperature ($T = 0$). }
\label{table:relaxationindirect}
\end{table}

\section{Conclusions}
\label{sec:conclusions}

In this work we introduce and analyze a method that harnesses the basic physics of STIRAP to realize an accelerated adiabatic geometric two-qubit gate.  Our approach is more flexible and resource efficient than previous proposals for STIRAP-based two-qubit gates in atomic platforms, and is especially well suited to platforms using an auxiliary system (transmission line, cavity, qubit) as a coupler.  We analyze in detail implementations of our basic idea in a system of two fluxonium superconducting qubits, considering both implementations based on direct modulation of coupling amplitudes, and based on modulation of the auxiliary system frequency.  Using realistic parameters, we find competitive gate fidelities and gate times.  We also discuss how our protocols have an inbuilt robustness to certain kinds of parameter variations, a feature derived from their connection to a purely adiabatic protocol.

In future work, it would be interesting to explore how the basic gate mechanism introduced here could be used in other systems, ranging from advanced superconducting qubits (e.g.,~$0$-$\pi$ qubits) to atomic platforms (e.g.,~neutral atom systems employing Rydberg levels).  It would also be interesting to study how our approach could be optimally employed in modular approaches to quantum computing, i.e., to realize remote gates in multiqubit systems coupled via a common bus mode.

\acknowledgments 
This work is financially supported by the Army Research Office under Grant Number W911NF-19-1-0328. We are grateful to the University of Chicago Research Computing Center
for computing resources to perform the calculations in this paper.

\appendix
\section{Physics of the STIRAP gate}
\label{sec:tripodgate}

In this section, we discuss the basic STIRAP geometric gate that utilizes a three-level $\Lambda$ configuration [Fig.~\ref{fig:qubitbus}(a)]. The STIRAP gate was proposed in Refs.~\cite{duan2001geometric,Kis2002Qubit}  for the adiabatic case and its accelerated version was subsequently formulated in Ref.~\cite{Ribeiro2019Accelerated}; we follow Ref.~\cite{Ribeiro2019Accelerated} for the discussion below. 

We begin by noting that Hamiltonian $\Hstirap(t)$ [Eq.~\eqref{eq:Hstirap}] of the $\Lambda$ system has an instantaneous zero-energy dark state
\begin{equation}
|\rmd(t)\rangle = \cos[\theta(t)]|q\rangle - e^{i\gamma(t)}\sin[\theta(t)]|a_2\rangle,
\end{equation}
which is a superposition of a qubit state $|q\rangle$ and the ancillary state $|a_2\rangle$, which are the two lower levels of the $\Lambda$ system. Note that $|\rmd(t)\rangle$ is orthogonal to the ancillary state $|a_1\rangle$, i.e., the upper level of the $\Lambda$ system. 
The STIRAP gate operates by utilizing a geometric evolution of the zero-energy dark state. 
By performing a ``double-STIRAP protocol", one realizes a cyclic adiabatic  evolution  $|q\rangle \rightarrow |a_2\rangle \rightarrow |q\rangle$. This results in a geometric phase being imprinted on the qubit state $|q\rangle$~\cite{duan2001geometric,Kis2002Qubit}. If $|q\rangle$ is chosen to be the logical qubit state $|11\rangle$, the protocol then realizes an arbitrary geometric controlled-phase gate~\cite{duan2001geometric}. Note that such a geometric gate can be performed without requiring precise timing of the pulses.

The double-STIRAP protocol relies on cyclically varying $\theta(t)$, where here we take
\begin{equation}\label{eq:theta}
\theta(t) = \begin{cases}
\displaystyle\frac{\pi}{2} P(t/t_g), & \displaystyle 0 \leq t \leq \frac{t_g}{2}\\[8pt]
\displaystyle\frac{\pi}{2} \left[1- P\left(\frac{t}{t_g} - \frac{1}{2} \right) \right], & \displaystyle \frac{t_g}{2}< t \leq t_g,
\end{cases}
\end{equation}
with $P(x)$ being a monotonic function that increases from $P(0) = 0$ to $P(1/2) = 1$. For a smooth on-and-off switch of the control fields, we pick a polynomial that satisfies $\dot{\theta} (0) = \dot{\theta}(t_g/2) = \dot{\theta}(t_g) = \ddot{\theta} (0) = \ddot{\theta}(t_g/2) = \ddot{\theta}(t_g) = 0$. Specifically, we take~\cite{Ribeiro2019Accelerated} 
\begin{equation}\label{eq:Px}
P(x) = 6 \left(2x\right)^5 - 15 \left(2x \right)^{4} + 10 \left(2x\right)^3.
\end{equation} 
To get a geometric phase $\gamma_0$, we choose the relative phase $\gamma(t)$ between the control fields as~\cite{Ribeiro2019Accelerated}
\begin{equation}
\gamma(t) = \gamma_0 \Theta \left(t - \frac{t_g}{2} \right),
\end{equation}
where $\Theta(t)$ denotes a Heaviside step function. 

For the adiabatic limit [$\dot{\theta}(t)/\Omega_0 \rightarrow 0$], one can show that~\cite{Ribeiro2019Accelerated} the evolution of the qubit subspace is decoupled from that of the ancillary-level subspace (the subspace spanned by the ancillary states $|a_1\rangle$ and $|a_2\rangle$) and the dark state $|\rmd(t)\rangle$ acquires a geometric phase $\gamma_0$ at $t = t_g$. The result is a geometric gate that is described by a unitary  $\Ugq$ [Eq.~\eqref{eq:UQubitAdiabatic}] in the qubit subspace. The full unitary evolution in the adiabatic limit is given by
$\hat{U}_{\mathrm{G}} = \Ugq\oplus \hat{U}_{\mathrm{G,anc}}$, where the $\hat{U}_{\mathrm{G.anc}}$ are the unitary operators in ancillary subspaces (see Ref.~\cite{Ribeiro2019Accelerated}). 

\section{SATD dressing for accelerated protocols}\label{sec:SATD}
 Here we provide a brief overview of the ``dressed state" approach of the STA protocols~\cite{Baksic2016Speeding,Ribeiro2019Accelerated}, used for accelerating adiabatic gates.       
The basic idea is to let the system evolve following a ``dressed" state that coincides with the adiabatic state at the beginning and end of the protocol.  This is accomplished by choosing a dressing function $\nu(t)$ that goes to zero at $t = 0$ and $t = t_g$. As in Refs.~\cite{Baksic2016Speeding,Ribeiro2019Accelerated}, we dress the $|\rmd(t)\rangle$ [Eq.~\eqref{eq:d2}] as $|\rmd_{\nu}(t)\rangle$, i.e.,
\begin{equation}\label{eq:dressingJ}
    |\rmd_\nu(t)\rangle = \exp \left[-i\nu(t)\hat{J}_{x}\right] |\rmd(t)\rangle,
\end{equation}
where $\hat{J}_x = (|\mathrm{b}_+(t)\rangle \langle  \rmd(t)| + |\mathrm{b}_-(t)\rangle \langle \rmd(t) | +\mathrm{H.c.})/\sqrt{2}$ with $|\mathrm{b}_\pm(t)\rangle$ being the bright eigenstates of the adiabatic Hamiltonian $\Hstirap(t)$ having eigenenergies $\pm \Omega_0/2$.  

As elaborated in Ref.~\cite{Ribeiro2019Accelerated}, to have the dark state acquire a purely geometric phase that is equal to the adiabatic geometric phase $\gamma_0$, we have to impose a constraint $\nu(t_g/2) =0$. To fulfill this constraint, we use the SATD~\cite{Baksic2016Speeding,Ribeiro2019Accelerated} dressing function
\begin{equation}\label{eq:nusatd}
\nu(t) = \nu_{\mathrm{SATD}}(t) \equiv \arctan \left[\frac{2\dot{\theta}(t)}{\Omega_0} \right].
\end{equation}
As shown in Refs.~\cite{Baksic2016Speeding,Ribeiro2019Accelerated}, the accelerated protocol obtained using the SATD dressing is implemented by changing the original pulse sequence following the \emph{analytical} formula given in Eq.~\eqref{eq:omegaSATD} of the main text. 

\section{Mitigation and cancellation of non-RWA errors}\label{sec:enhancedSATD}
Coherent errors also arise from nonresonant couplings that would be neglected within the RWA.  We can partially mitigate their effects by using the strategy introduced in Ref.~\cite{Setiawan2021Analytic}. To understand this approach, we start by writing the full system Hamiltonian including non-RWA terms as
 \begin{equation}\label{eq:Htot}
\hat{H}(t) = \Hstirap(t) + \Herr(t),
\end{equation}
where $\Hstirap(t)$ is the Hamiltonian for the ideal resonant (RWA) processes as in Eqs.~\eqref{eq:Hstirap} and $\Herr(t)$ describes all the unwanted nonresonant dynamics.

%%%%%%%%%%%%%%%%%%%%%%

Following Ref.~\cite{Setiawan2021Analytic}, we implement an enhanced version of the SATD protocol that mitigates the effects of non-RWA errors. This protocol has two key steps.

\subsection{Step 1: power optimization}
\label{subsec:RMSvoltage}

Given the degeneracy of perfect STA protocols in the RWA limit, we first mitigate unwanted effects from non-RWA processes described by $\Herr(t)$ [Eq.~\eqref{eq:Htot}], by choosing the SATD protocol that minimizes the rms value of the control field amplitude
\begin{equation}\label{eq:VRMS}
    g^{\mathrm{ac}}_{\mathrm{rms}} \equiv \sqrt{\frac{1}{t_g} \int_0^{t_g} (\left|g^{\mathrm{ac}}_A(t)\right|^2 + \left|g^{\mathrm{ac}}_B(t)\right|^2) dt}.
\end{equation}
Here, $\gac_j(t)$ is the amplitude of the modulation tone $j$ ($j=A,B$) (which includes the SATD correction);
see~Eqs.~\eqref{eq:gtac} and~\eqref{eq:gac} for the expression of $\gac_j(t)$ for the direct tunable couplings and indirect-modulated auxiliary mode scheme, respectively. This power minimization strategy can be understood from the fact that coherent errors due to non-RWA dynamics generally decrease as the pulse amplitude is reduced. One effect of the non-RWA processes is to induce time-dependent energy shifts of the computational levels whose magnitudes scale as $|g^{\mathrm{ac}}_{\mathrm{rms}}|^2$. The leading order of the energy shifts can be derived from second-order perturbation theory or a Magnus-expansion-based approach~\cite{Ribeiro2017Systematic,roque2020engineering} as
\begin{subequations}\label{eq:energyshift}
\begin{align}
    \delta\varepsilon_{k}(t) &= \sum_{\substack{j = A,B\\ \sigma = \pm}}     \sum_{l \,  |\Deltakl \neq 0}  \frac{|\gac_j(t) \langle k|\hat{n}_j(\hata^\dagger+\hata) |l\rangle|^2}{4\Deltakl},\\
		\delta\varepsilon_{k}(t) &= \sum_{\substack{j = A,B\\ \sigma = \pm}}     \sum_{l \,  |\Deltakl \neq 0}  \frac{|\gac_j(t) \langle k|\hata^\dagger\hata|l\rangle|^2}{4\Deltakl},
\end{align}
\end{subequations}
for the direct tunable couplings [method (I) of Fig.~\ref{fig:scheme}] and frequency-modulated auxiliary-mode scheme [Method (II) of Fig.~\ref{fig:scheme}], respectively. Here, $\Deltakljpm = \varepsilon_{k} - \varepsilon_{l} \pm \omegamodj$ is the detuning 
of the transition $|k\rangle \leftrightarrow |l\rangle$ from the modulation tone $\omegamodj$, while $\hat{n}_j(\hata^\dagger+\hata)$ and $\hata^\dagger\hata$ are the modulation operators for the direct tunable couplings  and frequency-modulated auxiliary-mode scheme, respectively.
The sums in Eqs.~\eqref{eq:energyshift} are evaluated for all nonresonant processes involving the computational state $|k\rangle$, which can be any of the four qubit states ($|gg,0\rangle$, $|ge,0\rangle$, $|eg,0\rangle$, $|ee,0\rangle$) or the two ancillary states in the $\Lambda$ system ($|ge,1\rangle$, $|gf,0\rangle$). The intermediate states $|l\rangle$ in 
Eqs.~(\ref{eq:energyshift}) could be either any of the six computational levels (i.e., resulting in crosstalk processes) or noncomputational states (i.e., ``leakage" levels). Since the energy shifts of the computational levels increase with the pulse amplitude, the SATD protocol that minimizes coherent errors has the minimum driving power $|\gac_{\mathrm{rms}}|^2$.
As derived in Ref.~\cite{Setiawan2021Analytic}, this power-optimal SATD protocol has a value of adiabatic pulse magnitude $\Omega_0 = \Omega_{\mathrm{opt}}$, where $\Omega_{\mathrm{opt}}/2\pi \simeq 1.135/t_g$.

%%%%%%%%%%%%%%%%%%%%%%%%%%%%%%%%%%%%%%%

\subsection{Step 2: cancellation of non-RWA errors via modification of accelerated protocol pulses}\label{sec:Magnuscorrection}
Having minimized the energy shifts $\delta \varepsilonk(t)$ [Eqs.~(\ref{eq:energyshift})] due to nonresonant processes by choosing the power-optimal SATD protocol, our next step to enhance the gate fidelity is to partially {\it cancel} the non-RWA errors by modifying the SATD pulses. To this end, we use an analytic correction strategy that follows Ref.~\cite{Setiawan2021Analytic}

Our correction technique is to cancel the undesired time-dependent non-RWA energy shifts by introducing a time-dependent variation of the two modulation frequencies  $\omegamodj$  ($j = {A,B}$) (i.e.,~a frequency chirp):
\begin{align}\label{eq:modomegaapp}
\omegamodj & \rightarrow \omegatmodj(t) = \omegamodj+  \delta\omegamodj(t)
\end{align}
with $\delta\omega_j(t)$ being
\begin{subequations}\label{eq:omegamod}
\begin{align}
\delta \omega_{\mathrm{mod},A}(t)&=  \delta\varepsilon_{ge1}(t) -   \delta\varepsilon_{ee0}(t), \\
\delta \omega_{\mathrm{mod},B}(t)&=  \delta\varepsilon_{ge1}(t) -   \delta\varepsilon_{gf0}(t),
\end{align}
\end{subequations}
and $\delta\varepsilon_{k}(t)$ given in Eqs.~(\ref{eq:energyshift}). These frequency shifts partially correct the non-RWA error, as they ensure that at every instant of time, each modulation tone is resonant (up to the leading order) with the corresponding transition it is supposed to modulate.

Our correction approach thus results in a two-step modification of the original pulse in Eqs.~(\ref{eq:omegadrive}). 
For a given gate time $t_g$, we first choose a power-optimal value of $\Omega_0$ as discussed in Sec.~\ref{subsec:RMSvoltage} and introduce the SATD correction to the original pulses as in Eqs.~(\ref{eq:omegaSATD}). Next, we chirp each of the two central tone frequencies as shown in 
Eq.~(\ref{eq:modomegaapp}). Thus, the total modification of the modulation tone $j$ ($j = A,B$) is given by
\begin{align}
\Omega_{j}(t) &\,\mathrm{exp}\left(i\omegamodj t\right)
\rightarrow \tilde{\Omega}_{j}(t) \,\mathrm{exp}\left(i\int_0^t\omegatmodj(t') dt' \right),
\end{align}
where the $\tilde{\Omega}_{j}(t)$ [see Eqs.~\eqref{eq:omegaSATD}] are the SATD envelopes of the modulation tones, and the $\omegatmodj(t)$ are the chirped modulation tone frequencies.

%%%%%%%%%%%%%%%%%%%%%%%%%%%%%%%%%%
\section{SATD pulse shape}\label{sec:pulse}
%%%%%%%%%%%%%%%%%%%%%%%%%%%%%%%%%%

\begin{figure}[t!]
\centering
\includegraphics[width=\linewidth]{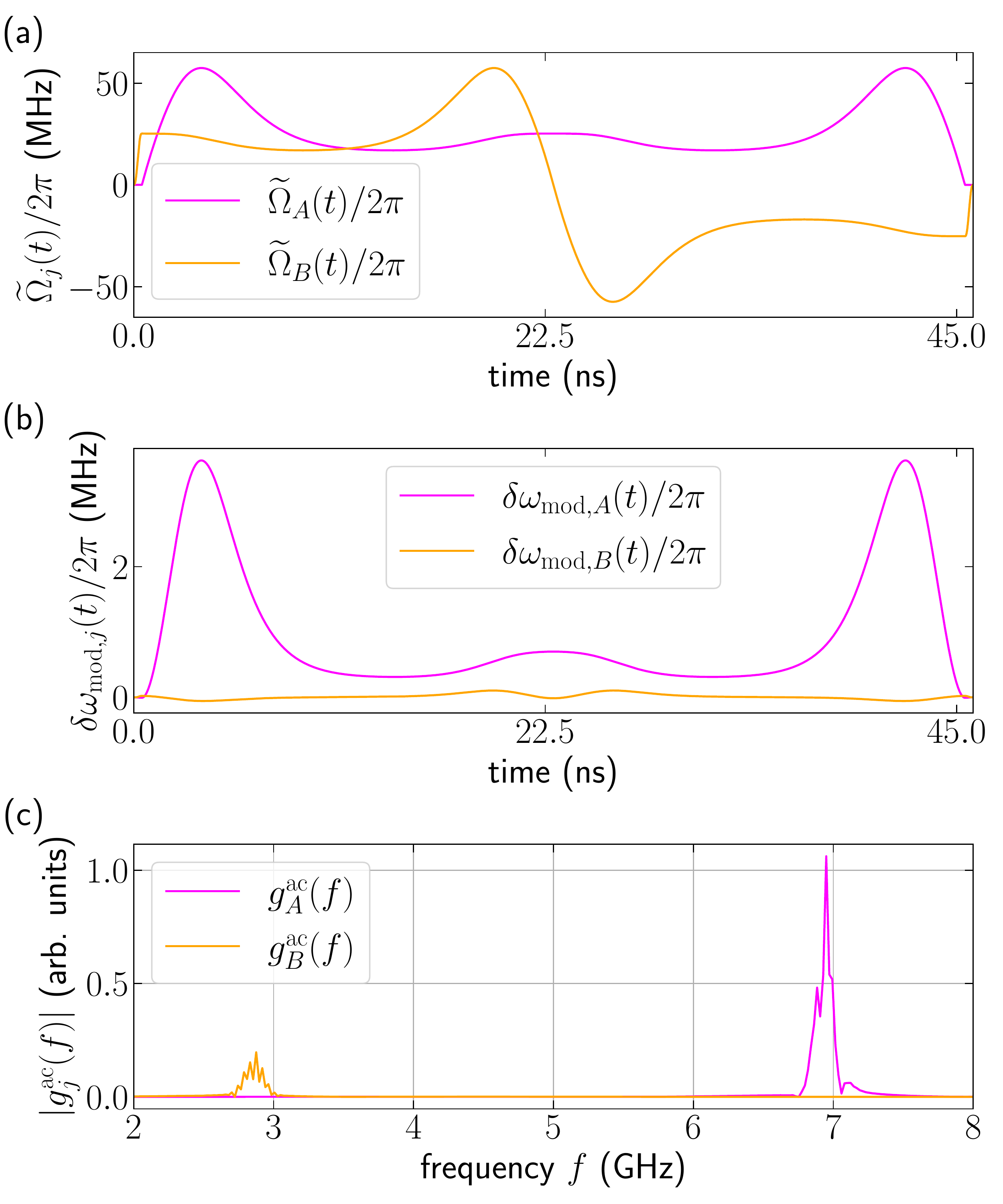}
\caption{ Modulation tones for realizing a CZ gate in a realistic superconducting setup consisting of two fluxonia, both coupled to a shared auxiliary mode via direct tunable couplings. (a) Time profiles of the pulse envelopes. (b) Time profiles of the frequency chirp corrections $\delta \omegamodj(t)$ [Eqs.~\eqref{eq:omegamod}]. (c) Fourier transform of $|\gac_j(f)|$ [Eqs.~\eqref{eq:gtac}]. Shown are the SATD pulse sequences given in Eqs.~\eqref{eq:realisticpulse} where the pulses are switched on (off) for a time duration $t_{\mathrm{ramp}}$ in the beginning (end) of the protocol with a smooth polynomial ramp function [Eq.~\eqref{eq:Px}]. 
The pulses are for $t_g=45$ ns, $t_{\mathrm{ramp}} = 0.45$ ns, and $\Omega_0/2\pi = \Omega_{\mathrm{opt}}/2\pi \simeq 25.22$ MHz. Circuit parameters are given in Table~\ref{table:paramsone}.}
\label{fig:pulse}
\end{figure}

The control field required to implement our two-qubit gate consists of two modulation tones with complex amplitudes $\tilde{\Omega}_A(t)$ and $\tilde{\Omega}_B(t)$ [see Eq.~\eqref{eq:omegadrive} and Eq.~\eqref{eq:omegaSATD} for uncorrected and SATD-corrected pulses, respectively].
    Figure~\ref{fig:pulse} shows plots of (a) amplitudes, (b) frequency chirp-corrections, and (c) Fourier components of the modulation tones corresponding to the case of the power-optimal SATD dynamics, each slightly modified by inclusion of pulse smoothing at the beginning and end---something that would be typically done in experiments.

Such modification is implemented by sandwiching the original pulses with a short ramp up (down), having a duration $t_{\mathrm{ramp}}$, at the beginning (end) of the protocol, during which $\tilde{\Omega}_{B}(t)$ is smoothly switched on (off). We can write the full modulation pulses as piecewise, smooth and continuous functions, split into three time intervals as
\begin{subequations}\label{eq:realisticpulse}
\begin{align}
\frac{\tilde{\Omega}_A(t)}{\Omega_0 } &= \begin{cases}
0 & \rm{(I)},\\
\displaystyle \sin[\theta(t_-)] +  \frac{\cos[\theta(t_-)]\ddot{\theta}(t_-)}{  \dot{\theta}^2(t_-)+\Omega_0^2/4} & \rm{(II)}, \\
0 & \rm{(III)},
\end{cases}\\
\frac{\tilde{\Omega}_B (t)}{\Omega_0 } &= \begin{cases}
\displaystyle P\left(\frac{t}{2\tramp}\right) & \rm{(I)},\\[10pt]
\displaystyle e^{i\gamma (t_-)} \left\{\cos[\theta(t_-)] -  \frac{\sin[\theta(t_-)]\ddot{\theta}(t_-)}{  \dot{\theta}^2(t_-)+\Omega_0^2/4}\right\}& \rm{(II)},\\[10pt]
\displaystyle 1- P\left(\frac{t_- - t_g}{2\tramp}\right) & \rm{(III)},
\end{cases}\label{eq:omegaaeturn}
\end{align}
\end{subequations}
with $t_- = t-t_{\mathrm{ramp}}$.
Here, we pick $\tramp = 0.01 t_g$ to be sufficiently shorter than the overall pulse length. This is to ensure that the dynamics due to the ramp will not substantially modify the dynamics of the overall protocol, but sufficiently long that there is no abrupt change in the pulse. The function $P(x)$ in Eq.~\eqref{eq:omegaaeturn} can be any smooth function [here we use a fifth-order polynomial as given in Eq.~\eqref{eq:Px}]. We stress that the results obtained with the inclusion of these ramps remain largely unchanged.

\bibliography{manuscript_1208}

\begin{thebibliography}{77}%
\makeatletter
\providecommand \@ifxundefined [1]{%
 \@ifx{#1\undefined}
}%
\providecommand \@ifnum [1]{%
 \ifnum #1\expandafter \@firstoftwo
 \else \expandafter \@secondoftwo
 \fi
}%
\providecommand \@ifx [1]{%
 \ifx #1\expandafter \@firstoftwo
 \else \expandafter \@secondoftwo
 \fi
}%
\providecommand \natexlab [1]{#1}%
\providecommand \enquote  [1]{``#1''}%
\providecommand \bibnamefont  [1]{#1}%
\providecommand \bibfnamefont [1]{#1}%
\providecommand \citenamefont [1]{#1}%
\providecommand \href@noop [0]{\@secondoftwo}%
\providecommand \href [0]{\begingroup \@sanitize@url \@href}%
\providecommand \@href[1]{\@@startlink{#1}\@@href}%
\providecommand \@@href[1]{\endgroup#1\@@endlink}%
\providecommand \@sanitize@url [0]{\catcode `\\12\catcode `\$12\catcode
  `\&12\catcode `\#12\catcode `\^12\catcode `\_12\catcode `\%12\relax}%
\providecommand \@@startlink[1]{}%
\providecommand \@@endlink[0]{}%
\providecommand \url  [0]{\begingroup\@sanitize@url \@url }%
\providecommand \@url [1]{\endgroup\@href {#1}{\urlprefix }}%
\providecommand \urlprefix  [0]{URL }%
\providecommand \Eprint [0]{\href }%
\providecommand \doibase [0]{https://doi.org/}%
\providecommand \selectlanguage [0]{\@gobble}%
\providecommand \bibinfo  [0]{\@secondoftwo}%
\providecommand \bibfield  [0]{\@secondoftwo}%
\providecommand \translation [1]{[#1]}%
\providecommand \BibitemOpen [0]{}%
\providecommand \bibitemStop [0]{}%
\providecommand \bibitemNoStop [0]{.\EOS\space}%
\providecommand \EOS [0]{\spacefactor3000\relax}%
\providecommand \BibitemShut  [1]{\csname bibitem#1\endcsname}%
\let\auto@bib@innerbib\@empty
%</preamble>
\bibitem [{\citenamefont {Zanardi}\ and\ \citenamefont
  {Rasetti}(1999)}]{Zanardi1999}%
  \BibitemOpen
  \bibfield  {author} {\bibinfo {author} {\bibfnamefont {P.}~\bibnamefont
  {Zanardi}}\ and\ \bibinfo {author} {\bibfnamefont {M.}~\bibnamefont
  {Rasetti}},\ }\bibfield  {title} {\bibinfo {title} {Holonomic quantum
  computation},\ }\href
  {https://doi.org/https://doi.org/10.1016/S0375-9601(99)00803-8} {\bibfield
  {journal} {\bibinfo  {journal} {Phys. Lett. A}\ }\textbf {\bibinfo {volume}
  {264}},\ \bibinfo {pages} {94} (\bibinfo {year} {1999})}\BibitemShut
  {NoStop}%
\bibitem [{\citenamefont {Pachos}\ \emph {et~al.}(1999)\citenamefont {Pachos},
  \citenamefont {Zanardi},\ and\ \citenamefont {Rasetti}}]{Pachos1999}%
  \BibitemOpen
  \bibfield  {author} {\bibinfo {author} {\bibfnamefont {J.}~\bibnamefont
  {Pachos}}, \bibinfo {author} {\bibfnamefont {P.}~\bibnamefont {Zanardi}},\
  and\ \bibinfo {author} {\bibfnamefont {M.}~\bibnamefont {Rasetti}},\
  }\bibfield  {title} {\bibinfo {title} {Non-abelian berry connections for
  quantum computation},\ }\href {https://doi.org/10.1103/PhysRevA.61.010305}
  {\bibfield  {journal} {\bibinfo  {journal} {Phys. Rev. A}\ }\textbf {\bibinfo
  {volume} {61}},\ \bibinfo {pages} {010305(R)} (\bibinfo {year}
  {1999})}\BibitemShut {NoStop}%
\bibitem [{\citenamefont {Unanyan}\ \emph {et~al.}(1999)\citenamefont
  {Unanyan}, \citenamefont {Shore},\ and\ \citenamefont
  {Bergmann}}]{Unayan1999Laser}%
  \BibitemOpen
  \bibfield  {author} {\bibinfo {author} {\bibfnamefont {R.~G.}\ \bibnamefont
  {Unanyan}}, \bibinfo {author} {\bibfnamefont {B.~W.}\ \bibnamefont {Shore}},\
  and\ \bibinfo {author} {\bibfnamefont {K.}~\bibnamefont {Bergmann}},\
  }\bibfield  {title} {\bibinfo {title} {Laser-driven population transfer in
  four-level atoms: Consequences of non-abelian geometrical adiabatic phase
  factors},\ }\href {https://doi.org/10.1103/PhysRevA.59.2910} {\bibfield
  {journal} {\bibinfo  {journal} {Phys. Rev. A}\ }\textbf {\bibinfo {volume}
  {59}},\ \bibinfo {pages} {2910} (\bibinfo {year} {1999})}\BibitemShut
  {NoStop}%
\bibitem [{\citenamefont {Duan}\ \emph {et~al.}(2001)\citenamefont {Duan},
  \citenamefont {Cirac},\ and\ \citenamefont {Zoller}}]{duan2001geometric}%
  \BibitemOpen
  \bibfield  {author} {\bibinfo {author} {\bibfnamefont {L.-M.}\ \bibnamefont
  {Duan}}, \bibinfo {author} {\bibfnamefont {J.~I.}\ \bibnamefont {Cirac}},\
  and\ \bibinfo {author} {\bibfnamefont {P.}~\bibnamefont {Zoller}},\
  }\bibfield  {title} {\bibinfo {title} {Geometric manipulation of trapped ions
  for quantum computation},\ }\href {https://doi.org/10.1126/science.1058835}
  {\bibfield  {journal} {\bibinfo  {journal} {Science}\ }\textbf {\bibinfo
  {volume} {292}},\ \bibinfo {pages} {1695} (\bibinfo {year}
  {2001})}\BibitemShut {NoStop}%
\bibitem [{\citenamefont {M\o{}ller}\ \emph {et~al.}(2008)\citenamefont
  {M\o{}ller}, \citenamefont {Madsen},\ and\ \citenamefont
  {M\o{}lmer}}]{Moller2008Quantum}%
  \BibitemOpen
  \bibfield  {author} {\bibinfo {author} {\bibfnamefont {D.}~\bibnamefont
  {M\o{}ller}}, \bibinfo {author} {\bibfnamefont {L.~B.}\ \bibnamefont
  {Madsen}},\ and\ \bibinfo {author} {\bibfnamefont {K.}~\bibnamefont
  {M\o{}lmer}},\ }\bibfield  {title} {\bibinfo {title} {Quantum gates and
  multiparticle entanglement by rydberg excitation blockade and adiabatic
  passage},\ }\href {https://doi.org/10.1103/PhysRevLett.100.170504} {\bibfield
   {journal} {\bibinfo  {journal} {Phys. Rev. Lett.}\ }\textbf {\bibinfo
  {volume} {100}},\ \bibinfo {pages} {170504} (\bibinfo {year}
  {2008})}\BibitemShut {NoStop}%
\bibitem [{\citenamefont {Kis}\ and\ \citenamefont
  {Renzoni}(2002)}]{Kis2002Qubit}%
  \BibitemOpen
  \bibfield  {author} {\bibinfo {author} {\bibfnamefont {Z.}~\bibnamefont
  {Kis}}\ and\ \bibinfo {author} {\bibfnamefont {F.}~\bibnamefont {Renzoni}},\
  }\bibfield  {title} {\bibinfo {title} {Qubit rotation by stimulated raman
  adiabatic passage},\ }\href {https://doi.org/10.1103/PhysRevA.65.032318}
  {\bibfield  {journal} {\bibinfo  {journal} {Phys. Rev. A}\ }\textbf {\bibinfo
  {volume} {65}},\ \bibinfo {pages} {032318} (\bibinfo {year}
  {2002})}\BibitemShut {NoStop}%
\bibitem [{\citenamefont {Faoro}\ \emph {et~al.}(2003)\citenamefont {Faoro},
  \citenamefont {Siewert},\ and\ \citenamefont {Fazio}}]{Faoro2003Non}%
  \BibitemOpen
  \bibfield  {author} {\bibinfo {author} {\bibfnamefont {L.}~\bibnamefont
  {Faoro}}, \bibinfo {author} {\bibfnamefont {J.}~\bibnamefont {Siewert}},\
  and\ \bibinfo {author} {\bibfnamefont {R.}~\bibnamefont {Fazio}},\ }\bibfield
   {title} {\bibinfo {title} {Non-abelian holonomies, charge pumping, and
  quantum computation with josephson junctions},\ }\href
  {https://doi.org/10.1103/PhysRevLett.90.028301} {\bibfield  {journal}
  {\bibinfo  {journal} {Phys. Rev. Lett.}\ }\textbf {\bibinfo {volume} {90}},\
  \bibinfo {pages} {028301} (\bibinfo {year} {2003})}\BibitemShut {NoStop}%
\bibitem [{\citenamefont {Solinas}\ \emph {et~al.}(2003)\citenamefont
  {Solinas}, \citenamefont {Zanardi}, \citenamefont {Zangh\`{\i}},\ and\
  \citenamefont {Rossi}}]{Solinas2003Holonomic}%
  \BibitemOpen
  \bibfield  {author} {\bibinfo {author} {\bibfnamefont {P.}~\bibnamefont
  {Solinas}}, \bibinfo {author} {\bibfnamefont {P.}~\bibnamefont {Zanardi}},
  \bibinfo {author} {\bibfnamefont {N.}~\bibnamefont {Zangh\`{\i}}},\ and\
  \bibinfo {author} {\bibfnamefont {F.}~\bibnamefont {Rossi}},\ }\bibfield
  {title} {\bibinfo {title} {Holonomic quantum gates: A semiconductor-based
  implementation},\ }\href {https://doi.org/10.1103/PhysRevA.67.062315}
  {\bibfield  {journal} {\bibinfo  {journal} {Phys. Rev. A}\ }\textbf {\bibinfo
  {volume} {67}},\ \bibinfo {pages} {062315} (\bibinfo {year}
  {2003})}\BibitemShut {NoStop}%
\bibitem [{\citenamefont {Frees}\ \emph {et~al.}(2019)\citenamefont {Frees},
  \citenamefont {Mehl}, \citenamefont {Gamble}, \citenamefont {Friesen},\ and\
  \citenamefont {Coppersmith}}]{frees2019adiabatic}%
  \BibitemOpen
  \bibfield  {author} {\bibinfo {author} {\bibfnamefont {A.}~\bibnamefont
  {Frees}}, \bibinfo {author} {\bibfnamefont {S.}~\bibnamefont {Mehl}},
  \bibinfo {author} {\bibfnamefont {J.~K.}\ \bibnamefont {Gamble}}, \bibinfo
  {author} {\bibfnamefont {M.}~\bibnamefont {Friesen}},\ and\ \bibinfo {author}
  {\bibfnamefont {S.}~\bibnamefont {Coppersmith}},\ }\bibfield  {title}
  {\bibinfo {title} {Adiabatic two-qubit gates in capacitively coupled quantum
  dot hybrid qubits},\ }\href {https://doi.org/10.1038/s41534-019-0190-7}
  {\bibfield  {journal} {\bibinfo  {journal} {npj Quantum Information}\
  }\textbf {\bibinfo {volume} {5}},\ \bibinfo {pages} {73} (\bibinfo {year}
  {2019})}\BibitemShut {NoStop}%
\bibitem [{\citenamefont {Zeng}\ \emph {et~al.}(2019)\citenamefont {Zeng},
  \citenamefont {Yang}, \citenamefont {Dzurak},\ and\ \citenamefont
  {Barnes}}]{Zeng2019Geometric}%
  \BibitemOpen
  \bibfield  {author} {\bibinfo {author} {\bibfnamefont {J.}~\bibnamefont
  {Zeng}}, \bibinfo {author} {\bibfnamefont {C.~H.}\ \bibnamefont {Yang}},
  \bibinfo {author} {\bibfnamefont {A.~S.}\ \bibnamefont {Dzurak}},\ and\
  \bibinfo {author} {\bibfnamefont {E.}~\bibnamefont {Barnes}},\ }\bibfield
  {title} {\bibinfo {title} {Geometric formalism for constructing arbitrary
  single-qubit dynamically corrected gates},\ }\href
  {https://doi.org/10.1103/PhysRevA.99.052321} {\bibfield  {journal} {\bibinfo
  {journal} {Phys. Rev. A}\ }\textbf {\bibinfo {volume} {99}},\ \bibinfo
  {pages} {052321} (\bibinfo {year} {2019})}\BibitemShut {NoStop}%
\bibitem [{\citenamefont {Dridi}\ \emph {et~al.}(2020)\citenamefont {Dridi},
  \citenamefont {Liu},\ and\ \citenamefont {Gu\'erin}}]{Dridi2020Optimal}%
  \BibitemOpen
  \bibfield  {author} {\bibinfo {author} {\bibfnamefont {G.}~\bibnamefont
  {Dridi}}, \bibinfo {author} {\bibfnamefont {K.}~\bibnamefont {Liu}},\ and\
  \bibinfo {author} {\bibfnamefont {S.}~\bibnamefont {Gu\'erin}},\ }\bibfield
  {title} {\bibinfo {title} {Optimal robust quantum control by inverse
  geometric optimization},\ }\href
  {https://doi.org/10.1103/PhysRevLett.125.250403} {\bibfield  {journal}
  {\bibinfo  {journal} {Phys. Rev. Lett.}\ }\textbf {\bibinfo {volume} {125}},\
  \bibinfo {pages} {250403} (\bibinfo {year} {2020})}\BibitemShut {NoStop}%
\bibitem [{\citenamefont {Laforgue}\ \emph
  {et~al.}(2022{\natexlab{a}})\citenamefont {Laforgue}, \citenamefont {Dridi},\
  and\ \citenamefont {Gu\'erin}}]{Laforgue2022Optimal}%
  \BibitemOpen
  \bibfield  {author} {\bibinfo {author} {\bibfnamefont {X.}~\bibnamefont
  {Laforgue}}, \bibinfo {author} {\bibfnamefont {G.}~\bibnamefont {Dridi}},\
  and\ \bibinfo {author} {\bibfnamefont {S.}~\bibnamefont {Gu\'erin}},\
  }\bibfield  {title} {\bibinfo {title} {Optimal robust stimulated raman exact
  passage by inverse optimization},\ }\href
  {https://doi.org/10.1103/PhysRevA.105.032807} {\bibfield  {journal} {\bibinfo
   {journal} {Phys. Rev. A}\ }\textbf {\bibinfo {volume} {105}},\ \bibinfo
  {pages} {032807} (\bibinfo {year} {2022}{\natexlab{a}})}\BibitemShut
  {NoStop}%
\bibitem [{\citenamefont {Laforgue}\ \emph
  {et~al.}(2022{\natexlab{b}})\citenamefont {Laforgue}, \citenamefont {Dridi},\
  and\ \citenamefont {Gu\'erin}}]{Laforgue2022Optimalb}%
  \BibitemOpen
  \bibfield  {author} {\bibinfo {author} {\bibfnamefont {X.}~\bibnamefont
  {Laforgue}}, \bibinfo {author} {\bibfnamefont {G.}~\bibnamefont {Dridi}},\
  and\ \bibinfo {author} {\bibfnamefont {S.}~\bibnamefont {Gu\'erin}},\
  }\bibfield  {title} {\bibinfo {title} {Optimal quantum control robust against
  pulse inhomogeneities: Analytic solutions},\ }\href
  {https://doi.org/10.1103/PhysRevA.106.052608} {\bibfield  {journal} {\bibinfo
   {journal} {Phys. Rev. A}\ }\textbf {\bibinfo {volume} {106}},\ \bibinfo
  {pages} {052608} (\bibinfo {year} {2022}{\natexlab{b}})}\BibitemShut
  {NoStop}%
\bibitem [{\citenamefont {Wu}\ \emph {et~al.}(2013)\citenamefont {Wu},
  \citenamefont {Gauger}, \citenamefont {George}, \citenamefont {M\"ott\"onen},
  \citenamefont {Riemann}, \citenamefont {Abrosimov}, \citenamefont {Becker},
  \citenamefont {Pohl}, \citenamefont {Itoh}, \citenamefont {Thewalt},\ and\
  \citenamefont {Morton}}]{Wu2013Geometric}%
  \BibitemOpen
  \bibfield  {author} {\bibinfo {author} {\bibfnamefont {H.}~\bibnamefont
  {Wu}}, \bibinfo {author} {\bibfnamefont {E.~M.}\ \bibnamefont {Gauger}},
  \bibinfo {author} {\bibfnamefont {R.~E.}\ \bibnamefont {George}}, \bibinfo
  {author} {\bibfnamefont {M.}~\bibnamefont {M\"ott\"onen}}, \bibinfo {author}
  {\bibfnamefont {H.}~\bibnamefont {Riemann}}, \bibinfo {author} {\bibfnamefont
  {N.~V.}\ \bibnamefont {Abrosimov}}, \bibinfo {author} {\bibfnamefont
  {P.}~\bibnamefont {Becker}}, \bibinfo {author} {\bibfnamefont {H.-J.}\
  \bibnamefont {Pohl}}, \bibinfo {author} {\bibfnamefont {K.~M.}\ \bibnamefont
  {Itoh}}, \bibinfo {author} {\bibfnamefont {M.~L.~W.}\ \bibnamefont
  {Thewalt}},\ and\ \bibinfo {author} {\bibfnamefont {J.~J.~L.}\ \bibnamefont
  {Morton}},\ }\bibfield  {title} {\bibinfo {title} {Geometric phase gates with
  adiabatic control in electron spin resonance},\ }\href
  {https://doi.org/10.1103/PhysRevA.87.032326} {\bibfield  {journal} {\bibinfo
  {journal} {Phys. Rev. A}\ }\textbf {\bibinfo {volume} {87}},\ \bibinfo
  {pages} {032326} (\bibinfo {year} {2013})}\BibitemShut {NoStop}%
\bibitem [{\citenamefont {Toyoda}\ \emph {et~al.}(2013)\citenamefont {Toyoda},
  \citenamefont {Uchida}, \citenamefont {Noguchi}, \citenamefont {Haze},\ and\
  \citenamefont {Urabe}}]{Toyoda2013Realization}%
  \BibitemOpen
  \bibfield  {author} {\bibinfo {author} {\bibfnamefont {K.}~\bibnamefont
  {Toyoda}}, \bibinfo {author} {\bibfnamefont {K.}~\bibnamefont {Uchida}},
  \bibinfo {author} {\bibfnamefont {A.}~\bibnamefont {Noguchi}}, \bibinfo
  {author} {\bibfnamefont {S.}~\bibnamefont {Haze}},\ and\ \bibinfo {author}
  {\bibfnamefont {S.}~\bibnamefont {Urabe}},\ }\bibfield  {title} {\bibinfo
  {title} {Realization of holonomic single-qubit operations},\ }\href
  {https://doi.org/10.1103/PhysRevA.87.052307} {\bibfield  {journal} {\bibinfo
  {journal} {Phys. Rev. A}\ }\textbf {\bibinfo {volume} {87}},\ \bibinfo
  {pages} {052307} (\bibinfo {year} {2013})}\BibitemShut {NoStop}%
\bibitem [{\citenamefont {Huang}\ \emph {et~al.}(2019)\citenamefont {Huang},
  \citenamefont {Wu}, \citenamefont {Wang}, \citenamefont {Hou}, \citenamefont
  {Wang}, \citenamefont {Zhang}, \citenamefont {Lian}, \citenamefont {Liu},
  \citenamefont {Wang}, \citenamefont {Zhang}, \citenamefont {He},
  \citenamefont {Chang}, \citenamefont {Xu},\ and\ \citenamefont
  {Duan}}]{Huang2019Experimental}%
  \BibitemOpen
  \bibfield  {author} {\bibinfo {author} {\bibfnamefont {Y.-Y.}\ \bibnamefont
  {Huang}}, \bibinfo {author} {\bibfnamefont {Y.-K.}\ \bibnamefont {Wu}},
  \bibinfo {author} {\bibfnamefont {F.}~\bibnamefont {Wang}}, \bibinfo {author}
  {\bibfnamefont {P.-Y.}\ \bibnamefont {Hou}}, \bibinfo {author} {\bibfnamefont
  {W.-B.}\ \bibnamefont {Wang}}, \bibinfo {author} {\bibfnamefont {W.-G.}\
  \bibnamefont {Zhang}}, \bibinfo {author} {\bibfnamefont {W.-Q.}\ \bibnamefont
  {Lian}}, \bibinfo {author} {\bibfnamefont {Y.-Q.}\ \bibnamefont {Liu}},
  \bibinfo {author} {\bibfnamefont {H.-Y.}\ \bibnamefont {Wang}}, \bibinfo
  {author} {\bibfnamefont {H.-Y.}\ \bibnamefont {Zhang}}, \bibinfo {author}
  {\bibfnamefont {L.}~\bibnamefont {He}}, \bibinfo {author} {\bibfnamefont
  {X.-Y.}\ \bibnamefont {Chang}}, \bibinfo {author} {\bibfnamefont
  {Y.}~\bibnamefont {Xu}},\ and\ \bibinfo {author} {\bibfnamefont {L.-M.}\
  \bibnamefont {Duan}},\ }\bibfield  {title} {\bibinfo {title} {Experimental
  realization of robust geometric quantum gates with solid-state spins},\
  }\href {https://doi.org/10.1103/PhysRevLett.122.010503} {\bibfield  {journal}
  {\bibinfo  {journal} {Phys. Rev. Lett.}\ }\textbf {\bibinfo {volume} {122}},\
  \bibinfo {pages} {010503} (\bibinfo {year} {2019})}\BibitemShut {NoStop}%
\bibitem [{\citenamefont {Vitanov}\ \emph {et~al.}(2017)\citenamefont
  {Vitanov}, \citenamefont {Rangelov}, \citenamefont {Shore},\ and\
  \citenamefont {Bergmann}}]{Vitanov2017Stimulated}%
  \BibitemOpen
  \bibfield  {author} {\bibinfo {author} {\bibfnamefont {N.~V.}\ \bibnamefont
  {Vitanov}}, \bibinfo {author} {\bibfnamefont {A.~A.}\ \bibnamefont
  {Rangelov}}, \bibinfo {author} {\bibfnamefont {B.~W.}\ \bibnamefont
  {Shore}},\ and\ \bibinfo {author} {\bibfnamefont {K.}~\bibnamefont
  {Bergmann}},\ }\bibfield  {title} {\bibinfo {title} {Stimulated raman
  adiabatic passage in physics, chemistry, and beyond},\ }\href
  {https://doi.org/10.1103/RevModPhys.89.015006} {\bibfield  {journal}
  {\bibinfo  {journal} {Rev. Mod. Phys.}\ }\textbf {\bibinfo {volume} {89}},\
  \bibinfo {pages} {015006} (\bibinfo {year} {2017})}\BibitemShut {NoStop}%
\bibitem [{\citenamefont {Ribeiro}\ and\ \citenamefont
  {Clerk}(2019)}]{Ribeiro2019Accelerated}%
  \BibitemOpen
  \bibfield  {author} {\bibinfo {author} {\bibfnamefont {H.}~\bibnamefont
  {Ribeiro}}\ and\ \bibinfo {author} {\bibfnamefont {A.~A.}\ \bibnamefont
  {Clerk}},\ }\bibfield  {title} {\bibinfo {title} {Accelerated adiabatic
  quantum gates: Optimizing speed versus robustness},\ }\href
  {https://doi.org/10.1103/PhysRevA.100.032323} {\bibfield  {journal} {\bibinfo
   {journal} {Phys. Rev. A}\ }\textbf {\bibinfo {volume} {100}},\ \bibinfo
  {pages} {032323} (\bibinfo {year} {2019})}\BibitemShut {NoStop}%
\bibitem [{\citenamefont {Setiawan}\ \emph {et~al.}(2021)\citenamefont
  {Setiawan}, \citenamefont {Groszkowski}, \citenamefont {Ribeiro},\ and\
  \citenamefont {Clerk}}]{Setiawan2021Analytic}%
  \BibitemOpen
  \bibfield  {author} {\bibinfo {author} {\bibfnamefont {F.}~\bibnamefont
  {Setiawan}}, \bibinfo {author} {\bibfnamefont {P.}~\bibnamefont
  {Groszkowski}}, \bibinfo {author} {\bibfnamefont {H.}~\bibnamefont
  {Ribeiro}},\ and\ \bibinfo {author} {\bibfnamefont {A.~A.}\ \bibnamefont
  {Clerk}},\ }\bibfield  {title} {\bibinfo {title} {Analytic design of
  accelerated adiabatic gates in realistic qubits: General theory and
  applications to superconducting circuits},\ }\href
  {https://doi.org/10.1103/PRXQuantum.2.030306} {\bibfield  {journal} {\bibinfo
   {journal} {PRX Quantum}\ }\textbf {\bibinfo {volume} {2}},\ \bibinfo {pages}
  {030306} (\bibinfo {year} {2021})}\BibitemShut {NoStop}%
\bibitem [{\citenamefont {Demirplak}\ and\ \citenamefont
  {Rice}(2003)}]{demirplak2003adiabatic}%
  \BibitemOpen
  \bibfield  {author} {\bibinfo {author} {\bibfnamefont {M.}~\bibnamefont
  {Demirplak}}\ and\ \bibinfo {author} {\bibfnamefont {S.~A.}\ \bibnamefont
  {Rice}},\ }\bibfield  {title} {\bibinfo {title} {Adiabatic population
  transfer with control fields},\ }\href {https://doi.org/10.1021/jp030708a}
  {\bibfield  {journal} {\bibinfo  {journal} {The Journal of Physical Chemistry
  A}\ }\textbf {\bibinfo {volume} {107}},\ \bibinfo {pages} {9937} (\bibinfo
  {year} {2003})}\BibitemShut {NoStop}%
\bibitem [{\citenamefont {Demirplak}\ and\ \citenamefont
  {Rice}(2005)}]{demirplak2005assisted}%
  \BibitemOpen
  \bibfield  {author} {\bibinfo {author} {\bibfnamefont {M.}~\bibnamefont
  {Demirplak}}\ and\ \bibinfo {author} {\bibfnamefont {S.~A.}\ \bibnamefont
  {Rice}},\ }\bibfield  {title} {\bibinfo {title} {Assisted adiabatic passage
  revisited},\ }\href {https://doi.org/10.1021/jp040647w} {\bibfield  {journal}
  {\bibinfo  {journal} {The Journal of Physical Chemistry B}\ }\textbf
  {\bibinfo {volume} {109}},\ \bibinfo {pages} {6838} (\bibinfo {year}
  {2005})}\BibitemShut {NoStop}%
\bibitem [{\citenamefont {Berry}(2009)}]{berry2009transitionless}%
  \BibitemOpen
  \bibfield  {author} {\bibinfo {author} {\bibfnamefont {M.~V.}\ \bibnamefont
  {Berry}},\ }\bibfield  {title} {\bibinfo {title} {Transitionless quantum
  driving},\ }\href {https://doi.org/10.1088/1751-8113/42/36/365303} {\bibfield
   {journal} {\bibinfo  {journal} {Journal of Physics A: Mathematical and
  Theoretical}\ }\textbf {\bibinfo {volume} {42}},\ \bibinfo {pages} {365303}
  (\bibinfo {year} {2009})}\BibitemShut {NoStop}%
\bibitem [{\citenamefont {Ib\'a\~nez}\ \emph {et~al.}(2012)\citenamefont
  {Ib\'a\~nez}, \citenamefont {Chen}, \citenamefont {Torrontegui},
  \citenamefont {Muga},\ and\ \citenamefont {Ruschhaupt}}]{Ibanez2012Multiple}%
  \BibitemOpen
  \bibfield  {author} {\bibinfo {author} {\bibfnamefont {S.}~\bibnamefont
  {Ib\'a\~nez}}, \bibinfo {author} {\bibfnamefont {X.}~\bibnamefont {Chen}},
  \bibinfo {author} {\bibfnamefont {E.}~\bibnamefont {Torrontegui}}, \bibinfo
  {author} {\bibfnamefont {J.~G.}\ \bibnamefont {Muga}},\ and\ \bibinfo
  {author} {\bibfnamefont {A.}~\bibnamefont {Ruschhaupt}},\ }\bibfield  {title}
  {\bibinfo {title} {Multiple schr\"odinger pictures and dynamics in shortcuts
  to adiabaticity},\ }\href {https://doi.org/10.1103/PhysRevLett.109.100403}
  {\bibfield  {journal} {\bibinfo  {journal} {Phys. Rev. Lett.}\ }\textbf
  {\bibinfo {volume} {109}},\ \bibinfo {pages} {100403} (\bibinfo {year}
  {2012})}\BibitemShut {NoStop}%
\bibitem [{\citenamefont {Gu\'ery-Odelin}\ \emph {et~al.}(2019)\citenamefont
  {Gu\'ery-Odelin}, \citenamefont {Ruschhaupt}, \citenamefont {Kiely},
  \citenamefont {Torrontegui}, \citenamefont {Mart\'{\i}nez-Garaot},\ and\
  \citenamefont {Muga}}]{Guery2019Shortcuts}%
  \BibitemOpen
  \bibfield  {author} {\bibinfo {author} {\bibfnamefont {D.}~\bibnamefont
  {Gu\'ery-Odelin}}, \bibinfo {author} {\bibfnamefont {A.}~\bibnamefont
  {Ruschhaupt}}, \bibinfo {author} {\bibfnamefont {A.}~\bibnamefont {Kiely}},
  \bibinfo {author} {\bibfnamefont {E.}~\bibnamefont {Torrontegui}}, \bibinfo
  {author} {\bibfnamefont {S.}~\bibnamefont {Mart\'{\i}nez-Garaot}},\ and\
  \bibinfo {author} {\bibfnamefont {J.~G.}\ \bibnamefont {Muga}},\ }\bibfield
  {title} {\bibinfo {title} {Shortcuts to adiabaticity: Concepts, methods, and
  applications},\ }\href {https://doi.org/10.1103/RevModPhys.91.045001}
  {\bibfield  {journal} {\bibinfo  {journal} {Rev. Mod. Phys.}\ }\textbf
  {\bibinfo {volume} {91}},\ \bibinfo {pages} {045001} (\bibinfo {year}
  {2019})}\BibitemShut {NoStop}%
\bibitem [{\citenamefont {Chen}\ \emph {et~al.}(2014)\citenamefont {Chen},
  \citenamefont {Neill}, \citenamefont {Roushan}, \citenamefont {Leung},
  \citenamefont {Fang}, \citenamefont {Barends}, \citenamefont {Kelly},
  \citenamefont {Campbell}, \citenamefont {Chen}, \citenamefont {Chiaro},
  \citenamefont {Dunsworth}, \citenamefont {Jeffrey}, \citenamefont {Megrant},
  \citenamefont {Mutus}, \citenamefont {O'Malley}, \citenamefont {Quintana},
  \citenamefont {Sank}, \citenamefont {Vainsencher}, \citenamefont {Wenner},
  \citenamefont {White}, \citenamefont {Geller}, \citenamefont {Cleland},\ and\
  \citenamefont {Martinis}}]{Chen2014Qubit}%
  \BibitemOpen
  \bibfield  {author} {\bibinfo {author} {\bibfnamefont {Y.}~\bibnamefont
  {Chen}}, \bibinfo {author} {\bibfnamefont {C.}~\bibnamefont {Neill}},
  \bibinfo {author} {\bibfnamefont {P.}~\bibnamefont {Roushan}}, \bibinfo
  {author} {\bibfnamefont {N.}~\bibnamefont {Leung}}, \bibinfo {author}
  {\bibfnamefont {M.}~\bibnamefont {Fang}}, \bibinfo {author} {\bibfnamefont
  {R.}~\bibnamefont {Barends}}, \bibinfo {author} {\bibfnamefont
  {J.}~\bibnamefont {Kelly}}, \bibinfo {author} {\bibfnamefont
  {B.}~\bibnamefont {Campbell}}, \bibinfo {author} {\bibfnamefont
  {Z.}~\bibnamefont {Chen}}, \bibinfo {author} {\bibfnamefont {B.}~\bibnamefont
  {Chiaro}}, \bibinfo {author} {\bibfnamefont {A.}~\bibnamefont {Dunsworth}},
  \bibinfo {author} {\bibfnamefont {E.}~\bibnamefont {Jeffrey}}, \bibinfo
  {author} {\bibfnamefont {A.}~\bibnamefont {Megrant}}, \bibinfo {author}
  {\bibfnamefont {J.~Y.}\ \bibnamefont {Mutus}}, \bibinfo {author}
  {\bibfnamefont {P.~J.~J.}\ \bibnamefont {O'Malley}}, \bibinfo {author}
  {\bibfnamefont {C.~M.}\ \bibnamefont {Quintana}}, \bibinfo {author}
  {\bibfnamefont {D.}~\bibnamefont {Sank}}, \bibinfo {author} {\bibfnamefont
  {A.}~\bibnamefont {Vainsencher}}, \bibinfo {author} {\bibfnamefont
  {J.}~\bibnamefont {Wenner}}, \bibinfo {author} {\bibfnamefont {T.~C.}\
  \bibnamefont {White}}, \bibinfo {author} {\bibfnamefont {M.~R.}\ \bibnamefont
  {Geller}}, \bibinfo {author} {\bibfnamefont {A.~N.}\ \bibnamefont
  {Cleland}},\ and\ \bibinfo {author} {\bibfnamefont {J.~M.}\ \bibnamefont
  {Martinis}},\ }\bibfield  {title} {\bibinfo {title} {Qubit architecture with
  high coherence and fast tunable coupling},\ }\href
  {https://doi.org/10.1103/PhysRevLett.113.220502} {\bibfield  {journal}
  {\bibinfo  {journal} {Phys. Rev. Lett.}\ }\textbf {\bibinfo {volume} {113}},\
  \bibinfo {pages} {220502} (\bibinfo {year} {2014})}\BibitemShut {NoStop}%
\bibitem [{\citenamefont {Zhong}\ \emph {et~al.}(2019)\citenamefont {Zhong},
  \citenamefont {Chang}, \citenamefont {Satzinger}, \citenamefont {Chou},
  \citenamefont {Bienfait}, \citenamefont {Conner}, \citenamefont {Dumur},
  \citenamefont {Grebel}, \citenamefont {Peairs}, \citenamefont {Povey} \emph
  {et~al.}}]{zhong2019violating}%
  \BibitemOpen
  \bibfield  {author} {\bibinfo {author} {\bibfnamefont {Y.}~\bibnamefont
  {Zhong}}, \bibinfo {author} {\bibfnamefont {H.-S.}\ \bibnamefont {Chang}},
  \bibinfo {author} {\bibfnamefont {K.}~\bibnamefont {Satzinger}}, \bibinfo
  {author} {\bibfnamefont {M.-H.}\ \bibnamefont {Chou}}, \bibinfo {author}
  {\bibfnamefont {A.}~\bibnamefont {Bienfait}}, \bibinfo {author}
  {\bibfnamefont {C.}~\bibnamefont {Conner}}, \bibinfo {author} {\bibfnamefont
  {{\'E}.}~\bibnamefont {Dumur}}, \bibinfo {author} {\bibfnamefont
  {J.}~\bibnamefont {Grebel}}, \bibinfo {author} {\bibfnamefont
  {G.}~\bibnamefont {Peairs}}, \bibinfo {author} {\bibfnamefont
  {R.}~\bibnamefont {Povey}}, \emph {et~al.},\ }\bibfield  {title} {\bibinfo
  {title} {Violating bell’s inequality with remotely connected
  superconducting qubits},\ }\href {https://doi.org/10.1038/s41567-019-0507-7}
  {\bibfield  {journal} {\bibinfo  {journal} {Nature Physics}\ }\textbf
  {\bibinfo {volume} {15}},\ \bibinfo {pages} {741} (\bibinfo {year}
  {2019})}\BibitemShut {NoStop}%
\bibitem [{\citenamefont {Magnard}\ \emph {et~al.}(2020)\citenamefont
  {Magnard}, \citenamefont {Storz}, \citenamefont {Kurpiers}, \citenamefont
  {Sch\"ar}, \citenamefont {Marxer}, \citenamefont {L\"utolf}, \citenamefont
  {Walter}, \citenamefont {Besse}, \citenamefont {Gabureac}, \citenamefont
  {Reuer}, \citenamefont {Akin}, \citenamefont {Royer}, \citenamefont {Blais},\
  and\ \citenamefont {Wallraff}}]{Magnard2020Microwave}%
  \BibitemOpen
  \bibfield  {author} {\bibinfo {author} {\bibfnamefont {P.}~\bibnamefont
  {Magnard}}, \bibinfo {author} {\bibfnamefont {S.}~\bibnamefont {Storz}},
  \bibinfo {author} {\bibfnamefont {P.}~\bibnamefont {Kurpiers}}, \bibinfo
  {author} {\bibfnamefont {J.}~\bibnamefont {Sch\"ar}}, \bibinfo {author}
  {\bibfnamefont {F.}~\bibnamefont {Marxer}}, \bibinfo {author} {\bibfnamefont
  {J.}~\bibnamefont {L\"utolf}}, \bibinfo {author} {\bibfnamefont
  {T.}~\bibnamefont {Walter}}, \bibinfo {author} {\bibfnamefont {J.-C.}\
  \bibnamefont {Besse}}, \bibinfo {author} {\bibfnamefont {M.}~\bibnamefont
  {Gabureac}}, \bibinfo {author} {\bibfnamefont {K.}~\bibnamefont {Reuer}},
  \bibinfo {author} {\bibfnamefont {A.}~\bibnamefont {Akin}}, \bibinfo {author}
  {\bibfnamefont {B.}~\bibnamefont {Royer}}, \bibinfo {author} {\bibfnamefont
  {A.}~\bibnamefont {Blais}},\ and\ \bibinfo {author} {\bibfnamefont
  {A.}~\bibnamefont {Wallraff}},\ }\bibfield  {title} {\bibinfo {title}
  {Microwave quantum link between superconducting circuits housed in spatially
  separated cryogenic systems},\ }\href
  {https://doi.org/10.1103/PhysRevLett.125.260502} {\bibfield  {journal}
  {\bibinfo  {journal} {Phys. Rev. Lett.}\ }\textbf {\bibinfo {volume} {125}},\
  \bibinfo {pages} {260502} (\bibinfo {year} {2020})}\BibitemShut {NoStop}%
\bibitem [{\citenamefont {Zhong}\ \emph {et~al.}(2021)\citenamefont {Zhong},
  \citenamefont {Chang}, \citenamefont {Bienfait}, \citenamefont {Dumur},
  \citenamefont {Chou}, \citenamefont {Conner}, \citenamefont {Grebel},
  \citenamefont {Povey}, \citenamefont {Yan}, \citenamefont {Schuster} \emph
  {et~al.}}]{zhong2021deterministic}%
  \BibitemOpen
  \bibfield  {author} {\bibinfo {author} {\bibfnamefont {Y.}~\bibnamefont
  {Zhong}}, \bibinfo {author} {\bibfnamefont {H.-S.}\ \bibnamefont {Chang}},
  \bibinfo {author} {\bibfnamefont {A.}~\bibnamefont {Bienfait}}, \bibinfo
  {author} {\bibfnamefont {{\'E}.}~\bibnamefont {Dumur}}, \bibinfo {author}
  {\bibfnamefont {M.-H.}\ \bibnamefont {Chou}}, \bibinfo {author}
  {\bibfnamefont {C.~R.}\ \bibnamefont {Conner}}, \bibinfo {author}
  {\bibfnamefont {J.}~\bibnamefont {Grebel}}, \bibinfo {author} {\bibfnamefont
  {R.~G.}\ \bibnamefont {Povey}}, \bibinfo {author} {\bibfnamefont
  {H.}~\bibnamefont {Yan}}, \bibinfo {author} {\bibfnamefont {D.~I.}\
  \bibnamefont {Schuster}}, \emph {et~al.},\ }\bibfield  {title} {\bibinfo
  {title} {Deterministic multi-qubit entanglement in a quantum network},\
  }\href {https://doi.org/10.1038/s41586-021-03288-7} {\bibfield  {journal}
  {\bibinfo  {journal} {Nature}\ }\textbf {\bibinfo {volume} {590}},\ \bibinfo
  {pages} {571} (\bibinfo {year} {2021})}\BibitemShut {NoStop}%
\bibitem [{\citenamefont {Yan}\ \emph {et~al.}(2022)\citenamefont {Yan},
  \citenamefont {Zhong}, \citenamefont {Chang}, \citenamefont {Bienfait},
  \citenamefont {Chou}, \citenamefont {Conner}, \citenamefont {Dumur},
  \citenamefont {Grebel}, \citenamefont {Povey},\ and\ \citenamefont
  {Cleland}}]{Yan2022Entanglement}%
  \BibitemOpen
  \bibfield  {author} {\bibinfo {author} {\bibfnamefont {H.}~\bibnamefont
  {Yan}}, \bibinfo {author} {\bibfnamefont {Y.}~\bibnamefont {Zhong}}, \bibinfo
  {author} {\bibfnamefont {H.-S.}\ \bibnamefont {Chang}}, \bibinfo {author}
  {\bibfnamefont {A.}~\bibnamefont {Bienfait}}, \bibinfo {author}
  {\bibfnamefont {M.-H.}\ \bibnamefont {Chou}}, \bibinfo {author}
  {\bibfnamefont {C.~R.}\ \bibnamefont {Conner}}, \bibinfo {author}
  {\bibfnamefont {E.}~\bibnamefont {Dumur}}, \bibinfo {author} {\bibfnamefont
  {J.}~\bibnamefont {Grebel}}, \bibinfo {author} {\bibfnamefont {R.~G.}\
  \bibnamefont {Povey}},\ and\ \bibinfo {author} {\bibfnamefont {A.~N.}\
  \bibnamefont {Cleland}},\ }\bibfield  {title} {\bibinfo {title} {Entanglement
  purification and protection in a superconducting quantum network},\ }\href
  {https://doi.org/10.1103/PhysRevLett.128.080504} {\bibfield  {journal}
  {\bibinfo  {journal} {Phys. Rev. Lett.}\ }\textbf {\bibinfo {volume} {128}},\
  \bibinfo {pages} {080504} (\bibinfo {year} {2022})}\BibitemShut {NoStop}%
\bibitem [{\citenamefont {McKay}\ \emph {et~al.}(2016)\citenamefont {McKay},
  \citenamefont {Filipp}, \citenamefont {Mezzacapo}, \citenamefont {Magesan},
  \citenamefont {Chow},\ and\ \citenamefont {Gambetta}}]{McKay2016Universal}%
  \BibitemOpen
  \bibfield  {author} {\bibinfo {author} {\bibfnamefont {D.~C.}\ \bibnamefont
  {McKay}}, \bibinfo {author} {\bibfnamefont {S.}~\bibnamefont {Filipp}},
  \bibinfo {author} {\bibfnamefont {A.}~\bibnamefont {Mezzacapo}}, \bibinfo
  {author} {\bibfnamefont {E.}~\bibnamefont {Magesan}}, \bibinfo {author}
  {\bibfnamefont {J.~M.}\ \bibnamefont {Chow}},\ and\ \bibinfo {author}
  {\bibfnamefont {J.~M.}\ \bibnamefont {Gambetta}},\ }\bibfield  {title}
  {\bibinfo {title} {Universal gate for fixed-frequency qubits via a tunable
  bus},\ }\href {https://doi.org/10.1103/PhysRevApplied.6.064007} {\bibfield
  {journal} {\bibinfo  {journal} {Phys. Rev. Applied}\ }\textbf {\bibinfo
  {volume} {6}},\ \bibinfo {pages} {064007} (\bibinfo {year}
  {2016})}\BibitemShut {NoStop}%
\bibitem [{\citenamefont {Reagor}\ \emph {et~al.}(2018)\citenamefont {Reagor},
  \citenamefont {Osborn}, \citenamefont {Tezak}, \citenamefont {Staley},
  \citenamefont {Prawiroatmodjo}, \citenamefont {Scheer}, \citenamefont
  {Alidoust}, \citenamefont {Sete}, \citenamefont {Didier}, \citenamefont
  {da~Silva} \emph {et~al.}}]{reagor2018demonstration}%
  \BibitemOpen
  \bibfield  {author} {\bibinfo {author} {\bibfnamefont {M.}~\bibnamefont
  {Reagor}}, \bibinfo {author} {\bibfnamefont {C.~B.}\ \bibnamefont {Osborn}},
  \bibinfo {author} {\bibfnamefont {N.}~\bibnamefont {Tezak}}, \bibinfo
  {author} {\bibfnamefont {A.}~\bibnamefont {Staley}}, \bibinfo {author}
  {\bibfnamefont {G.}~\bibnamefont {Prawiroatmodjo}}, \bibinfo {author}
  {\bibfnamefont {M.}~\bibnamefont {Scheer}}, \bibinfo {author} {\bibfnamefont
  {N.}~\bibnamefont {Alidoust}}, \bibinfo {author} {\bibfnamefont {E.~A.}\
  \bibnamefont {Sete}}, \bibinfo {author} {\bibfnamefont {N.}~\bibnamefont
  {Didier}}, \bibinfo {author} {\bibfnamefont {M.~P.}\ \bibnamefont
  {da~Silva}}, \emph {et~al.},\ }\bibfield  {title} {\bibinfo {title}
  {Demonstration of universal parametric entangling gates on a multi-qubit
  lattice},\ }\href {https://doi.org/10.1126/sciadv.aao3603} {\bibfield
  {journal} {\bibinfo  {journal} {Science advances}\ }\textbf {\bibinfo
  {volume} {4}},\ \bibinfo {pages} {eaao3603} (\bibinfo {year}
  {2018})}\BibitemShut {NoStop}%
\bibitem [{\citenamefont {Mundada}\ \emph {et~al.}(2019)\citenamefont
  {Mundada}, \citenamefont {Zhang}, \citenamefont {Hazard},\ and\ \citenamefont
  {Houck}}]{Mundada2019Suppression}%
  \BibitemOpen
  \bibfield  {author} {\bibinfo {author} {\bibfnamefont {P.}~\bibnamefont
  {Mundada}}, \bibinfo {author} {\bibfnamefont {G.}~\bibnamefont {Zhang}},
  \bibinfo {author} {\bibfnamefont {T.}~\bibnamefont {Hazard}},\ and\ \bibinfo
  {author} {\bibfnamefont {A.}~\bibnamefont {Houck}},\ }\bibfield  {title}
  {\bibinfo {title} {Suppression of qubit crosstalk in a tunable coupling
  superconducting circuit},\ }\href
  {https://doi.org/10.1103/PhysRevApplied.12.054023} {\bibfield  {journal}
  {\bibinfo  {journal} {Phys. Rev. Applied}\ }\textbf {\bibinfo {volume}
  {12}},\ \bibinfo {pages} {054023} (\bibinfo {year} {2019})}\BibitemShut
  {NoStop}%
\bibitem [{\citenamefont {Ganzhorn}\ \emph {et~al.}(2019)\citenamefont
  {Ganzhorn}, \citenamefont {Egger}, \citenamefont {Barkoutsos}, \citenamefont
  {Ollitrault}, \citenamefont {Salis}, \citenamefont {Moll}, \citenamefont
  {Roth}, \citenamefont {Fuhrer}, \citenamefont {Mueller}, \citenamefont
  {Woerner}, \citenamefont {Tavernelli},\ and\ \citenamefont
  {Filipp}}]{Ganzhorn2019Gate}%
  \BibitemOpen
  \bibfield  {author} {\bibinfo {author} {\bibfnamefont {M.}~\bibnamefont
  {Ganzhorn}}, \bibinfo {author} {\bibfnamefont {D.}~\bibnamefont {Egger}},
  \bibinfo {author} {\bibfnamefont {P.}~\bibnamefont {Barkoutsos}}, \bibinfo
  {author} {\bibfnamefont {P.}~\bibnamefont {Ollitrault}}, \bibinfo {author}
  {\bibfnamefont {G.}~\bibnamefont {Salis}}, \bibinfo {author} {\bibfnamefont
  {N.}~\bibnamefont {Moll}}, \bibinfo {author} {\bibfnamefont {M.}~\bibnamefont
  {Roth}}, \bibinfo {author} {\bibfnamefont {A.}~\bibnamefont {Fuhrer}},
  \bibinfo {author} {\bibfnamefont {P.}~\bibnamefont {Mueller}}, \bibinfo
  {author} {\bibfnamefont {S.}~\bibnamefont {Woerner}}, \bibinfo {author}
  {\bibfnamefont {I.}~\bibnamefont {Tavernelli}},\ and\ \bibinfo {author}
  {\bibfnamefont {S.}~\bibnamefont {Filipp}},\ }\bibfield  {title} {\bibinfo
  {title} {Gate-efficient simulation of molecular eigenstates on a quantum
  computer},\ }\href {https://doi.org/10.1103/PhysRevApplied.11.044092}
  {\bibfield  {journal} {\bibinfo  {journal} {Phys. Rev. Applied}\ }\textbf
  {\bibinfo {volume} {11}},\ \bibinfo {pages} {044092} (\bibinfo {year}
  {2019})}\BibitemShut {NoStop}%
\bibitem [{\citenamefont {Xu}\ \emph {et~al.}(2020)\citenamefont {Xu},
  \citenamefont {Chu}, \citenamefont {Yuan}, \citenamefont {Qiu}, \citenamefont
  {Zhou}, \citenamefont {Zhang}, \citenamefont {Tan}, \citenamefont {Yu},
  \citenamefont {Liu}, \citenamefont {Li}, \citenamefont {Yan},\ and\
  \citenamefont {Yu}}]{Yuan2020High}%
  \BibitemOpen
  \bibfield  {author} {\bibinfo {author} {\bibfnamefont {Y.}~\bibnamefont
  {Xu}}, \bibinfo {author} {\bibfnamefont {J.}~\bibnamefont {Chu}}, \bibinfo
  {author} {\bibfnamefont {J.}~\bibnamefont {Yuan}}, \bibinfo {author}
  {\bibfnamefont {J.}~\bibnamefont {Qiu}}, \bibinfo {author} {\bibfnamefont
  {Y.}~\bibnamefont {Zhou}}, \bibinfo {author} {\bibfnamefont {L.}~\bibnamefont
  {Zhang}}, \bibinfo {author} {\bibfnamefont {X.}~\bibnamefont {Tan}}, \bibinfo
  {author} {\bibfnamefont {Y.}~\bibnamefont {Yu}}, \bibinfo {author}
  {\bibfnamefont {S.}~\bibnamefont {Liu}}, \bibinfo {author} {\bibfnamefont
  {J.}~\bibnamefont {Li}}, \bibinfo {author} {\bibfnamefont {F.}~\bibnamefont
  {Yan}},\ and\ \bibinfo {author} {\bibfnamefont {D.}~\bibnamefont {Yu}},\
  }\bibfield  {title} {\bibinfo {title} {High-fidelity, high-scalability
  two-qubit gate scheme for superconducting qubits},\ }\href
  {https://doi.org/10.1103/PhysRevLett.125.240503} {\bibfield  {journal}
  {\bibinfo  {journal} {Phys. Rev. Lett.}\ }\textbf {\bibinfo {volume} {125}},\
  \bibinfo {pages} {240503} (\bibinfo {year} {2020})}\BibitemShut {NoStop}%
\bibitem [{\citenamefont {Ganzhorn}\ \emph {et~al.}(2020)\citenamefont
  {Ganzhorn}, \citenamefont {Salis}, \citenamefont {Egger}, \citenamefont
  {Fuhrer}, \citenamefont {Mergenthaler}, \citenamefont {M\"uller},
  \citenamefont {M\"uller}, \citenamefont {Paredes}, \citenamefont {Pechal},
  \citenamefont {Werninghaus},\ and\ \citenamefont
  {Filipp}}]{Ganzhorn2020Benchmarking}%
  \BibitemOpen
  \bibfield  {author} {\bibinfo {author} {\bibfnamefont {M.}~\bibnamefont
  {Ganzhorn}}, \bibinfo {author} {\bibfnamefont {G.}~\bibnamefont {Salis}},
  \bibinfo {author} {\bibfnamefont {D.~J.}\ \bibnamefont {Egger}}, \bibinfo
  {author} {\bibfnamefont {A.}~\bibnamefont {Fuhrer}}, \bibinfo {author}
  {\bibfnamefont {M.}~\bibnamefont {Mergenthaler}}, \bibinfo {author}
  {\bibfnamefont {C.}~\bibnamefont {M\"uller}}, \bibinfo {author}
  {\bibfnamefont {P.}~\bibnamefont {M\"uller}}, \bibinfo {author}
  {\bibfnamefont {S.}~\bibnamefont {Paredes}}, \bibinfo {author} {\bibfnamefont
  {M.}~\bibnamefont {Pechal}}, \bibinfo {author} {\bibfnamefont
  {M.}~\bibnamefont {Werninghaus}},\ and\ \bibinfo {author} {\bibfnamefont
  {S.}~\bibnamefont {Filipp}},\ }\bibfield  {title} {\bibinfo {title}
  {Benchmarking the noise sensitivity of different parametric two-qubit gates
  in a single superconducting quantum computing platform},\ }\href
  {https://doi.org/10.1103/PhysRevResearch.2.033447} {\bibfield  {journal}
  {\bibinfo  {journal} {Phys. Rev. Research}\ }\textbf {\bibinfo {volume}
  {2}},\ \bibinfo {pages} {033447} (\bibinfo {year} {2020})}\BibitemShut
  {NoStop}%
\bibitem [{\citenamefont {Abrams}\ \emph {et~al.}(2020)\citenamefont {Abrams},
  \citenamefont {Didier}, \citenamefont {Johnson}, \citenamefont {Silva},\ and\
  \citenamefont {Ryan}}]{abrams2020implementation}%
  \BibitemOpen
  \bibfield  {author} {\bibinfo {author} {\bibfnamefont {D.~M.}\ \bibnamefont
  {Abrams}}, \bibinfo {author} {\bibfnamefont {N.}~\bibnamefont {Didier}},
  \bibinfo {author} {\bibfnamefont {B.~R.}\ \bibnamefont {Johnson}}, \bibinfo
  {author} {\bibfnamefont {M.~P.~d.}\ \bibnamefont {Silva}},\ and\ \bibinfo
  {author} {\bibfnamefont {C.~A.}\ \bibnamefont {Ryan}},\ }\bibfield  {title}
  {\bibinfo {title} {Implementation of xy entangling gates with a single
  calibrated pulse},\ }\href {https://doi.org/10.1038/s41928-020-00498-1}
  {\bibfield  {journal} {\bibinfo  {journal} {Nature Electronics}\ }\textbf
  {\bibinfo {volume} {3}},\ \bibinfo {pages} {744} (\bibinfo {year}
  {2020})}\BibitemShut {NoStop}%
\bibitem [{\citenamefont {Stehlik}\ \emph {et~al.}(2021)\citenamefont
  {Stehlik}, \citenamefont {Zajac}, \citenamefont {Underwood}, \citenamefont
  {Phung}, \citenamefont {Blair}, \citenamefont {Carnevale}, \citenamefont
  {Klaus}, \citenamefont {Keefe}, \citenamefont {Carniol}, \citenamefont
  {Kumph}, \citenamefont {Steffen},\ and\ \citenamefont
  {Dial}}]{Stehlik2021Tunable}%
  \BibitemOpen
  \bibfield  {author} {\bibinfo {author} {\bibfnamefont {J.}~\bibnamefont
  {Stehlik}}, \bibinfo {author} {\bibfnamefont {D.~M.}\ \bibnamefont {Zajac}},
  \bibinfo {author} {\bibfnamefont {D.~L.}\ \bibnamefont {Underwood}}, \bibinfo
  {author} {\bibfnamefont {T.}~\bibnamefont {Phung}}, \bibinfo {author}
  {\bibfnamefont {J.}~\bibnamefont {Blair}}, \bibinfo {author} {\bibfnamefont
  {S.}~\bibnamefont {Carnevale}}, \bibinfo {author} {\bibfnamefont
  {D.}~\bibnamefont {Klaus}}, \bibinfo {author} {\bibfnamefont {G.~A.}\
  \bibnamefont {Keefe}}, \bibinfo {author} {\bibfnamefont {A.}~\bibnamefont
  {Carniol}}, \bibinfo {author} {\bibfnamefont {M.}~\bibnamefont {Kumph}},
  \bibinfo {author} {\bibfnamefont {M.}~\bibnamefont {Steffen}},\ and\ \bibinfo
  {author} {\bibfnamefont {O.~E.}\ \bibnamefont {Dial}},\ }\bibfield  {title}
  {\bibinfo {title} {Tunable coupling architecture for fixed-frequency transmon
  superconducting qubits},\ }\href
  {https://doi.org/10.1103/PhysRevLett.127.080505} {\bibfield  {journal}
  {\bibinfo  {journal} {Phys. Rev. Lett.}\ }\textbf {\bibinfo {volume} {127}},\
  \bibinfo {pages} {080505} (\bibinfo {year} {2021})}\BibitemShut {NoStop}%
\bibitem [{\citenamefont {Sung}\ \emph {et~al.}(2021)\citenamefont {Sung},
  \citenamefont {Ding}, \citenamefont {Braum\"uller}, \citenamefont
  {Veps\"al\"ainen}, \citenamefont {Kannan}, \citenamefont {Kjaergaard},
  \citenamefont {Greene}, \citenamefont {Samach}, \citenamefont {McNally},
  \citenamefont {Kim}, \citenamefont {Melville}, \citenamefont {Niedzielski},
  \citenamefont {Schwartz}, \citenamefont {Yoder}, \citenamefont {Orlando},
  \citenamefont {Gustavsson},\ and\ \citenamefont
  {Oliver}}]{Sung2021Realization}%
  \BibitemOpen
  \bibfield  {author} {\bibinfo {author} {\bibfnamefont {Y.}~\bibnamefont
  {Sung}}, \bibinfo {author} {\bibfnamefont {L.}~\bibnamefont {Ding}}, \bibinfo
  {author} {\bibfnamefont {J.}~\bibnamefont {Braum\"uller}}, \bibinfo {author}
  {\bibfnamefont {A.}~\bibnamefont {Veps\"al\"ainen}}, \bibinfo {author}
  {\bibfnamefont {B.}~\bibnamefont {Kannan}}, \bibinfo {author} {\bibfnamefont
  {M.}~\bibnamefont {Kjaergaard}}, \bibinfo {author} {\bibfnamefont
  {A.}~\bibnamefont {Greene}}, \bibinfo {author} {\bibfnamefont {G.~O.}\
  \bibnamefont {Samach}}, \bibinfo {author} {\bibfnamefont {C.}~\bibnamefont
  {McNally}}, \bibinfo {author} {\bibfnamefont {D.}~\bibnamefont {Kim}},
  \bibinfo {author} {\bibfnamefont {A.}~\bibnamefont {Melville}}, \bibinfo
  {author} {\bibfnamefont {B.~M.}\ \bibnamefont {Niedzielski}}, \bibinfo
  {author} {\bibfnamefont {M.~E.}\ \bibnamefont {Schwartz}}, \bibinfo {author}
  {\bibfnamefont {J.~L.}\ \bibnamefont {Yoder}}, \bibinfo {author}
  {\bibfnamefont {T.~P.}\ \bibnamefont {Orlando}}, \bibinfo {author}
  {\bibfnamefont {S.}~\bibnamefont {Gustavsson}},\ and\ \bibinfo {author}
  {\bibfnamefont {W.~D.}\ \bibnamefont {Oliver}},\ }\bibfield  {title}
  {\bibinfo {title} {Realization of high-fidelity cz and $zz$-free iswap gates
  with a tunable coupler},\ }\href {https://doi.org/10.1103/PhysRevX.11.021058}
  {\bibfield  {journal} {\bibinfo  {journal} {Phys. Rev. X}\ }\textbf {\bibinfo
  {volume} {11}},\ \bibinfo {pages} {021058} (\bibinfo {year}
  {2021})}\BibitemShut {NoStop}%
\bibitem [{\citenamefont {Leung}\ \emph {et~al.}(2019)\citenamefont {Leung},
  \citenamefont {Lu}, \citenamefont {Chakram}, \citenamefont {Naik},
  \citenamefont {Earnest}, \citenamefont {Ma}, \citenamefont {Jacobs},
  \citenamefont {Cleland},\ and\ \citenamefont
  {Schuster}}]{leung2019deterministic}%
  \BibitemOpen
  \bibfield  {author} {\bibinfo {author} {\bibfnamefont {N.}~\bibnamefont
  {Leung}}, \bibinfo {author} {\bibfnamefont {Y.}~\bibnamefont {Lu}}, \bibinfo
  {author} {\bibfnamefont {S.}~\bibnamefont {Chakram}}, \bibinfo {author}
  {\bibfnamefont {R.}~\bibnamefont {Naik}}, \bibinfo {author} {\bibfnamefont
  {N.}~\bibnamefont {Earnest}}, \bibinfo {author} {\bibfnamefont
  {R.}~\bibnamefont {Ma}}, \bibinfo {author} {\bibfnamefont {K.}~\bibnamefont
  {Jacobs}}, \bibinfo {author} {\bibfnamefont {A.}~\bibnamefont {Cleland}},\
  and\ \bibinfo {author} {\bibfnamefont {D.}~\bibnamefont {Schuster}},\
  }\bibfield  {title} {\bibinfo {title} {Deterministic bidirectional
  communication and remote entanglement generation between superconducting
  qubits},\ }\href {https://doi.org/10.1038/s41534-019-0128-0} {\bibfield
  {journal} {\bibinfo  {journal} {npj Quantum Information}\ }\textbf {\bibinfo
  {volume} {5}},\ \bibinfo {pages} {18} (\bibinfo {year} {2019})}\BibitemShut
  {NoStop}%
\bibitem [{\citenamefont {Hong}\ \emph {et~al.}(2020)\citenamefont {Hong},
  \citenamefont {Papageorge}, \citenamefont {Sivarajah}, \citenamefont
  {Crossman}, \citenamefont {Didier}, \citenamefont {Polloreno}, \citenamefont
  {Sete}, \citenamefont {Turkowski}, \citenamefont {da~Silva},\ and\
  \citenamefont {Johnson}}]{Hong2020Demonstration}%
  \BibitemOpen
  \bibfield  {author} {\bibinfo {author} {\bibfnamefont {S.~S.}\ \bibnamefont
  {Hong}}, \bibinfo {author} {\bibfnamefont {A.~T.}\ \bibnamefont
  {Papageorge}}, \bibinfo {author} {\bibfnamefont {P.}~\bibnamefont
  {Sivarajah}}, \bibinfo {author} {\bibfnamefont {G.}~\bibnamefont {Crossman}},
  \bibinfo {author} {\bibfnamefont {N.}~\bibnamefont {Didier}}, \bibinfo
  {author} {\bibfnamefont {A.~M.}\ \bibnamefont {Polloreno}}, \bibinfo {author}
  {\bibfnamefont {E.~A.}\ \bibnamefont {Sete}}, \bibinfo {author}
  {\bibfnamefont {S.~W.}\ \bibnamefont {Turkowski}}, \bibinfo {author}
  {\bibfnamefont {M.~P.}\ \bibnamefont {da~Silva}},\ and\ \bibinfo {author}
  {\bibfnamefont {B.~R.}\ \bibnamefont {Johnson}},\ }\bibfield  {title}
  {\bibinfo {title} {Demonstration of a parametrically activated entangling
  gate protected from flux noise},\ }\href
  {https://doi.org/10.1103/PhysRevA.101.012302} {\bibfield  {journal} {\bibinfo
   {journal} {Phys. Rev. A}\ }\textbf {\bibinfo {volume} {101}},\ \bibinfo
  {pages} {012302} (\bibinfo {year} {2020})}\BibitemShut {NoStop}%
\bibitem [{\citenamefont {Manucharyan}\ \emph {et~al.}(2009)\citenamefont
  {Manucharyan}, \citenamefont {Koch}, \citenamefont {Glazman},\ and\
  \citenamefont {Devoret}}]{manucharyan2009fluxonium}%
  \BibitemOpen
  \bibfield  {author} {\bibinfo {author} {\bibfnamefont {V.~E.}\ \bibnamefont
  {Manucharyan}}, \bibinfo {author} {\bibfnamefont {J.}~\bibnamefont {Koch}},
  \bibinfo {author} {\bibfnamefont {L.~I.}\ \bibnamefont {Glazman}},\ and\
  \bibinfo {author} {\bibfnamefont {M.~H.}\ \bibnamefont {Devoret}},\
  }\bibfield  {title} {\bibinfo {title} {Fluxonium: Single cooper-pair circuit
  free of charge offsets},\ }\href {https://doi.org/10.1126/science.1175552}
  {\bibfield  {journal} {\bibinfo  {journal} {Science}\ }\textbf {\bibinfo
  {volume} {326}},\ \bibinfo {pages} {113} (\bibinfo {year}
  {2009})}\BibitemShut {NoStop}%
\bibitem [{\citenamefont {Earnest}\ \emph {et~al.}(2018)\citenamefont
  {Earnest}, \citenamefont {Chakram}, \citenamefont {Lu}, \citenamefont
  {Irons}, \citenamefont {Naik}, \citenamefont {Leung}, \citenamefont {Ocola},
  \citenamefont {Czaplewski}, \citenamefont {Baker}, \citenamefont {Lawrence},
  \citenamefont {Koch},\ and\ \citenamefont
  {Schuster}}]{Earnest2018Realization}%
  \BibitemOpen
  \bibfield  {author} {\bibinfo {author} {\bibfnamefont {N.}~\bibnamefont
  {Earnest}}, \bibinfo {author} {\bibfnamefont {S.}~\bibnamefont {Chakram}},
  \bibinfo {author} {\bibfnamefont {Y.}~\bibnamefont {Lu}}, \bibinfo {author}
  {\bibfnamefont {N.}~\bibnamefont {Irons}}, \bibinfo {author} {\bibfnamefont
  {R.~K.}\ \bibnamefont {Naik}}, \bibinfo {author} {\bibfnamefont
  {N.}~\bibnamefont {Leung}}, \bibinfo {author} {\bibfnamefont
  {L.}~\bibnamefont {Ocola}}, \bibinfo {author} {\bibfnamefont {D.~A.}\
  \bibnamefont {Czaplewski}}, \bibinfo {author} {\bibfnamefont
  {B.}~\bibnamefont {Baker}}, \bibinfo {author} {\bibfnamefont
  {J.}~\bibnamefont {Lawrence}}, \bibinfo {author} {\bibfnamefont
  {J.}~\bibnamefont {Koch}},\ and\ \bibinfo {author} {\bibfnamefont {D.~I.}\
  \bibnamefont {Schuster}},\ }\bibfield  {title} {\bibinfo {title} {Realization
  of a $\mathrm{\ensuremath{\Lambda}}$ system with metastable states of a
  capacitively shunted fluxonium},\ }\href
  {https://doi.org/10.1103/PhysRevLett.120.150504} {\bibfield  {journal}
  {\bibinfo  {journal} {Phys. Rev. Lett.}\ }\textbf {\bibinfo {volume} {120}},\
  \bibinfo {pages} {150504} (\bibinfo {year} {2018})}\BibitemShut {NoStop}%
\bibitem [{\citenamefont {Nguyen}\ \emph {et~al.}(2019)\citenamefont {Nguyen},
  \citenamefont {Lin}, \citenamefont {Somoroff}, \citenamefont {Mencia},
  \citenamefont {Grabon},\ and\ \citenamefont {Manucharyan}}]{Nguyen2019High}%
  \BibitemOpen
  \bibfield  {author} {\bibinfo {author} {\bibfnamefont {L.~B.}\ \bibnamefont
  {Nguyen}}, \bibinfo {author} {\bibfnamefont {Y.-H.}\ \bibnamefont {Lin}},
  \bibinfo {author} {\bibfnamefont {A.}~\bibnamefont {Somoroff}}, \bibinfo
  {author} {\bibfnamefont {R.}~\bibnamefont {Mencia}}, \bibinfo {author}
  {\bibfnamefont {N.}~\bibnamefont {Grabon}},\ and\ \bibinfo {author}
  {\bibfnamefont {V.~E.}\ \bibnamefont {Manucharyan}},\ }\bibfield  {title}
  {\bibinfo {title} {High-coherence fluxonium qubit},\ }\href
  {https://doi.org/10.1103/PhysRevX.9.041041} {\bibfield  {journal} {\bibinfo
  {journal} {Phys. Rev. X}\ }\textbf {\bibinfo {volume} {9}},\ \bibinfo {pages}
  {041041} (\bibinfo {year} {2019})}\BibitemShut {NoStop}%
\bibitem [{\citenamefont {Zhang}\ \emph {et~al.}(2021)\citenamefont {Zhang},
  \citenamefont {Chakram}, \citenamefont {Roy}, \citenamefont {Earnest},
  \citenamefont {Lu}, \citenamefont {Huang}, \citenamefont {Weiss},
  \citenamefont {Koch},\ and\ \citenamefont {Schuster}}]{Helin2021Universal}%
  \BibitemOpen
  \bibfield  {author} {\bibinfo {author} {\bibfnamefont {H.}~\bibnamefont
  {Zhang}}, \bibinfo {author} {\bibfnamefont {S.}~\bibnamefont {Chakram}},
  \bibinfo {author} {\bibfnamefont {T.}~\bibnamefont {Roy}}, \bibinfo {author}
  {\bibfnamefont {N.}~\bibnamefont {Earnest}}, \bibinfo {author} {\bibfnamefont
  {Y.}~\bibnamefont {Lu}}, \bibinfo {author} {\bibfnamefont {Z.}~\bibnamefont
  {Huang}}, \bibinfo {author} {\bibfnamefont {D.~K.}\ \bibnamefont {Weiss}},
  \bibinfo {author} {\bibfnamefont {J.}~\bibnamefont {Koch}},\ and\ \bibinfo
  {author} {\bibfnamefont {D.~I.}\ \bibnamefont {Schuster}},\ }\bibfield
  {title} {\bibinfo {title} {Universal fast-flux control of a coherent,
  low-frequency qubit},\ }\href {https://doi.org/10.1103/PhysRevX.11.011010}
  {\bibfield  {journal} {\bibinfo  {journal} {Phys. Rev. X}\ }\textbf {\bibinfo
  {volume} {11}},\ \bibinfo {pages} {011010} (\bibinfo {year}
  {2021})}\BibitemShut {NoStop}%
\bibitem [{\citenamefont {Yan}\ \emph {et~al.}(2018)\citenamefont {Yan},
  \citenamefont {Krantz}, \citenamefont {Sung}, \citenamefont {Kjaergaard},
  \citenamefont {Campbell}, \citenamefont {Orlando}, \citenamefont
  {Gustavsson},\ and\ \citenamefont {Oliver}}]{Yan2018Tunable}%
  \BibitemOpen
  \bibfield  {author} {\bibinfo {author} {\bibfnamefont {F.}~\bibnamefont
  {Yan}}, \bibinfo {author} {\bibfnamefont {P.}~\bibnamefont {Krantz}},
  \bibinfo {author} {\bibfnamefont {Y.}~\bibnamefont {Sung}}, \bibinfo {author}
  {\bibfnamefont {M.}~\bibnamefont {Kjaergaard}}, \bibinfo {author}
  {\bibfnamefont {D.~L.}\ \bibnamefont {Campbell}}, \bibinfo {author}
  {\bibfnamefont {T.~P.}\ \bibnamefont {Orlando}}, \bibinfo {author}
  {\bibfnamefont {S.}~\bibnamefont {Gustavsson}},\ and\ \bibinfo {author}
  {\bibfnamefont {W.~D.}\ \bibnamefont {Oliver}},\ }\bibfield  {title}
  {\bibinfo {title} {Tunable coupling scheme for implementing high-fidelity
  two-qubit gates},\ }\href {https://doi.org/10.1103/PhysRevApplied.10.054062}
  {\bibfield  {journal} {\bibinfo  {journal} {Phys. Rev. Applied}\ }\textbf
  {\bibinfo {volume} {10}},\ \bibinfo {pages} {054062} (\bibinfo {year}
  {2018})}\BibitemShut {NoStop}%
\bibitem [{\citenamefont {Foxen}\ \emph {et~al.}(2020)\citenamefont {Foxen},
  \citenamefont {Neill}, \citenamefont {Dunsworth}, \citenamefont {Roushan},
  \citenamefont {Chiaro}, \citenamefont {Megrant}, \citenamefont {Kelly},
  \citenamefont {Chen}, \citenamefont {Satzinger}, \citenamefont {Barends},
  \citenamefont {Arute}, \citenamefont {Arya}, \citenamefont {Babbush},
  \citenamefont {Bacon}, \citenamefont {Bardin}, \citenamefont {Boixo},
  \citenamefont {Buell}, \citenamefont {Burkett}, \citenamefont {Chen},
  \citenamefont {Collins}, \citenamefont {Farhi}, \citenamefont {Fowler},
  \citenamefont {Gidney}, \citenamefont {Giustina}, \citenamefont {Graff},
  \citenamefont {Harrigan}, \citenamefont {Huang}, \citenamefont {Isakov},
  \citenamefont {Jeffrey}, \citenamefont {Jiang}, \citenamefont {Kafri},
  \citenamefont {Kechedzhi}, \citenamefont {Klimov}, \citenamefont {Korotkov},
  \citenamefont {Kostritsa}, \citenamefont {Landhuis}, \citenamefont {Lucero},
  \citenamefont {McClean}, \citenamefont {McEwen}, \citenamefont {Mi},
  \citenamefont {Mohseni}, \citenamefont {Mutus}, \citenamefont {Naaman},
  \citenamefont {Neeley}, \citenamefont {Niu}, \citenamefont {Petukhov},
  \citenamefont {Quintana}, \citenamefont {Rubin}, \citenamefont {Sank},
  \citenamefont {Smelyanskiy}, \citenamefont {Vainsencher}, \citenamefont
  {White}, \citenamefont {Yao}, \citenamefont {Yeh}, \citenamefont {Zalcman},
  \citenamefont {Neven},\ and\ \citenamefont
  {Martinis}}]{Foxen2020Demonstrating}%
  \BibitemOpen
  \bibfield  {author} {\bibinfo {author} {\bibfnamefont {B.}~\bibnamefont
  {Foxen}}, \bibinfo {author} {\bibfnamefont {C.}~\bibnamefont {Neill}},
  \bibinfo {author} {\bibfnamefont {A.}~\bibnamefont {Dunsworth}}, \bibinfo
  {author} {\bibfnamefont {P.}~\bibnamefont {Roushan}}, \bibinfo {author}
  {\bibfnamefont {B.}~\bibnamefont {Chiaro}}, \bibinfo {author} {\bibfnamefont
  {A.}~\bibnamefont {Megrant}}, \bibinfo {author} {\bibfnamefont
  {J.}~\bibnamefont {Kelly}}, \bibinfo {author} {\bibfnamefont
  {Z.}~\bibnamefont {Chen}}, \bibinfo {author} {\bibfnamefont {K.}~\bibnamefont
  {Satzinger}}, \bibinfo {author} {\bibfnamefont {R.}~\bibnamefont {Barends}},
  \bibinfo {author} {\bibfnamefont {F.}~\bibnamefont {Arute}}, \bibinfo
  {author} {\bibfnamefont {K.}~\bibnamefont {Arya}}, \bibinfo {author}
  {\bibfnamefont {R.}~\bibnamefont {Babbush}}, \bibinfo {author} {\bibfnamefont
  {D.}~\bibnamefont {Bacon}}, \bibinfo {author} {\bibfnamefont {J.~C.}\
  \bibnamefont {Bardin}}, \bibinfo {author} {\bibfnamefont {S.}~\bibnamefont
  {Boixo}}, \bibinfo {author} {\bibfnamefont {D.}~\bibnamefont {Buell}},
  \bibinfo {author} {\bibfnamefont {B.}~\bibnamefont {Burkett}}, \bibinfo
  {author} {\bibfnamefont {Y.}~\bibnamefont {Chen}}, \bibinfo {author}
  {\bibfnamefont {R.}~\bibnamefont {Collins}}, \bibinfo {author} {\bibfnamefont
  {E.}~\bibnamefont {Farhi}}, \bibinfo {author} {\bibfnamefont
  {A.}~\bibnamefont {Fowler}}, \bibinfo {author} {\bibfnamefont
  {C.}~\bibnamefont {Gidney}}, \bibinfo {author} {\bibfnamefont
  {M.}~\bibnamefont {Giustina}}, \bibinfo {author} {\bibfnamefont
  {R.}~\bibnamefont {Graff}}, \bibinfo {author} {\bibfnamefont
  {M.}~\bibnamefont {Harrigan}}, \bibinfo {author} {\bibfnamefont
  {T.}~\bibnamefont {Huang}}, \bibinfo {author} {\bibfnamefont {S.~V.}\
  \bibnamefont {Isakov}}, \bibinfo {author} {\bibfnamefont {E.}~\bibnamefont
  {Jeffrey}}, \bibinfo {author} {\bibfnamefont {Z.}~\bibnamefont {Jiang}},
  \bibinfo {author} {\bibfnamefont {D.}~\bibnamefont {Kafri}}, \bibinfo
  {author} {\bibfnamefont {K.}~\bibnamefont {Kechedzhi}}, \bibinfo {author}
  {\bibfnamefont {P.}~\bibnamefont {Klimov}}, \bibinfo {author} {\bibfnamefont
  {A.}~\bibnamefont {Korotkov}}, \bibinfo {author} {\bibfnamefont
  {F.}~\bibnamefont {Kostritsa}}, \bibinfo {author} {\bibfnamefont
  {D.}~\bibnamefont {Landhuis}}, \bibinfo {author} {\bibfnamefont
  {E.}~\bibnamefont {Lucero}}, \bibinfo {author} {\bibfnamefont
  {J.}~\bibnamefont {McClean}}, \bibinfo {author} {\bibfnamefont
  {M.}~\bibnamefont {McEwen}}, \bibinfo {author} {\bibfnamefont
  {X.}~\bibnamefont {Mi}}, \bibinfo {author} {\bibfnamefont {M.}~\bibnamefont
  {Mohseni}}, \bibinfo {author} {\bibfnamefont {J.~Y.}\ \bibnamefont {Mutus}},
  \bibinfo {author} {\bibfnamefont {O.}~\bibnamefont {Naaman}}, \bibinfo
  {author} {\bibfnamefont {M.}~\bibnamefont {Neeley}}, \bibinfo {author}
  {\bibfnamefont {M.}~\bibnamefont {Niu}}, \bibinfo {author} {\bibfnamefont
  {A.}~\bibnamefont {Petukhov}}, \bibinfo {author} {\bibfnamefont
  {C.}~\bibnamefont {Quintana}}, \bibinfo {author} {\bibfnamefont
  {N.}~\bibnamefont {Rubin}}, \bibinfo {author} {\bibfnamefont
  {D.}~\bibnamefont {Sank}}, \bibinfo {author} {\bibfnamefont {V.}~\bibnamefont
  {Smelyanskiy}}, \bibinfo {author} {\bibfnamefont {A.}~\bibnamefont
  {Vainsencher}}, \bibinfo {author} {\bibfnamefont {T.~C.}\ \bibnamefont
  {White}}, \bibinfo {author} {\bibfnamefont {Z.}~\bibnamefont {Yao}}, \bibinfo
  {author} {\bibfnamefont {P.}~\bibnamefont {Yeh}}, \bibinfo {author}
  {\bibfnamefont {A.}~\bibnamefont {Zalcman}}, \bibinfo {author} {\bibfnamefont
  {H.}~\bibnamefont {Neven}},\ and\ \bibinfo {author} {\bibfnamefont {J.~M.}\
  \bibnamefont {Martinis}} (\bibinfo {collaboration} {Google AI Quantum}),\
  }\bibfield  {title} {\bibinfo {title} {Demonstrating a continuous set of
  two-qubit gates for near-term quantum algorithms},\ }\href
  {https://doi.org/10.1103/PhysRevLett.125.120504} {\bibfield  {journal}
  {\bibinfo  {journal} {Phys. Rev. Lett.}\ }\textbf {\bibinfo {volume} {125}},\
  \bibinfo {pages} {120504} (\bibinfo {year} {2020})}\BibitemShut {NoStop}%
\bibitem [{\citenamefont {Li}\ \emph {et~al.}(2020)\citenamefont {Li},
  \citenamefont {Cai}, \citenamefont {Yan}, \citenamefont {Wang}, \citenamefont
  {Pan}, \citenamefont {Ma}, \citenamefont {Cai}, \citenamefont {Han},
  \citenamefont {Hua}, \citenamefont {Han}, \citenamefont {Wu}, \citenamefont
  {Zhang}, \citenamefont {Wang}, \citenamefont {Song}, \citenamefont {Duan},\
  and\ \citenamefont {Sun}}]{Li2020Tunable}%
  \BibitemOpen
  \bibfield  {author} {\bibinfo {author} {\bibfnamefont {X.}~\bibnamefont
  {Li}}, \bibinfo {author} {\bibfnamefont {T.}~\bibnamefont {Cai}}, \bibinfo
  {author} {\bibfnamefont {H.}~\bibnamefont {Yan}}, \bibinfo {author}
  {\bibfnamefont {Z.}~\bibnamefont {Wang}}, \bibinfo {author} {\bibfnamefont
  {X.}~\bibnamefont {Pan}}, \bibinfo {author} {\bibfnamefont {Y.}~\bibnamefont
  {Ma}}, \bibinfo {author} {\bibfnamefont {W.}~\bibnamefont {Cai}}, \bibinfo
  {author} {\bibfnamefont {J.}~\bibnamefont {Han}}, \bibinfo {author}
  {\bibfnamefont {Z.}~\bibnamefont {Hua}}, \bibinfo {author} {\bibfnamefont
  {X.}~\bibnamefont {Han}}, \bibinfo {author} {\bibfnamefont {Y.}~\bibnamefont
  {Wu}}, \bibinfo {author} {\bibfnamefont {H.}~\bibnamefont {Zhang}}, \bibinfo
  {author} {\bibfnamefont {H.}~\bibnamefont {Wang}}, \bibinfo {author}
  {\bibfnamefont {Y.}~\bibnamefont {Song}}, \bibinfo {author} {\bibfnamefont
  {L.}~\bibnamefont {Duan}},\ and\ \bibinfo {author} {\bibfnamefont
  {L.}~\bibnamefont {Sun}},\ }\bibfield  {title} {\bibinfo {title} {Tunable
  coupler for realizing a controlled-phase gate with dynamically decoupled
  regime in a superconducting circuit},\ }\href
  {https://doi.org/10.1103/PhysRevApplied.14.024070} {\bibfield  {journal}
  {\bibinfo  {journal} {Phys. Rev. Applied}\ }\textbf {\bibinfo {volume}
  {14}},\ \bibinfo {pages} {024070} (\bibinfo {year} {2020})}\BibitemShut
  {NoStop}%
\bibitem [{\citenamefont {Moskalenko}\ \emph {et~al.}(2022)\citenamefont
  {Moskalenko}, \citenamefont {Simakov}, \citenamefont {Abramov}, \citenamefont
  {Moskalev}, \citenamefont {Pishchimova}, \citenamefont {Smirnov},
  \citenamefont {Zikiy}, \citenamefont {Rodionov},\ and\ \citenamefont
  {Besedin}}]{moskalenko2022high}%
  \BibitemOpen
  \bibfield  {author} {\bibinfo {author} {\bibfnamefont {I.~N.}\ \bibnamefont
  {Moskalenko}}, \bibinfo {author} {\bibfnamefont {I.~A.}\ \bibnamefont
  {Simakov}}, \bibinfo {author} {\bibfnamefont {N.~N.}\ \bibnamefont
  {Abramov}}, \bibinfo {author} {\bibfnamefont {D.~O.}\ \bibnamefont
  {Moskalev}}, \bibinfo {author} {\bibfnamefont {A.~A.}\ \bibnamefont
  {Pishchimova}}, \bibinfo {author} {\bibfnamefont {N.~S.}\ \bibnamefont
  {Smirnov}}, \bibinfo {author} {\bibfnamefont {E.~V.}\ \bibnamefont {Zikiy}},
  \bibinfo {author} {\bibfnamefont {I.~A.}\ \bibnamefont {Rodionov}},\ and\
  \bibinfo {author} {\bibfnamefont {I.~S.}\ \bibnamefont {Besedin}},\
  }\bibfield  {title} {\bibinfo {title} {High fidelity two-qubit gates on
  fluxoniums using a tunable coupler},\ }\href
  {https://arxiv.org/abs/2203.16302} {\bibfield  {journal} {\bibinfo  {journal}
  {arXiv:2203.16302}\ } (\bibinfo {year} {2022})}\BibitemShut {NoStop}%
\bibitem [{\citenamefont {Baksic}\ \emph {et~al.}(2016)\citenamefont {Baksic},
  \citenamefont {Ribeiro},\ and\ \citenamefont {Clerk}}]{Baksic2016Speeding}%
  \BibitemOpen
  \bibfield  {author} {\bibinfo {author} {\bibfnamefont {A.}~\bibnamefont
  {Baksic}}, \bibinfo {author} {\bibfnamefont {H.}~\bibnamefont {Ribeiro}},\
  and\ \bibinfo {author} {\bibfnamefont {A.~A.}\ \bibnamefont {Clerk}},\
  }\bibfield  {title} {\bibinfo {title} {Speeding up adiabatic quantum state
  transfer by using dressed states},\ }\href
  {https://doi.org/10.1103/PhysRevLett.116.230503} {\bibfield  {journal}
  {\bibinfo  {journal} {Phys. Rev. Lett.}\ }\textbf {\bibinfo {volume} {116}},\
  \bibinfo {pages} {230503} (\bibinfo {year} {2016})}\BibitemShut {NoStop}%
\bibitem [{\citenamefont {Demirplak}\ and\ \citenamefont
  {Rice}(2008)}]{demirplak2008consistency}%
  \BibitemOpen
  \bibfield  {author} {\bibinfo {author} {\bibfnamefont {M.}~\bibnamefont
  {Demirplak}}\ and\ \bibinfo {author} {\bibfnamefont {S.~A.}\ \bibnamefont
  {Rice}},\ }\bibfield  {title} {\bibinfo {title} {On the consistency,
  extremal, and global properties of counterdiabatic fields},\ }\href
  {https://doi.org/10.1063/1.2992152} {\bibfield  {journal} {\bibinfo
  {journal} {The Journal of chemical physics}\ }\textbf {\bibinfo {volume}
  {129}},\ \bibinfo {pages} {154111} (\bibinfo {year} {2008})}\BibitemShut
  {NoStop}%
\bibitem [{\citenamefont {Zhou}\ \emph {et~al.}(2017)\citenamefont {Zhou},
  \citenamefont {Baksic}, \citenamefont {Ribeiro}, \citenamefont {Yale},
  \citenamefont {Heremans}, \citenamefont {Jerger}, \citenamefont {Auer},
  \citenamefont {Burkard}, \citenamefont {Clerk},\ and\ \citenamefont
  {Awschalom}}]{Zhou2017Accelerated}%
  \BibitemOpen
  \bibfield  {author} {\bibinfo {author} {\bibfnamefont {B.~B.}\ \bibnamefont
  {Zhou}}, \bibinfo {author} {\bibfnamefont {A.}~\bibnamefont {Baksic}},
  \bibinfo {author} {\bibfnamefont {H.}~\bibnamefont {Ribeiro}}, \bibinfo
  {author} {\bibfnamefont {C.~G.}\ \bibnamefont {Yale}}, \bibinfo {author}
  {\bibfnamefont {F.~J.}\ \bibnamefont {Heremans}}, \bibinfo {author}
  {\bibfnamefont {P.~C.}\ \bibnamefont {Jerger}}, \bibinfo {author}
  {\bibfnamefont {A.}~\bibnamefont {Auer}}, \bibinfo {author} {\bibfnamefont
  {G.}~\bibnamefont {Burkard}}, \bibinfo {author} {\bibfnamefont {A.~A.}\
  \bibnamefont {Clerk}},\ and\ \bibinfo {author} {\bibfnamefont {D.~D.}\
  \bibnamefont {Awschalom}},\ }\bibfield  {title} {\bibinfo {title}
  {Accelerated quantum control using superadiabatic dynamics in a solid-state
  lambda system},\ }\href {https://doi.org/10.1038/nphys3967} {\bibfield
  {journal} {\bibinfo  {journal} {Nature Physics}\ }\textbf {\bibinfo {volume}
  {13}},\ \bibinfo {pages} {330} (\bibinfo {year} {2017})}\BibitemShut
  {NoStop}%
\bibitem [{\citenamefont {Ribeiro}\ \emph {et~al.}(2017)\citenamefont
  {Ribeiro}, \citenamefont {Baksic},\ and\ \citenamefont
  {Clerk}}]{Ribeiro2017Systematic}%
  \BibitemOpen
  \bibfield  {author} {\bibinfo {author} {\bibfnamefont {H.}~\bibnamefont
  {Ribeiro}}, \bibinfo {author} {\bibfnamefont {A.}~\bibnamefont {Baksic}},\
  and\ \bibinfo {author} {\bibfnamefont {A.~A.}\ \bibnamefont {Clerk}},\
  }\bibfield  {title} {\bibinfo {title} {Systematic magnus-based approach for
  suppressing leakage and nonadiabatic errors in quantum dynamics},\ }\href
  {https://doi.org/10.1103/PhysRevX.7.011021} {\bibfield  {journal} {\bibinfo
  {journal} {Phys. Rev. X}\ }\textbf {\bibinfo {volume} {7}},\ \bibinfo {pages}
  {011021} (\bibinfo {year} {2017})}\BibitemShut {NoStop}%
\bibitem [{\citenamefont {Roque}\ \emph {et~al.}(2021)\citenamefont {Roque},
  \citenamefont {Clerk},\ and\ \citenamefont {Ribeiro}}]{roque2020engineering}%
  \BibitemOpen
  \bibfield  {author} {\bibinfo {author} {\bibfnamefont {T.~F.}\ \bibnamefont
  {Roque}}, \bibinfo {author} {\bibfnamefont {A.~A.}\ \bibnamefont {Clerk}},\
  and\ \bibinfo {author} {\bibfnamefont {H.}~\bibnamefont {Ribeiro}},\
  }\bibfield  {title} {\bibinfo {title} {Engineering fast high-fidelity quantum
  operations with constrained interactions},\ }\href
  {https://doi.org/10.1038/s41534-020-00349-z} {\bibfield  {journal} {\bibinfo
  {journal} {npj Quantum Information}\ }\textbf {\bibinfo {volume} {7}},\
  \bibinfo {pages} {28} (\bibinfo {year} {2021})}\BibitemShut {NoStop}%
\bibitem [{\citenamefont {Nielsen}(2002)}]{nielsen2002simple}%
  \BibitemOpen
  \bibfield  {author} {\bibinfo {author} {\bibfnamefont {M.~A.}\ \bibnamefont
  {Nielsen}},\ }\bibfield  {title} {\bibinfo {title} {A simple formula for the
  average gate fidelity of a quantum dynamical operation},\ }\href
  {https://doi.org/10.1016/S0375-9601(02)01272-0} {\bibfield  {journal}
  {\bibinfo  {journal} {Physics Letters A}\ }\textbf {\bibinfo {volume}
  {303}},\ \bibinfo {pages} {249} (\bibinfo {year} {2002})}\BibitemShut
  {NoStop}%
\bibitem [{\citenamefont {Cabrera}\ and\ \citenamefont
  {Baylis}(2007)}]{cabrera2007average}%
  \BibitemOpen
  \bibfield  {author} {\bibinfo {author} {\bibfnamefont {R.}~\bibnamefont
  {Cabrera}}\ and\ \bibinfo {author} {\bibfnamefont {W.}~\bibnamefont
  {Baylis}},\ }\bibfield  {title} {\bibinfo {title} {Average fidelity in
  n-qubit systems},\ }\href {https://doi.org/10.1016/j.physleta.2007.03.068}
  {\bibfield  {journal} {\bibinfo  {journal} {Physics Letters A}\ }\textbf
  {\bibinfo {volume} {368}},\ \bibinfo {pages} {25} (\bibinfo {year}
  {2007})}\BibitemShut {NoStop}%
\bibitem [{\citenamefont {Johansson}\ \emph {et~al.}(2012)\citenamefont
  {Johansson}, \citenamefont {Nation},\ and\ \citenamefont
  {Nori}}]{johansson2012qutip}%
  \BibitemOpen
  \bibfield  {author} {\bibinfo {author} {\bibfnamefont {J.~R.}\ \bibnamefont
  {Johansson}}, \bibinfo {author} {\bibfnamefont {P.~D.}\ \bibnamefont
  {Nation}},\ and\ \bibinfo {author} {\bibfnamefont {F.}~\bibnamefont {Nori}},\
  }\bibfield  {title} {\bibinfo {title} {Qutip: An open-source python framework
  for the dynamics of open quantum systems},\ }\href
  {https://doi.org/10.1016/j.cpc.2012.02.021} {\bibfield  {journal} {\bibinfo
  {journal} {Computer Physics Communications}\ }\textbf {\bibinfo {volume}
  {183}},\ \bibinfo {pages} {1760} (\bibinfo {year} {2012})}\BibitemShut
  {NoStop}%
\bibitem [{\citenamefont {Johansson}\ \emph {et~al.}(2013)\citenamefont
  {Johansson}, \citenamefont {Nation},\ and\ \citenamefont
  {Nori}}]{johansson2012qutip2}%
  \BibitemOpen
  \bibfield  {author} {\bibinfo {author} {\bibfnamefont {J.~R.}\ \bibnamefont
  {Johansson}}, \bibinfo {author} {\bibfnamefont {P.~D.}\ \bibnamefont
  {Nation}},\ and\ \bibinfo {author} {\bibfnamefont {F.}~\bibnamefont {Nori}},\
  }\bibfield  {title} {\bibinfo {title} {Qutip 2: A python framework for the
  dynamics of open quantum systems},\ }\href
  {https://doi.org/10.1016/j.cpc.2012.11.019} {\bibfield  {journal} {\bibinfo
  {journal} {Computer Physics Communications}\ }\textbf {\bibinfo {volume}
  {184}},\ \bibinfo {pages} {1234} (\bibinfo {year} {2013})}\BibitemShut
  {NoStop}%
\bibitem [{\citenamefont {Chen}\ \emph {et~al.}(2021)\citenamefont {Chen},
  \citenamefont {Nesterov}, \citenamefont {Manucharyan},\ and\ \citenamefont
  {Vavilov}}]{chen2021fast}%
  \BibitemOpen
  \bibfield  {author} {\bibinfo {author} {\bibfnamefont {Y.}~\bibnamefont
  {Chen}}, \bibinfo {author} {\bibfnamefont {K.~N.}\ \bibnamefont {Nesterov}},
  \bibinfo {author} {\bibfnamefont {V.~E.}\ \bibnamefont {Manucharyan}},\ and\
  \bibinfo {author} {\bibfnamefont {M.~G.}\ \bibnamefont {Vavilov}},\
  }\bibfield  {title} {\bibinfo {title} {Fast flux entangling gate for
  fluxonium circuits},\ }\href {https://arxiv.org/abs/2110.00632v1} {\bibfield
  {journal} {\bibinfo  {journal} {arXiv:2110.00632}\ } (\bibinfo {year}
  {2021})}\BibitemShut {NoStop}%
\bibitem [{\citenamefont {Nesterov}\ \emph {et~al.}(2021)\citenamefont
  {Nesterov}, \citenamefont {Ficheux}, \citenamefont {Manucharyan},\ and\
  \citenamefont {Vavilov}}]{Nesterov2021Proposal}%
  \BibitemOpen
  \bibfield  {author} {\bibinfo {author} {\bibfnamefont {K.~N.}\ \bibnamefont
  {Nesterov}}, \bibinfo {author} {\bibfnamefont {Q.}~\bibnamefont {Ficheux}},
  \bibinfo {author} {\bibfnamefont {V.~E.}\ \bibnamefont {Manucharyan}},\ and\
  \bibinfo {author} {\bibfnamefont {M.~G.}\ \bibnamefont {Vavilov}},\
  }\bibfield  {title} {\bibinfo {title} {Proposal for entangling gates on
  fluxonium qubits via a two-photon transition},\ }\href
  {https://doi.org/10.1103/PRXQuantum.2.020345} {\bibfield  {journal} {\bibinfo
   {journal} {PRX Quantum}\ }\textbf {\bibinfo {volume} {2}},\ \bibinfo {pages}
  {020345} (\bibinfo {year} {2021})}\BibitemShut {NoStop}%
\bibitem [{\citenamefont {Moskalenko}\ \emph {et~al.}(2021)\citenamefont
  {Moskalenko}, \citenamefont {Besedin}, \citenamefont {Simakov},\ and\
  \citenamefont {Ustinov}}]{moskalenko2021tunable}%
  \BibitemOpen
  \bibfield  {author} {\bibinfo {author} {\bibfnamefont {I.}~\bibnamefont
  {Moskalenko}}, \bibinfo {author} {\bibfnamefont {I.}~\bibnamefont {Besedin}},
  \bibinfo {author} {\bibfnamefont {I.}~\bibnamefont {Simakov}},\ and\ \bibinfo
  {author} {\bibfnamefont {A.}~\bibnamefont {Ustinov}},\ }\bibfield  {title}
  {\bibinfo {title} {Tunable coupling scheme for implementing two-qubit gates
  on fluxonium qubits},\ }\href {https://doi.org/10.1063/5.0064800} {\bibfield
  {journal} {\bibinfo  {journal} {Applied Physics Letters}\ }\textbf {\bibinfo
  {volume} {119}},\ \bibinfo {pages} {194001} (\bibinfo {year}
  {2021})}\BibitemShut {NoStop}%
\bibitem [{\citenamefont {Nesterov}\ \emph {et~al.}(2022)\citenamefont
  {Nesterov}, \citenamefont {Wang}, \citenamefont {Manucharyan},\ and\
  \citenamefont {Vavilov}}]{nesterov2022controlled}%
  \BibitemOpen
  \bibfield  {author} {\bibinfo {author} {\bibfnamefont {K.~N.}\ \bibnamefont
  {Nesterov}}, \bibinfo {author} {\bibfnamefont {C.}~\bibnamefont {Wang}},
  \bibinfo {author} {\bibfnamefont {V.~E.}\ \bibnamefont {Manucharyan}},\ and\
  \bibinfo {author} {\bibfnamefont {M.~G.}\ \bibnamefont {Vavilov}},\
  }\bibfield  {title} {\bibinfo {title} {cnot gates for fluxonium qubits via
  selective darkening of transitions},\ }\href
  {https://doi.org/10.1103/PhysRevApplied.18.034063} {\bibfield  {journal}
  {\bibinfo  {journal} {Phys. Rev. Appl.}\ }\textbf {\bibinfo {volume} {18}},\
  \bibinfo {pages} {034063} (\bibinfo {year} {2022})}\BibitemShut {NoStop}%
\bibitem [{\citenamefont {Cai}\ \emph {et~al.}(2021)\citenamefont {Cai},
  \citenamefont {Wang}, \citenamefont {Wang}, \citenamefont {Han},
  \citenamefont {Wu}, \citenamefont {Song},\ and\ \citenamefont
  {Duan}}]{Cai2021All}%
  \BibitemOpen
  \bibfield  {author} {\bibinfo {author} {\bibfnamefont {T.-Q.}\ \bibnamefont
  {Cai}}, \bibinfo {author} {\bibfnamefont {J.-H.}\ \bibnamefont {Wang}},
  \bibinfo {author} {\bibfnamefont {Z.-L.}\ \bibnamefont {Wang}}, \bibinfo
  {author} {\bibfnamefont {X.-Y.}\ \bibnamefont {Han}}, \bibinfo {author}
  {\bibfnamefont {Y.-K.}\ \bibnamefont {Wu}}, \bibinfo {author} {\bibfnamefont
  {Y.-P.}\ \bibnamefont {Song}},\ and\ \bibinfo {author} {\bibfnamefont
  {L.-M.}\ \bibnamefont {Duan}},\ }\bibfield  {title} {\bibinfo {title}
  {All-microwave nonadiabatic multiqubit geometric phase gate for
  superconducting qubits},\ }\href
  {https://doi.org/10.1103/PhysRevResearch.3.043071} {\bibfield  {journal}
  {\bibinfo  {journal} {Phys. Rev. Research}\ }\textbf {\bibinfo {volume}
  {3}},\ \bibinfo {pages} {043071} (\bibinfo {year} {2021})}\BibitemShut
  {NoStop}%
\bibitem [{\citenamefont {Nesterov}\ \emph {et~al.}(2018)\citenamefont
  {Nesterov}, \citenamefont {Pechenezhskiy}, \citenamefont {Wang},
  \citenamefont {Manucharyan},\ and\ \citenamefont
  {Vavilov}}]{Nesterov2018Microwave}%
  \BibitemOpen
  \bibfield  {author} {\bibinfo {author} {\bibfnamefont {K.~N.}\ \bibnamefont
  {Nesterov}}, \bibinfo {author} {\bibfnamefont {I.~V.}\ \bibnamefont
  {Pechenezhskiy}}, \bibinfo {author} {\bibfnamefont {C.}~\bibnamefont {Wang}},
  \bibinfo {author} {\bibfnamefont {V.~E.}\ \bibnamefont {Manucharyan}},\ and\
  \bibinfo {author} {\bibfnamefont {M.~G.}\ \bibnamefont {Vavilov}},\
  }\bibfield  {title} {\bibinfo {title} {Microwave-activated controlled-$z$
  gate for fixed-frequency fluxonium qubits},\ }\href
  {https://doi.org/10.1103/PhysRevA.98.030301} {\bibfield  {journal} {\bibinfo
  {journal} {Phys. Rev. A}\ }\textbf {\bibinfo {volume} {98}},\ \bibinfo
  {pages} {030301} (\bibinfo {year} {2018})}\BibitemShut {NoStop}%
\bibitem [{\citenamefont {Nguyen}\ \emph {et~al.}(2022)\citenamefont {Nguyen},
  \citenamefont {Koolstra}, \citenamefont {Kim}, \citenamefont {Morvan},
  \citenamefont {Chistolini}, \citenamefont {Singh}, \citenamefont {Nesterov},
  \citenamefont {J\"unger}, \citenamefont {Chen}, \citenamefont {Pedramrazi},
  \citenamefont {Mitchell}, \citenamefont {Kreikebaum}, \citenamefont {Puri},
  \citenamefont {Santiago},\ and\ \citenamefont
  {Siddiqi}}]{nguyen2022scalable}%
  \BibitemOpen
  \bibfield  {author} {\bibinfo {author} {\bibfnamefont {L.~B.}\ \bibnamefont
  {Nguyen}}, \bibinfo {author} {\bibfnamefont {G.}~\bibnamefont {Koolstra}},
  \bibinfo {author} {\bibfnamefont {Y.}~\bibnamefont {Kim}}, \bibinfo {author}
  {\bibfnamefont {A.}~\bibnamefont {Morvan}}, \bibinfo {author} {\bibfnamefont
  {T.}~\bibnamefont {Chistolini}}, \bibinfo {author} {\bibfnamefont
  {S.}~\bibnamefont {Singh}}, \bibinfo {author} {\bibfnamefont {K.~N.}\
  \bibnamefont {Nesterov}}, \bibinfo {author} {\bibfnamefont {C.}~\bibnamefont
  {J\"unger}}, \bibinfo {author} {\bibfnamefont {L.}~\bibnamefont {Chen}},
  \bibinfo {author} {\bibfnamefont {Z.}~\bibnamefont {Pedramrazi}}, \bibinfo
  {author} {\bibfnamefont {B.~K.}\ \bibnamefont {Mitchell}}, \bibinfo {author}
  {\bibfnamefont {J.~M.}\ \bibnamefont {Kreikebaum}}, \bibinfo {author}
  {\bibfnamefont {S.}~\bibnamefont {Puri}}, \bibinfo {author} {\bibfnamefont
  {D.~I.}\ \bibnamefont {Santiago}},\ and\ \bibinfo {author} {\bibfnamefont
  {I.}~\bibnamefont {Siddiqi}},\ }\bibfield  {title} {\bibinfo {title}
  {Blueprint for a high-performance fluxonium quantum processor},\ }\href
  {https://doi.org/10.1103/PRXQuantum.3.037001} {\bibfield  {journal} {\bibinfo
   {journal} {PRX Quantum}\ }\textbf {\bibinfo {volume} {3}},\ \bibinfo {pages}
  {037001} (\bibinfo {year} {2022})}\BibitemShut {NoStop}%
\bibitem [{\citenamefont {Weiss}\ \emph {et~al.}(2022)\citenamefont {Weiss},
  \citenamefont {Zhang}, \citenamefont {Ding}, \citenamefont {Ma},
  \citenamefont {Schuster},\ and\ \citenamefont {Koch}}]{weiss2022fast}%
  \BibitemOpen
  \bibfield  {author} {\bibinfo {author} {\bibfnamefont {D.}~\bibnamefont
  {Weiss}}, \bibinfo {author} {\bibfnamefont {H.}~\bibnamefont {Zhang}},
  \bibinfo {author} {\bibfnamefont {C.}~\bibnamefont {Ding}}, \bibinfo {author}
  {\bibfnamefont {Y.}~\bibnamefont {Ma}}, \bibinfo {author} {\bibfnamefont
  {D.~I.}\ \bibnamefont {Schuster}},\ and\ \bibinfo {author} {\bibfnamefont
  {J.}~\bibnamefont {Koch}},\ }\bibfield  {title} {\bibinfo {title} {Fast
  high-fidelity gates for galvanically-coupled fluxonium qubits using strong
  flux modulation},\ }\href {https://doi.org/10.1103/PRXQuantum.3.040336}
  {\bibfield  {journal} {\bibinfo  {journal} {PRX Quantum}\ }\textbf {\bibinfo
  {volume} {3}},\ \bibinfo {pages} {040336} (\bibinfo {year}
  {2022})}\BibitemShut {NoStop}%
\bibitem [{\citenamefont {Dogan}\ \emph {et~al.}(2022)\citenamefont {Dogan},
  \citenamefont {Rosenstock}, \citenamefont {Guevel}, \citenamefont {Xiong},
  \citenamefont {Mencia}, \citenamefont {Somoroff}, \citenamefont {Nesterov},
  \citenamefont {Vavilov}, \citenamefont {Manucharyan},\ and\ \citenamefont
  {Wang}}]{dogan2022demonstration}%
  \BibitemOpen
  \bibfield  {author} {\bibinfo {author} {\bibfnamefont {E.}~\bibnamefont
  {Dogan}}, \bibinfo {author} {\bibfnamefont {D.}~\bibnamefont {Rosenstock}},
  \bibinfo {author} {\bibfnamefont {L.~L.}\ \bibnamefont {Guevel}}, \bibinfo
  {author} {\bibfnamefont {H.}~\bibnamefont {Xiong}}, \bibinfo {author}
  {\bibfnamefont {R.~A.}\ \bibnamefont {Mencia}}, \bibinfo {author}
  {\bibfnamefont {A.}~\bibnamefont {Somoroff}}, \bibinfo {author}
  {\bibfnamefont {K.~N.}\ \bibnamefont {Nesterov}}, \bibinfo {author}
  {\bibfnamefont {M.~G.}\ \bibnamefont {Vavilov}}, \bibinfo {author}
  {\bibfnamefont {V.~E.}\ \bibnamefont {Manucharyan}},\ and\ \bibinfo {author}
  {\bibfnamefont {C.}~\bibnamefont {Wang}},\ }\bibfield  {title} {\bibinfo
  {title} {Demonstration of the two-fluxonium cross-resonance gate},\ }\href
  {https://arxiv.org/abs/2204.11829} {\bibfield  {journal} {\bibinfo  {journal}
  {arXiv:2204.11829}\ } (\bibinfo {year} {2022})}\BibitemShut {NoStop}%
\bibitem [{\citenamefont {Ficheux}\ \emph {et~al.}(2021)\citenamefont
  {Ficheux}, \citenamefont {Nguyen}, \citenamefont {Somoroff}, \citenamefont
  {Xiong}, \citenamefont {Nesterov}, \citenamefont {Vavilov},\ and\
  \citenamefont {Manucharyan}}]{Ficheux2021Fast}%
  \BibitemOpen
  \bibfield  {author} {\bibinfo {author} {\bibfnamefont {Q.}~\bibnamefont
  {Ficheux}}, \bibinfo {author} {\bibfnamefont {L.~B.}\ \bibnamefont {Nguyen}},
  \bibinfo {author} {\bibfnamefont {A.}~\bibnamefont {Somoroff}}, \bibinfo
  {author} {\bibfnamefont {H.}~\bibnamefont {Xiong}}, \bibinfo {author}
  {\bibfnamefont {K.~N.}\ \bibnamefont {Nesterov}}, \bibinfo {author}
  {\bibfnamefont {M.~G.}\ \bibnamefont {Vavilov}},\ and\ \bibinfo {author}
  {\bibfnamefont {V.~E.}\ \bibnamefont {Manucharyan}},\ }\bibfield  {title}
  {\bibinfo {title} {Fast logic with slow qubits: Microwave-activated
  controlled-z gate on low-frequency fluxoniums},\ }\href
  {https://doi.org/10.1103/PhysRevX.11.021026} {\bibfield  {journal} {\bibinfo
  {journal} {Phys. Rev. X}\ }\textbf {\bibinfo {volume} {11}},\ \bibinfo
  {pages} {021026} (\bibinfo {year} {2021})}\BibitemShut {NoStop}%
\bibitem [{\citenamefont {Xiong}\ \emph {et~al.}(2022)\citenamefont {Xiong},
  \citenamefont {Ficheux}, \citenamefont {Somoroff}, \citenamefont {Nguyen},
  \citenamefont {Dogan}, \citenamefont {Rosenstock}, \citenamefont {Wang},
  \citenamefont {Nesterov}, \citenamefont {Vavilov},\ and\ \citenamefont
  {Manucharyan}}]{Xiong2022Arbitrary}%
  \BibitemOpen
  \bibfield  {author} {\bibinfo {author} {\bibfnamefont {H.}~\bibnamefont
  {Xiong}}, \bibinfo {author} {\bibfnamefont {Q.}~\bibnamefont {Ficheux}},
  \bibinfo {author} {\bibfnamefont {A.}~\bibnamefont {Somoroff}}, \bibinfo
  {author} {\bibfnamefont {L.~B.}\ \bibnamefont {Nguyen}}, \bibinfo {author}
  {\bibfnamefont {E.}~\bibnamefont {Dogan}}, \bibinfo {author} {\bibfnamefont
  {D.}~\bibnamefont {Rosenstock}}, \bibinfo {author} {\bibfnamefont
  {C.}~\bibnamefont {Wang}}, \bibinfo {author} {\bibfnamefont {K.~N.}\
  \bibnamefont {Nesterov}}, \bibinfo {author} {\bibfnamefont {M.~G.}\
  \bibnamefont {Vavilov}},\ and\ \bibinfo {author} {\bibfnamefont {V.~E.}\
  \bibnamefont {Manucharyan}},\ }\bibfield  {title} {\bibinfo {title}
  {Arbitrary controlled-phase gate on fluxonium qubits using differential ac
  stark shifts},\ }\href {https://doi.org/10.1103/PhysRevResearch.4.023040}
  {\bibfield  {journal} {\bibinfo  {journal} {Phys. Rev. Research}\ }\textbf
  {\bibinfo {volume} {4}},\ \bibinfo {pages} {023040} (\bibinfo {year}
  {2022})}\BibitemShut {NoStop}%
\bibitem [{\citenamefont {Bao}\ \emph {et~al.}(2022)\citenamefont {Bao},
  \citenamefont {Deng}, \citenamefont {Ding}, \citenamefont {Gao},
  \citenamefont {Gao}, \citenamefont {Huang}, \citenamefont {Jiang},
  \citenamefont {Ku}, \citenamefont {Li}, \citenamefont {Ma}, \citenamefont
  {Ni}, \citenamefont {Qin}, \citenamefont {Song}, \citenamefont {Sun},
  \citenamefont {Tang}, \citenamefont {Wang}, \citenamefont {Wu}, \citenamefont
  {Xia}, \citenamefont {Yu}, \citenamefont {Zhang}, \citenamefont {Zhang},
  \citenamefont {Zhang}, \citenamefont {Zhou}, \citenamefont {Zhu},
  \citenamefont {Shi}, \citenamefont {Chen}, \citenamefont {Zhao},\ and\
  \citenamefont {Deng}}]{Bao2022Fluxonium}%
  \BibitemOpen
  \bibfield  {author} {\bibinfo {author} {\bibfnamefont {F.}~\bibnamefont
  {Bao}}, \bibinfo {author} {\bibfnamefont {H.}~\bibnamefont {Deng}}, \bibinfo
  {author} {\bibfnamefont {D.}~\bibnamefont {Ding}}, \bibinfo {author}
  {\bibfnamefont {R.}~\bibnamefont {Gao}}, \bibinfo {author} {\bibfnamefont
  {X.}~\bibnamefont {Gao}}, \bibinfo {author} {\bibfnamefont {C.}~\bibnamefont
  {Huang}}, \bibinfo {author} {\bibfnamefont {X.}~\bibnamefont {Jiang}},
  \bibinfo {author} {\bibfnamefont {H.-S.}\ \bibnamefont {Ku}}, \bibinfo
  {author} {\bibfnamefont {Z.}~\bibnamefont {Li}}, \bibinfo {author}
  {\bibfnamefont {X.}~\bibnamefont {Ma}}, \bibinfo {author} {\bibfnamefont
  {X.}~\bibnamefont {Ni}}, \bibinfo {author} {\bibfnamefont {J.}~\bibnamefont
  {Qin}}, \bibinfo {author} {\bibfnamefont {Z.}~\bibnamefont {Song}}, \bibinfo
  {author} {\bibfnamefont {H.}~\bibnamefont {Sun}}, \bibinfo {author}
  {\bibfnamefont {C.}~\bibnamefont {Tang}}, \bibinfo {author} {\bibfnamefont
  {T.}~\bibnamefont {Wang}}, \bibinfo {author} {\bibfnamefont {F.}~\bibnamefont
  {Wu}}, \bibinfo {author} {\bibfnamefont {T.}~\bibnamefont {Xia}}, \bibinfo
  {author} {\bibfnamefont {W.}~\bibnamefont {Yu}}, \bibinfo {author}
  {\bibfnamefont {F.}~\bibnamefont {Zhang}}, \bibinfo {author} {\bibfnamefont
  {G.}~\bibnamefont {Zhang}}, \bibinfo {author} {\bibfnamefont
  {X.}~\bibnamefont {Zhang}}, \bibinfo {author} {\bibfnamefont
  {J.}~\bibnamefont {Zhou}}, \bibinfo {author} {\bibfnamefont {X.}~\bibnamefont
  {Zhu}}, \bibinfo {author} {\bibfnamefont {Y.}~\bibnamefont {Shi}}, \bibinfo
  {author} {\bibfnamefont {J.}~\bibnamefont {Chen}}, \bibinfo {author}
  {\bibfnamefont {H.-H.}\ \bibnamefont {Zhao}},\ and\ \bibinfo {author}
  {\bibfnamefont {C.}~\bibnamefont {Deng}},\ }\bibfield  {title} {\bibinfo
  {title} {Fluxonium: An alternative qubit platform for high-fidelity
  operations},\ }\href {https://doi.org/10.1103/PhysRevLett.129.010502}
  {\bibfield  {journal} {\bibinfo  {journal} {Phys. Rev. Lett.}\ }\textbf
  {\bibinfo {volume} {129}},\ \bibinfo {pages} {010502} (\bibinfo {year}
  {2022})}\BibitemShut {NoStop}%
\bibitem [{\citenamefont {Lin}\ \emph {et~al.}(2018)\citenamefont {Lin},
  \citenamefont {Nguyen}, \citenamefont {Grabon}, \citenamefont {San~Miguel},
  \citenamefont {Pankratova},\ and\ \citenamefont
  {Manucharyan}}]{Lin2018Demonstration}%
  \BibitemOpen
  \bibfield  {author} {\bibinfo {author} {\bibfnamefont {Y.-H.}\ \bibnamefont
  {Lin}}, \bibinfo {author} {\bibfnamefont {L.~B.}\ \bibnamefont {Nguyen}},
  \bibinfo {author} {\bibfnamefont {N.}~\bibnamefont {Grabon}}, \bibinfo
  {author} {\bibfnamefont {J.}~\bibnamefont {San~Miguel}}, \bibinfo {author}
  {\bibfnamefont {N.}~\bibnamefont {Pankratova}},\ and\ \bibinfo {author}
  {\bibfnamefont {V.~E.}\ \bibnamefont {Manucharyan}},\ }\bibfield  {title}
  {\bibinfo {title} {Demonstration of protection of a superconducting qubit
  from energy decay},\ }\href {https://doi.org/10.1103/PhysRevLett.120.150503}
  {\bibfield  {journal} {\bibinfo  {journal} {Phys. Rev. Lett.}\ }\textbf
  {\bibinfo {volume} {120}},\ \bibinfo {pages} {150503} (\bibinfo {year}
  {2018})}\BibitemShut {NoStop}%
\bibitem [{\citenamefont {Somoroff}\ \emph {et~al.}(2021)\citenamefont
  {Somoroff}, \citenamefont {Ficheux}, \citenamefont {Mencia}, \citenamefont
  {Xiong}, \citenamefont {Kuzmin},\ and\ \citenamefont
  {Manucharyan}}]{somoroff2021millisecond}%
  \BibitemOpen
  \bibfield  {author} {\bibinfo {author} {\bibfnamefont {A.}~\bibnamefont
  {Somoroff}}, \bibinfo {author} {\bibfnamefont {Q.}~\bibnamefont {Ficheux}},
  \bibinfo {author} {\bibfnamefont {R.~A.}\ \bibnamefont {Mencia}}, \bibinfo
  {author} {\bibfnamefont {H.}~\bibnamefont {Xiong}}, \bibinfo {author}
  {\bibfnamefont {R.~V.}\ \bibnamefont {Kuzmin}},\ and\ \bibinfo {author}
  {\bibfnamefont {V.~E.}\ \bibnamefont {Manucharyan}},\ }\bibfield  {title}
  {\bibinfo {title} {Millisecond coherence in a superconducting qubit},\ }\href
  {https://arxiv.org/abs/2103.08578} {\bibfield  {journal} {\bibinfo  {journal}
  {arXiv:2103.08578}\ } (\bibinfo {year} {2021})}\BibitemShut {NoStop}%
\bibitem [{\citenamefont {Huang}\ \emph {et~al.}(2018)\citenamefont {Huang},
  \citenamefont {Lu}, \citenamefont {Kapit}, \citenamefont {Schuster},\ and\
  \citenamefont {Koch}}]{Huang2018Universal}%
  \BibitemOpen
  \bibfield  {author} {\bibinfo {author} {\bibfnamefont {Z.}~\bibnamefont
  {Huang}}, \bibinfo {author} {\bibfnamefont {Y.}~\bibnamefont {Lu}}, \bibinfo
  {author} {\bibfnamefont {E.}~\bibnamefont {Kapit}}, \bibinfo {author}
  {\bibfnamefont {D.~I.}\ \bibnamefont {Schuster}},\ and\ \bibinfo {author}
  {\bibfnamefont {J.}~\bibnamefont {Koch}},\ }\bibfield  {title} {\bibinfo
  {title} {Universal stabilization of single-qubit states using a tunable
  coupler},\ }\href {https://doi.org/10.1103/PhysRevA.97.062345} {\bibfield
  {journal} {\bibinfo  {journal} {Phys. Rev. A}\ }\textbf {\bibinfo {volume}
  {97}},\ \bibinfo {pages} {062345} (\bibinfo {year} {2018})}\BibitemShut
  {NoStop}%
\bibitem [{\citenamefont {Groszkowski}\ and\ \citenamefont
  {Koch}(2021)}]{groszkowski2021scqubits}%
  \BibitemOpen
  \bibfield  {author} {\bibinfo {author} {\bibfnamefont {P.}~\bibnamefont
  {Groszkowski}}\ and\ \bibinfo {author} {\bibfnamefont {J.}~\bibnamefont
  {Koch}},\ }\bibfield  {title} {\bibinfo {title} {Scqubits: a python package
  for superconducting qubits},\ }\href
  {https://doi.org/10.22331/q-2021-11-17-583} {\bibfield  {journal} {\bibinfo
  {journal} {Quantum}\ }\textbf {\bibinfo {volume} {5}},\ \bibinfo {pages}
  {583} (\bibinfo {year} {2021})}\BibitemShut {NoStop}%
\bibitem [{\citenamefont {Lu}\ \emph {et~al.}(2017)\citenamefont {Lu},
  \citenamefont {Chakram}, \citenamefont {Leung}, \citenamefont {Earnest},
  \citenamefont {Naik}, \citenamefont {Huang}, \citenamefont {Groszkowski},
  \citenamefont {Kapit}, \citenamefont {Koch},\ and\ \citenamefont
  {Schuster}}]{Yao2017Universal}%
  \BibitemOpen
  \bibfield  {author} {\bibinfo {author} {\bibfnamefont {Y.}~\bibnamefont
  {Lu}}, \bibinfo {author} {\bibfnamefont {S.}~\bibnamefont {Chakram}},
  \bibinfo {author} {\bibfnamefont {N.}~\bibnamefont {Leung}}, \bibinfo
  {author} {\bibfnamefont {N.}~\bibnamefont {Earnest}}, \bibinfo {author}
  {\bibfnamefont {R.~K.}\ \bibnamefont {Naik}}, \bibinfo {author}
  {\bibfnamefont {Z.}~\bibnamefont {Huang}}, \bibinfo {author} {\bibfnamefont
  {P.}~\bibnamefont {Groszkowski}}, \bibinfo {author} {\bibfnamefont
  {E.}~\bibnamefont {Kapit}}, \bibinfo {author} {\bibfnamefont
  {J.}~\bibnamefont {Koch}},\ and\ \bibinfo {author} {\bibfnamefont {D.~I.}\
  \bibnamefont {Schuster}},\ }\bibfield  {title} {\bibinfo {title} {Universal
  stabilization of a parametrically coupled qubit},\ }\href
  {https://doi.org/10.1103/PhysRevLett.119.150502} {\bibfield  {journal}
  {\bibinfo  {journal} {Phys. Rev. Lett.}\ }\textbf {\bibinfo {volume} {119}},\
  \bibinfo {pages} {150502} (\bibinfo {year} {2017})}\BibitemShut {NoStop}%
\bibitem [{\citenamefont {Pop}\ \emph {et~al.}(2014)\citenamefont {Pop},
  \citenamefont {Geerlings}, \citenamefont {Catelani}, \citenamefont
  {Schoelkopf}, \citenamefont {Glazman},\ and\ \citenamefont
  {Devoret}}]{pop2014coherent}%
  \BibitemOpen
  \bibfield  {author} {\bibinfo {author} {\bibfnamefont {I.~M.}\ \bibnamefont
  {Pop}}, \bibinfo {author} {\bibfnamefont {K.}~\bibnamefont {Geerlings}},
  \bibinfo {author} {\bibfnamefont {G.}~\bibnamefont {Catelani}}, \bibinfo
  {author} {\bibfnamefont {R.~J.}\ \bibnamefont {Schoelkopf}}, \bibinfo
  {author} {\bibfnamefont {L.~I.}\ \bibnamefont {Glazman}},\ and\ \bibinfo
  {author} {\bibfnamefont {M.~H.}\ \bibnamefont {Devoret}},\ }\bibfield
  {title} {\bibinfo {title} {Coherent suppression of electromagnetic
  dissipation due to superconducting quasiparticles},\ }\href
  {https://doi.org/10.1038/nature13017} {\bibfield  {journal} {\bibinfo
  {journal} {Nature}\ }\textbf {\bibinfo {volume} {508}},\ \bibinfo {pages}
  {369} (\bibinfo {year} {2014})}\BibitemShut {NoStop}%
\bibitem [{\citenamefont {Smith}\ \emph {et~al.}(2020)\citenamefont {Smith},
  \citenamefont {Kou}, \citenamefont {Xiao}, \citenamefont {Vool},\ and\
  \citenamefont {Devoret}}]{smith2020superconducting}%
  \BibitemOpen
  \bibfield  {author} {\bibinfo {author} {\bibfnamefont {W.}~\bibnamefont
  {Smith}}, \bibinfo {author} {\bibfnamefont {A.}~\bibnamefont {Kou}}, \bibinfo
  {author} {\bibfnamefont {X.}~\bibnamefont {Xiao}}, \bibinfo {author}
  {\bibfnamefont {U.}~\bibnamefont {Vool}},\ and\ \bibinfo {author}
  {\bibfnamefont {M.}~\bibnamefont {Devoret}},\ }\bibfield  {title} {\bibinfo
  {title} {Superconducting circuit protected by two-cooper-pair tunneling},\
  }\href {https://doi.org/10.1038/s41534-019-0231-2} {\bibfield  {journal}
  {\bibinfo  {journal} {npj Quantum Information}\ }\textbf {\bibinfo {volume}
  {6}},\ \bibinfo {pages} {8} (\bibinfo {year} {2020})}\BibitemShut {NoStop}%
\bibitem [{\citenamefont {Wang}\ \emph {et~al.}(2019)\citenamefont {Wang},
  \citenamefont {Shankar}, \citenamefont {Minev}, \citenamefont
  {Campagne-Ibarcq}, \citenamefont {Narla},\ and\ \citenamefont
  {Devoret}}]{Wang2019Cavity}%
  \BibitemOpen
  \bibfield  {author} {\bibinfo {author} {\bibfnamefont {Z.}~\bibnamefont
  {Wang}}, \bibinfo {author} {\bibfnamefont {S.}~\bibnamefont {Shankar}},
  \bibinfo {author} {\bibfnamefont {Z.~K.}\ \bibnamefont {Minev}}, \bibinfo
  {author} {\bibfnamefont {P.}~\bibnamefont {Campagne-Ibarcq}}, \bibinfo
  {author} {\bibfnamefont {A.}~\bibnamefont {Narla}},\ and\ \bibinfo {author}
  {\bibfnamefont {M.~H.}\ \bibnamefont {Devoret}},\ }\bibfield  {title}
  {\bibinfo {title} {Cavity attenuators for superconducting qubits},\ }\href
  {https://doi.org/10.1103/PhysRevApplied.11.014031} {\bibfield  {journal}
  {\bibinfo  {journal} {Phys. Rev. Applied}\ }\textbf {\bibinfo {volume}
  {11}},\ \bibinfo {pages} {014031} (\bibinfo {year} {2019})}\BibitemShut
  {NoStop}%
\end{thebibliography}%

\end{document}